\documentclass[%
onecolumn,
nobibnotes,
bibnotes,
 amsmath,
 amssymb,
aps,
pra,
11pt,
longbibliography,
]{revtex4-1}

\usepackage{amsmath,amsfonts}
\usepackage{graphicx}
\usepackage{epstopdf, epsfig}
\usepackage[english]{babel}
\usepackage[utf8]{inputenc}
\usepackage{xcolor}
\usepackage{float}

\usepackage{bbold}
\usepackage{bbm}
\usepackage[colorlinks=true, allcolors=blue]{hyperref}
\usepackage{float}
\usepackage{caption}
\usepackage{comment}
\usepackage{booktabs}
\usepackage{multirow}
\usepackage{tikz}
\usetikzlibrary{arrows, patterns}
\tikzset{>=latex}

\providecommand\bnabla{\boldsymbol{\nabla}}
\providecommand\xx{\mathbf{x}}

\providecommand\tvv{\tilde{\mathbf{v}}}

\providecommand\uu{\mathbf{u}}
\providecommand\ouu{\overline{\mathbf{u}}}

\providecommand\tuu{\tilde{\mathbf{u}}}

\providecommand\tuua{\tilde{\mathbf{u}}^{\dagger}}

\providecommand\op{\overline{p}}
\providecommand\tp{\tilde{p}}
\providecommand\tpa{\tilde{p}^{\dagger}}

\providecommand\qq{\mathbf{q}}

\providecommand\tqq{\tilde{\mathbf{q}}}

\providecommand\tnu{\tilde{\nu}}

\providecommand\tnua{\tilde{\nu}^{\dagger}}

\providecommand\off{\overline{\mathbf{f}}}

\providecommand\tff{\tilde{\mathbf{f}}}

\providecommand\tffx{\tilde{\mathbf{f}}_{\mathbf{u}}}

\providecommand\tffnu{\tilde{{f}}_{\tilde{\nu}}}

\providecommand\omm{\overline{\mathbf{m}}}
\providecommand\tmm{\tilde{\mathbf{m}}}

\providecommand\tFF{\tilde{\mathbf{F}}}
\providecommand\tGG{\tilde{\mathbf{G}}}

\begin{document}

\author{Lucas Franceschini}
\author{Denis Sipp}
\author{Olivier Marquet}
\affiliation{ONERA/DAAA, Universit\'{e} Paris Saclay, 8 rue des Vertugadins, 92190 Meudon, France}

\title{Mean-flow Data Assimilation based on minimal correction of turbulence models: application to turbulent high-Reynolds number backward-facing step}

\date{\today}

\begin{abstract}
In this article, we provide a methodology to reconstruct high-Reynolds number turbulent mean-flows from few time-averaged measurements. A turbulent flow over a backward-facing step at $Re=28275$ is considered to illustrate the potential of the approach. The data-assimilation procedure, based on a variational approach, consists in correcting a given baseline model by tuning space-dependent source terms such that the corresponding solution matches available measurements (obtained here from direct-numerical simulations). The baseline model chosen here consists in Reynolds-Averaged-Navier--Stokes (RANS) equations closed with the turbulence Spalart--Allmaras model.  We investigate two possible tuning functions: a source term in the momentum equations, which is able to compensate for the deficiencies in the modeling of the Reynolds stresses by the Boussinesq approximation and a source term in the turbulence equation, which modifies the balance between the eddy-viscosity production and dissipation.
The quality of the mean-flow reconstruction strongly depends on the baseline model and on the quantity of measurements.
In the case of many measurements, very accurate reconstructions of the mean-flow are obtained with the model corrected by the source term in the momentum equations, while the reconstruction is more approximate when tuning the source term in the turbulence model. In the case of few measurements, this ``rigidity" of the corrected turbulence model is favourably used and allows the best  mean-flow reconstruction. The flexibility / rigidity of a model is further discussed in the light of a singular-value decomposition of the linear input / output operator between source term and measurements.
\end{abstract}

\maketitle

\section{Introduction}

Numerical simulations for turbulent aerodynamic flows are widely used in the everyday life of engineers and researchers. However, those techniques are prone to errors for several reasons: inaccurate turbulence models, erroneous numerical setup (including geometry and boundary conditions), imprecise upstream turbulence intensities, over-diffusive numerical methods, among others. For those reasons, experimental results are commonly used to validate numerical simulations. Yet, experimental techniques are also prone to uncertainties caused by different reasons, for example measurement biases, defects on the physical model or wind-tunnel (deformations, vibrations, inhomogeneous upstream conditions). 

Data assimilation aims at intimately combining experiment results and numerical simulations to produce a compromise, an improved picture, by tuning uncertain parameters in numerical simulations to match measurement data within experimental uncertainty bounds.
The compromise is driven by a balance between measurement accuracy (measurement covariance matrix) and model uncertainties (state covariance matrix).
Several approaches have been designed to efficiently determine such a compromise. These can be classified into two categories: gradient-based optimization techniques, also called 3D/4D Var, where uncertain parameters in the numerical model are optimized to minimize a cost-functional involving measurement mismatch and model accuracy \citep{LeDimet86}; ensemble based-techniques, where ensembles of members representing uncertainties are propagated with the model and corrected using measurements \citep{Evensen09}.

Data assimilation historically emerged in the field of meteorological forecasts (\citet{Lorenc86}, \citet{Liu08}), where the lack of accurate models, uncertain initial and boundary conditions yield poor predictions. In other areas of fluid mechanics, motivations for the use of data-assimilation techniques were multifold: estimation of 
initial or inlet boundary conditions in open flows \citep{Harader13,Mons16}, interpolation of velocity fields between sequences of images \citep{Heitz10,Yang15}, identification of pollutant release location in urban areas \citep{Mons17} and even as a theoretical tool to investigate aspects of the decay of large scales in homogeneous isotropic turbulence \citep{Mons14}. 
In aerodynamics, predicting time-averaged quantities is the predominant interest in many industrial applications. Indeed, most of the measurements performed in industrial wind-tunnels are time-averaged pressure distributions and forces. Furthermore, the numerical simulations of turbulent flows over complex geometries are mostly achieved with the Reynolds--Averaged Navier--Stokes (RANS) equations and turbulence models, thus promoting the low computational cost over the accuracy.
In so far, we will therefore favour a time-averaged approach based on RANS equations and the data assimilation problem should rather be considered in the framework of inverse problems 
(see \citet{Foures14,Symon17} for the gradient-based optimization approach and \citet{iglesias2013ensemble,kato2013approach} for the ensemble-based framework).
Concerning the gradient-based optimization approach, \citet{Foures14} started with a low-Reynolds cylinder flow ($Re=O(10^2)$) exhibiting vortex shedding. They tuned a volume-force (modeling the force associated to the Reynolds-stress) acting in the steady Navier--Stokes equations such that its corresponding solution best matches velocity measurements, mimicking a real experimental situation where such measurements are provided by a Particle Image Velocimetry setup.
The same procedure was applied by \citet{Symon17} to reconstruct the mean-flow around an idealized airfoil at a higher Reynolds number ($Re=O(10^4)$). This time, additional difficulties related to the well-posedness of the steady Navier--Stokes equations at such high Reynolds numbers had to be faced.
At even higher Reynolds numbers, {the RANS equations supplemented with a turbulence model} is a reasonable choice for the baseline model, since they usually provide solutions that {aim at approximating the turbulent mean flow. For instance}, \citet{Li17} {optimized} a set of coefficients in a $k-\omega$ RANS model to match as closely as possible some given higher-fidelity-data.
Yet, such an approach is strongly constrained by the structure of the {turbulence} model and does not allow a high-enough flexibility to adjust the model, {especially in flow regions  where such models are known to be deficient. To overcome that limitation,}  \citet{duraisamy2019turbulence} employed gradient-based optimization techniques to tune spatially dependent production terms in {turbulence} models, so as to recover mean-flow data obtained by DNS or experiments. For example, \citet{Singh16} tuned space-dependent functions modulating the strength of the eddy-viscosity production term {in the Spalart--Allmaras model}. 
This procedure, called field-inversion by the authors, has been presented in more details in \citet{parish2016paradigm}. More recently, \citet{He19} applied this technique to reconstruct a mean-flow based on PIV measurement, which was then used to for a resolvent analysis (see also \citet{symon2019tale}). 
Concerning ensemble approaches, \citet{kato2013approach,kato2015data} varied arbitrarily the coefficients of a Spalart--Allmaras model to build the members of an ensemble and a Kalman filtering method then led to the tuning of optimal values of these coefficients with respect to given measurements.

The choice of the baseline RANS model is of importance in data-assimilation. Indeed, the better the model, the smaller the correction. Most of the turbulence models used in industrial design offices model Reynolds stresses with a Boussinesq approximation and an eddy-viscosity (as done in the most common turbulence models). Although this is known to be a strong weakness, the numerical procedure to find solutions with such models is far superior than for the more accurate Reynolds-Stress-Models, which remain only seldomly used today by engineers.
The robustness issue is even more critical for data-assimilation since tuning source term correction functions involves solving stiff nonlinear state equations, that is RANS equations closed with a turbulence model and driven by an additional forcing.

In this article, we extend the time-averaged gradient-based optimization approach introduced by \citet{Foures14,Symon17}
to higher Reynolds number turbulent flows
by using RANS equations with a turbulence model as baseline model for the data-assimilation procedure.
In the present work, we choose the Spalart--Allmaras model (SA) \citep{Spalart94} for its numerical simplicity and robustness, even though the procedure may be extended to any other turbulence model in principle. We explore two correction possibilities.

The first consists in a volume-force acting in the momentum equations (with the turbulence model still active). This force is supposed to stand for an optimal correction of the stresses induced by the eddy-viscosity term, whose spatial distribution is governed by the SA model. This allows in particular the turbulent stresses to escape the strong constraint linked to the Boussinesq assumption, which is known to be only well adapted in shear dominated regions.
This approach may be considered as an extension of \citet{Foures14} and \citet{Symon17}, the difference being the consideration of a background eddy-viscosity that evolves with the data-assimilation process according to solution of the full corrected model. This modeling is better suited for turbulent flows: for example, the uncorrected solution (for which the correction term is null), which is for example required to initialize the optimization algorithm, now corresponds to the RANS-SA solution. Such a solution may easily be obtained with standard numerical algorithms and is already much closer to the targeted mean-flow than an initial guess based on a laminar model. The assimilation procedure therefore only needs to compensate the {\it expected rather small} imperfections of the turbulence model, while using a laminar model for turbulent flows implies compensating for the whole discrepancy between the laminar steady solution and the targeted mean-flow.

The second correction is a source term acting on the equation governing the strength of the eddy-viscosity. This time the correction only modifies the production and dissipation terms which drive the strength of the eddy-viscosity. Obviously, the Reynolds stresses cannot escape the constraint linked with the Boussinesq-approximation and a smaller set of reachable velocity-pressure fields is therefore expected with this second correction. {Since a reference mean-flow (obtained for instance from DNS) might in fact not be compatible with a Boussinesq constraint, one expects less accurate mean-flow reconstruction using this second baseline model.}  Yet, we will see that such a choice may still have some advantages, both from a numerical point of view (robustness to find solutions) and from a modeling point of view. In particular, a crucial point {when assessing a baseline model for data-assimilation} lies in the quantity of available measurements. Complete mean velocity fields may for example be provided by advanced optical measurement methods, such as Particle-Image-Velocimetry (PIV) measurements.
However, such optical measurements are difficult to implement in industrial wind-tunnel facilities. Therefore, {it is worth} considering only few point-wise velocity measurements such as those given by Pitot probes or hot-wire measurements along a line to extract cross-stream velocity profiles. 
We will see that both the available quantity of measurements and the used correction-term in the baseline model have strong impacts on the quality of the {mean-flow} reconstruction. \citet{Foures14} and \citet{Symon17} have already examined point-wise measurements in the case of a laminar model at lower Reynolds numbers: in particular, \citet{Symon17} showed that point-wise measurements may lead to noisy reconstructions and that a measurement operator based on a spatial averaging of the measure over a finite region could improve the quality of the reconstruction by smoothing the gradients of the reconstructed velocity field in the vicinity of the measurements. We will show that we can handle noisiness associated to point-wise measurements in different ways, either by selecting a good correction term in the model or by penalizing small-scale features in the correction term. {The latter can also be regarded as a covariance matrix modeling uncertainties of the correction term.}.

The article is organised as follows. First (\S \ref{sec:config}), we will describe the physical configuration of interest, by showing the reference solution (obtained by DNS) and the corresponding (uncorrected) RANS-SA solution, pointing out the differences between them and motivating the need for data-assimilation. The flow configuration is a rounded Backward-Facing Step (BFS, \citet{Dandois07}), for which the SA model overestimates the recirculation length. Then (\S \ref{sec:theory}), we will discuss in more details the two above mentioned correction functions together with the gradient-based optimization procedure underlying the data-assimilation procedure. Finally (\S \ref{sec:results}), we will present the data-assimilation results on the backward-facing step configuration, by comparing to the reference mean-flow the assimilated flowfields obtained with the two baseline models, using first dense and then sparse velocity measurements. {The dense velocity measurement case is an important theoretical step when addressing the performance of data-assimilation to reconstruct flow fields. Indeed,  it allows to determine the best reconstructed flow-field for a given correction term, to which the flow reconstructed using sparse data can then be compared. Such ideal situation is of limited interest from an experimental point of view, but is more interesting in data-driven turbulence modeling \citep{duraisamy2019turbulence}. For instance, in the field inversion and machine learning approach proposed by  \citet{Parish16}, \citet{Singh16}, the inverse modeling is first applied on a flow configuration to extract the (spatially) optimal correction terms for a given turbulence model. Machine learning is then used in a second step to transform the corrections terms computed for several flow configurations into into corrective model forms.}. The performance of the reconstruction as a function of the baseline model and measurement-data sparsity is finally discussed in light of an observability Gramian analysis of the linearized baseline model.   
%
\section{Flow Configuration and numerical solutions} \label{sec:config}

\begin{figure}
\centering
\begin{tabular}{ll}
(a) & (b) \\
\includegraphics[trim={1cm 1cm 1cm 7cm},clip,width=7cm]{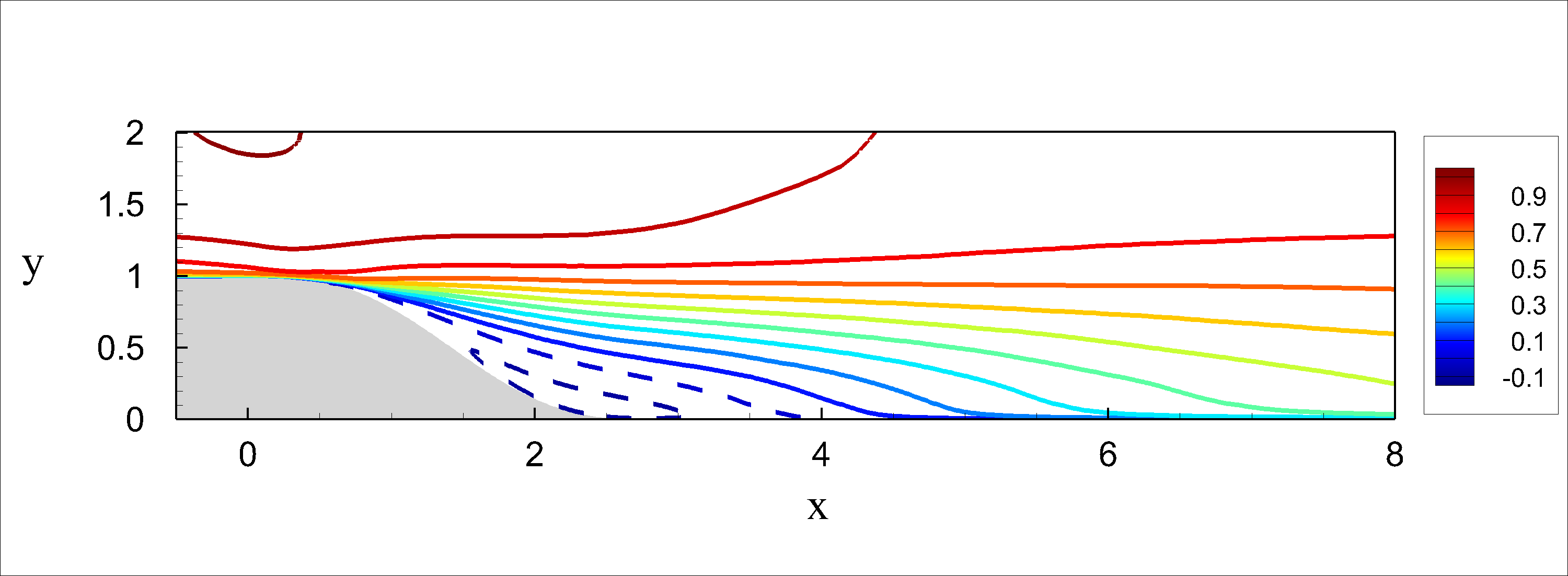}   &  \includegraphics[trim={1cm 1cm 1cm 7cm},clip,width=7cm]{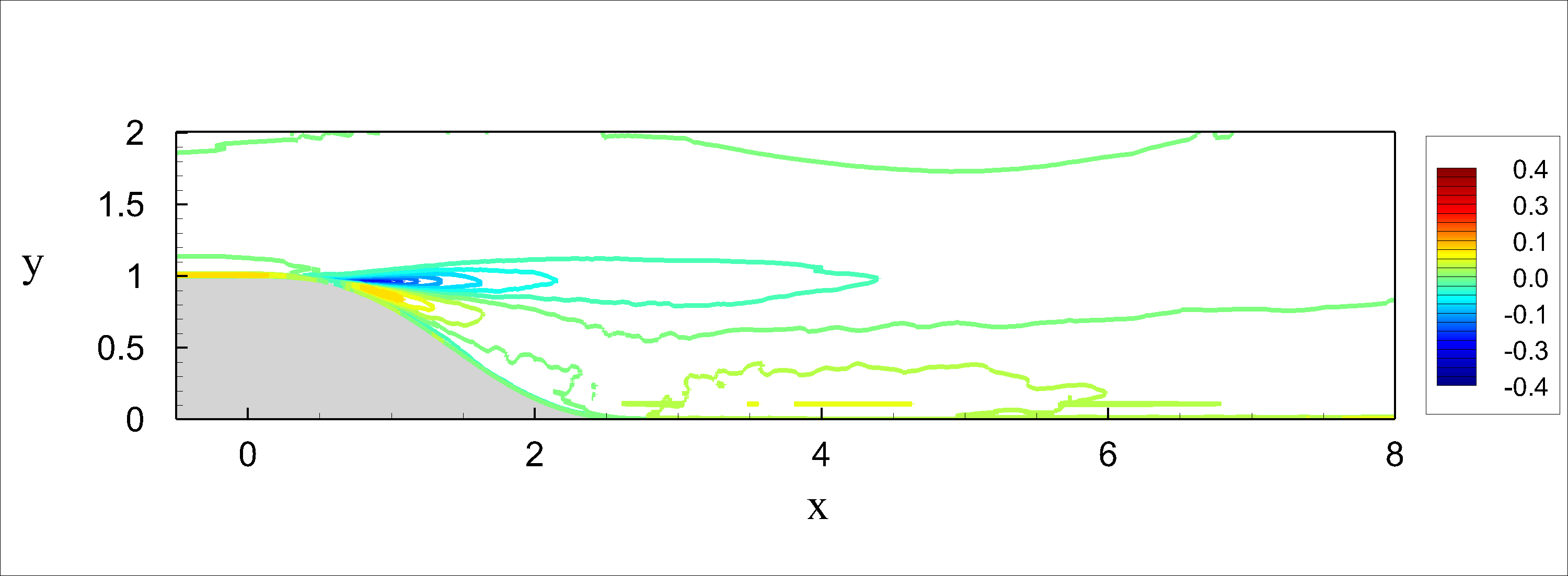} 
\end{tabular}
\begin{tabular}{cc}
(c)    & \\
& \includegraphics[trim={1cm 0.4cm 1cm 0.4cm},clip,width=7cm]{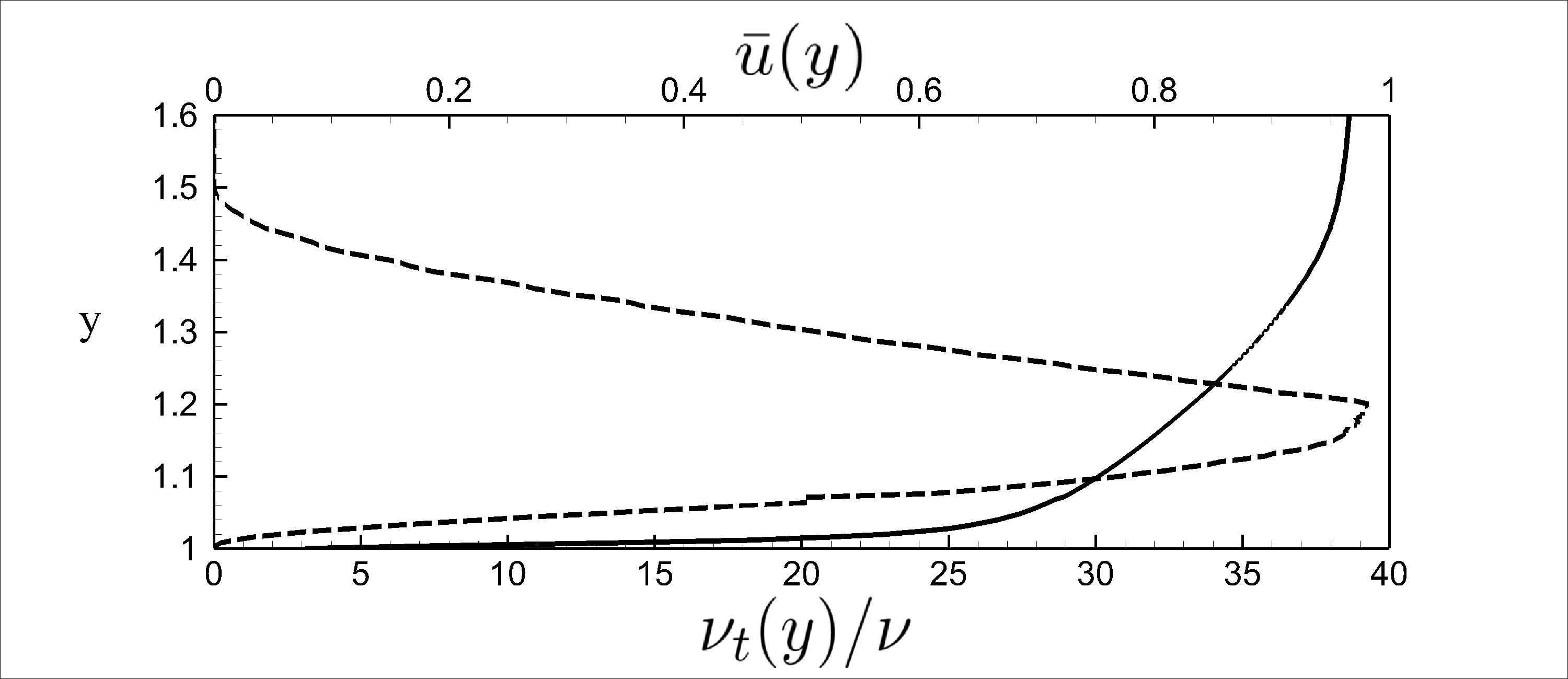}
\end{tabular}
\caption{Reference mean-flow solution obtained from DNS  \citep{Dandois07} at $\textit{Re}=28275$. (a) Streamwise velocity field $\bar{u}$, positive and negative isocontours being displayed with solid and dashed lines, respectively. (b) Streamwise component of the mean force $\bar{f}_{x}$ induced by the Reynolds stress tensor. (c) Profiles of the streamwise velocity $\bar{u} $ (solid lines) and the eddy-viscosity ratio $\nu_{t}/\nu$ (dashed lines) at $x= -2.5$.}\label{fig:DNS}
\end{figure}

We investigate the turbulent flow above a Backward-Facing rounded Step defined by $y(x) = [\sin ( a \pi x ) - a \pi x]/(2 \pi) + 1, \;0 \leq x \leq 2/a$ with $a=0.703$. Direct Numerical Simulations (DNS) and Large Eddy Simulations (LES) were performed by \citet{Dandois07} on this configuration at a Reynolds number $\textit{Re}=28 275$, based on the height of the step and on the inflow velocity. In the following, those quantities are used to make all variables dimensionless. As detailed in \citet{Dandois07}, the Direct Numerical Simulation (DNS) was fed at the inlet boundary with fluctuations generated from a time-dependent zero-pressure-gradient turbulence simulation (see \citet{Lund98}). This allows to describe a situation where the incoming boundary layer is fully turbulent. The streamwise component of the time- and spanwise-averaged velocity field is displayed in Figure \ref{fig:DNS}(a). The negative streamwise velocity (dashed line) indicates the existence of a mean re-circulation region, extending from the separation point of the turbulent boundary layer at $x_d = 0.53$ to the re-attachement point at $x_r = 3.93$. The mean (time- and spanwise-averaged) velocity $\ouu$ and pressure $\op$ fields satisfy the Reynolds-Averaged Navier--Stokes (RANS) equations:
\begin{equation}\label{eqn:RANS}
   \ouu \cdot \nabla \ouu + \nabla \op - \nabla \cdot ( \nu \nabla_s \ouu ) = - \nabla \cdot \underline{\tau} \;\;, \nabla \cdot \ouu = 0,
\end{equation}
where $\nabla_s= \left( \nabla + \nabla^T\right)$ and  $\underline{\tau}=\overline{\uu' \otimes \uu'}$ is the Reynolds-stress tensor that  
contains all nonlinear interactions of the fluctuations $\uu'$. {It induces a force on the mean flow, $\off = - \nabla \cdot \underline{\tau}$, whose streamwise component 
is displayed in figure \ref{fig:DNS}(b).}\\
Turbulence modeling consists in closing the above system of equations by approximating the Reynolds stress as a function of the mean-flow. Among the several models existing in the literature, we will consider the one-equation Spalart--Allmaras model (see \citep{Spalart94}). The mean-flow velocity and pressure $(\tuu,\tp)$, that  approximate the time- and spanwise-averaged flow $(\ouu,\op)$, are then solutions of the following  closed system of equations:
\begin{eqnarray}\label{eqn:RANS-SA}
       \tuu \cdot \nabla \tuu + \nabla \tp - \nabla \cdot \left( (\nu+\nu_t(\tnu)) \nabla_s \tuu \right) & =& \mathbf{0}, \;\;\;\; \nabla \cdot \tuu  = 0    \label{eqn:RANS-SAb} \\
        \tuu \cdot \nabla \tnu - \nabla \cdot ( \eta(\tnu{}) \nabla \tnu ) & =&  s(\tnu,\nabla \tnu, \nabla \tuu).       \label{eqn:SA-tnu}
\end{eqnarray}
The equation on the second line is  the Spalart--Allmaras model, that governs the eddy-viscosity-related variable $\tnu$. Its  diffusivity is denoted $\eta(\tnu)$ and the source term is defined as $s(\tnu,\nabla \tnu, \nabla \tuu) = P(\tnu,\nabla \tuu) + D(\tnu,\nabla \tuu) + C(\nabla \tnu)$, that is the sum of the production $P(\tnu,\nabla \tuu)$,
destruction $D(\tnu,\nabla \tuu)$ and cross diffusion $C(\nabla \tnu)$ terms. More details about the exact definitions of $\tnu$  and these source terms are given in Appendix \ref{apd:SA}. \\

\begin{figure}
\centering
\includegraphics[trim={0.5cm 2.5cm 0.3cm 3.5cm},clip,width=8cm]{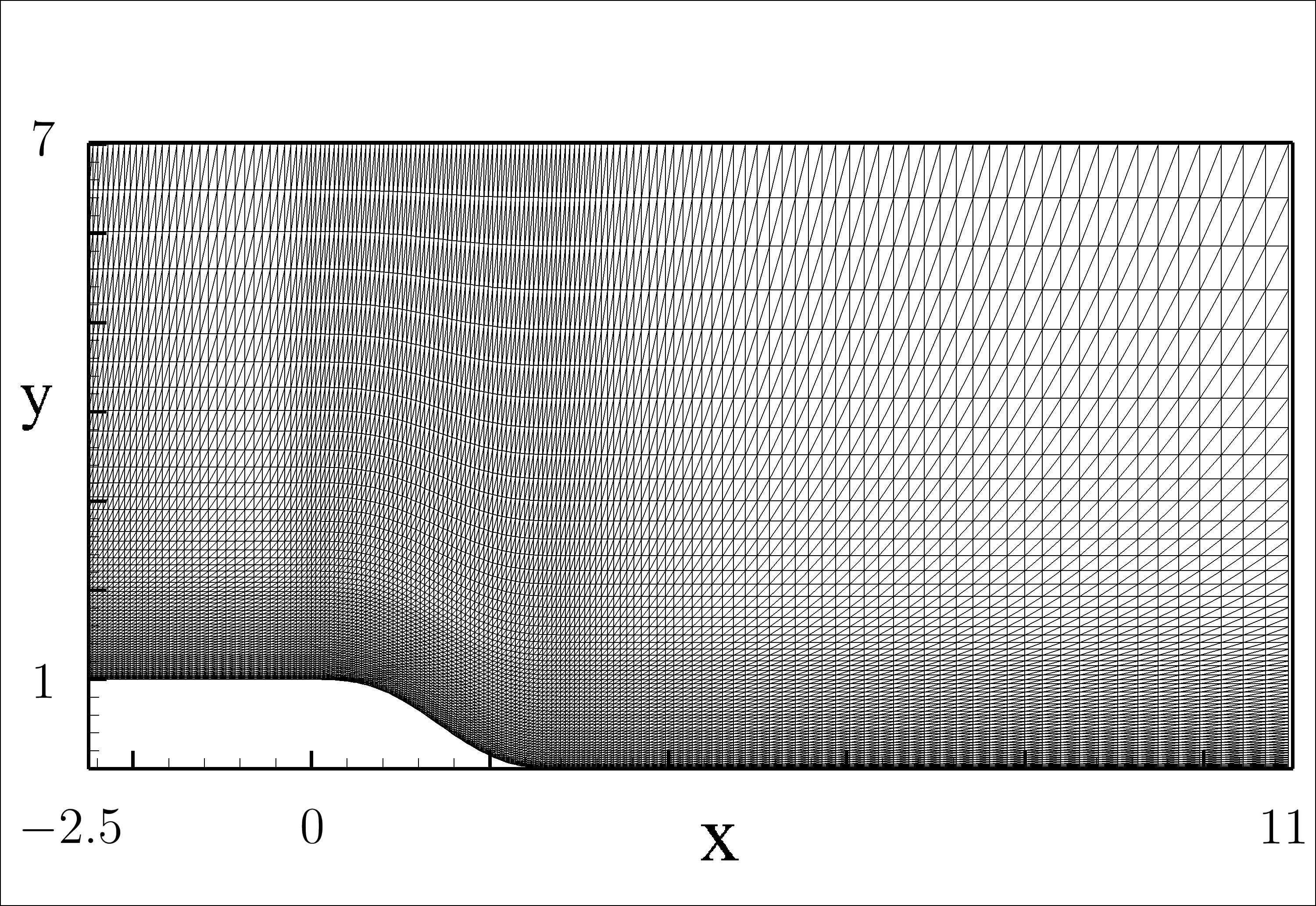}
\caption{Example of (coarsened) mesh used in the present study. The mesh contains 103200 elements and, with a Finite-Element-Discretisation involving $\mathcal{P}^1_b$ elements for the velocity and eddy-viscosity fields and $\mathcal{P}^1$ elements for the pressure, resulting in around 518000 degrees of freedom.}\label{fig:SketchDomain}
\end{figure}

A Streamline-Upwind Petrov-Galerkin (SUPG) finite-element method is used for the spatial discretization of (\ref{eqn:RANS-SAb}) and  (\ref{eqn:SA-tnu}), instead of the classical Galerkin finite-element method used by \citet{Foures14} (see Appendix \ref{apd:SUPG})). For high Reynolds numbers flows, this allows stabilisation of the convection operator. The computational domain is sketched in figure \ref{fig:SketchDomain}. No-slip boundary conditions $\tuu=0$ are imposed at the lower wall, and symmetry boundary conditions $(\partial_{y} \tilde{u}_{x},\tilde{u}_{y},\partial_{y} \tilde{\nu})=(0,0,0)$ at the top boundary located at $y=7$. At the outlet $x=11$, a classical outflow condition $ (\nu+\nu_t) \nabla_s \tuu \cdot \mathbf{n} + \tp \mathbf{n} =0  $ is enforced for the momentum equations while a Neuman boundary condition $\partial_x \tnu=0$ is used for the $\tilde{\nu}$ equation.

\begin{figure}
\centering
\begin{tabular}{cccc}
& baseline REF & & baseline CST \\ 
(a) & \includegraphics[trim={0.5cm 0.5cm 0.5cm 0.5cm},clip,width=7cm]{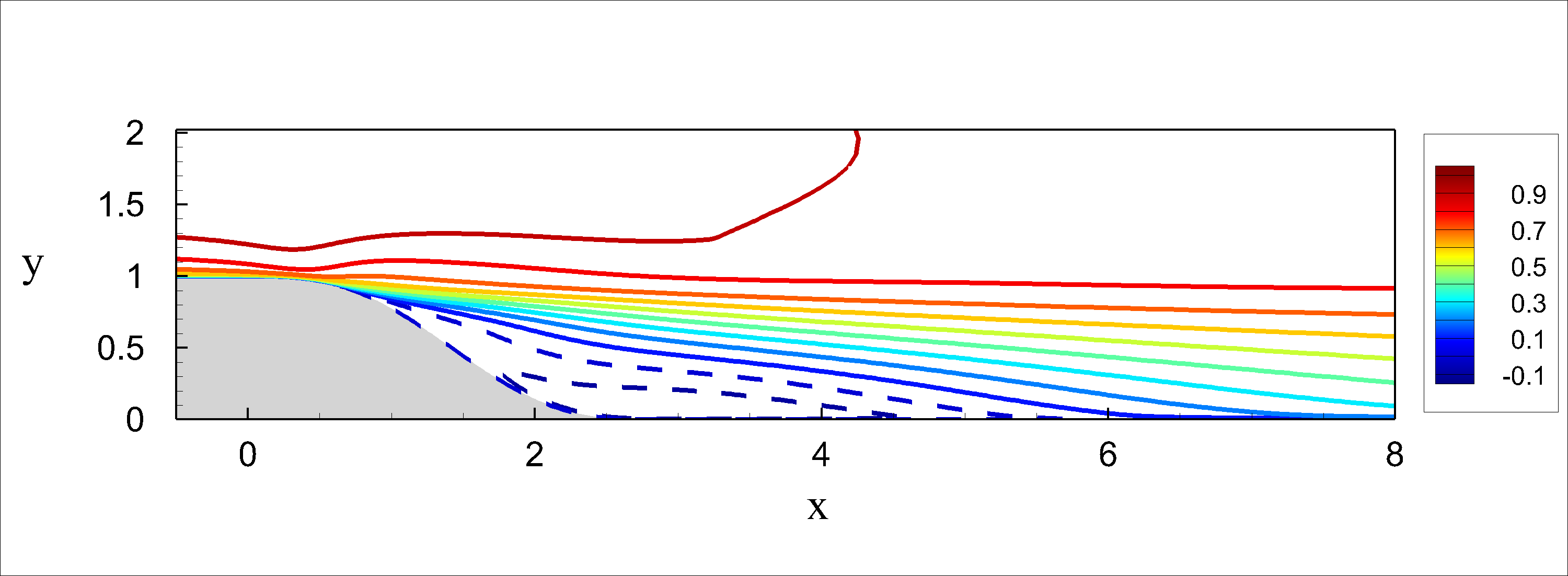} & (b) & \includegraphics[trim={0.5cm 0.5cm 0.5cm 0.5cm},clip,width=7cm]{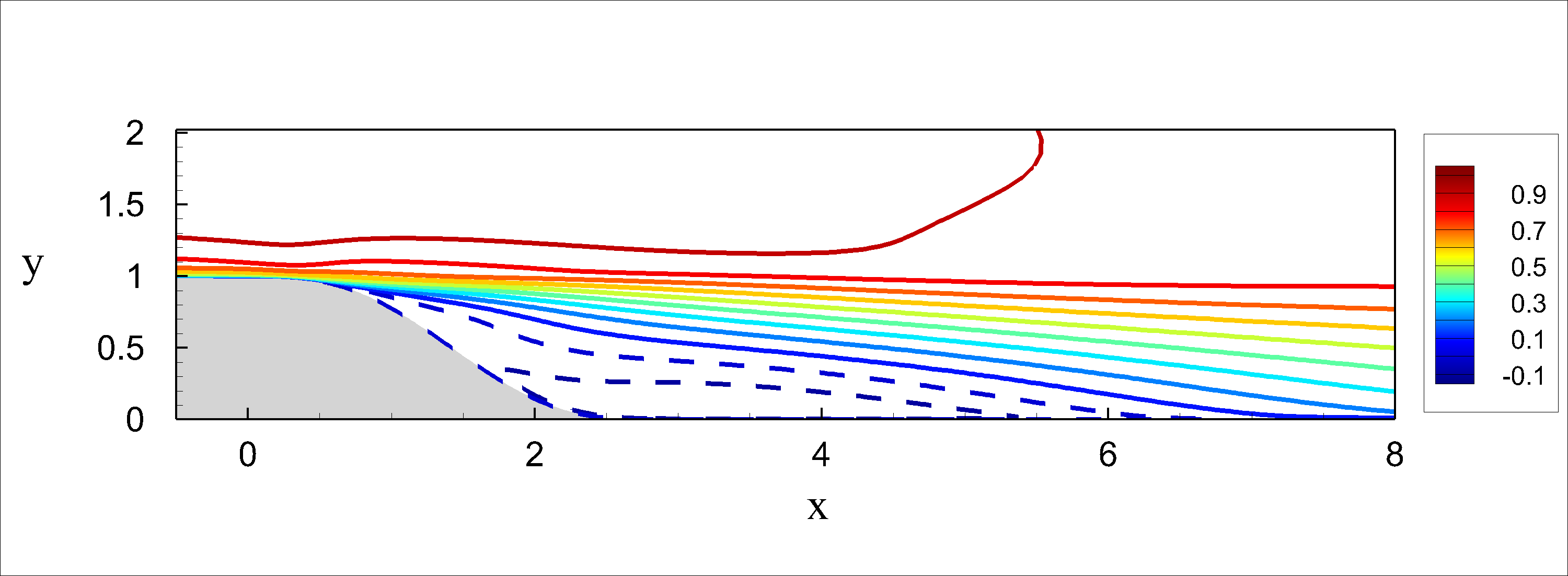} \\
(c) & \includegraphics[trim={0.5cm 0.5cm 0.5cm 0.5cm},clip,width=7cm]{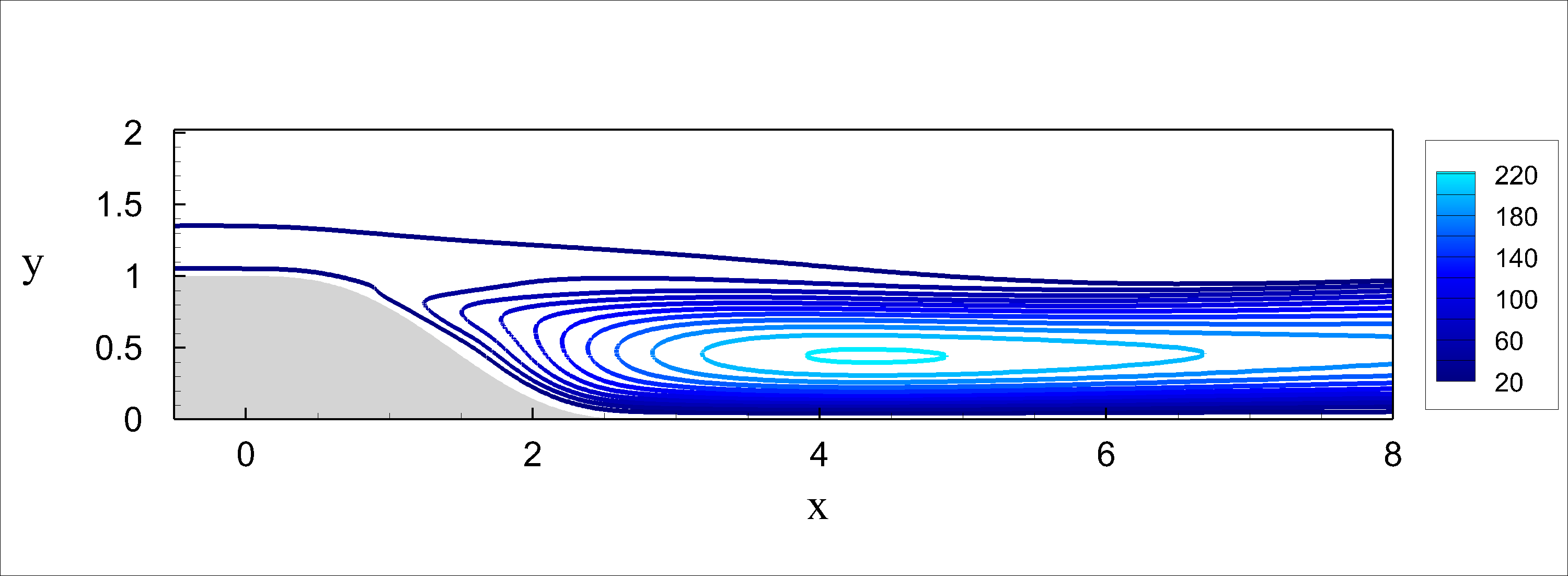} & (d) & \includegraphics[trim={0.5cm 0.5cm 0.5cm 0.5cm},clip,width=7cm]{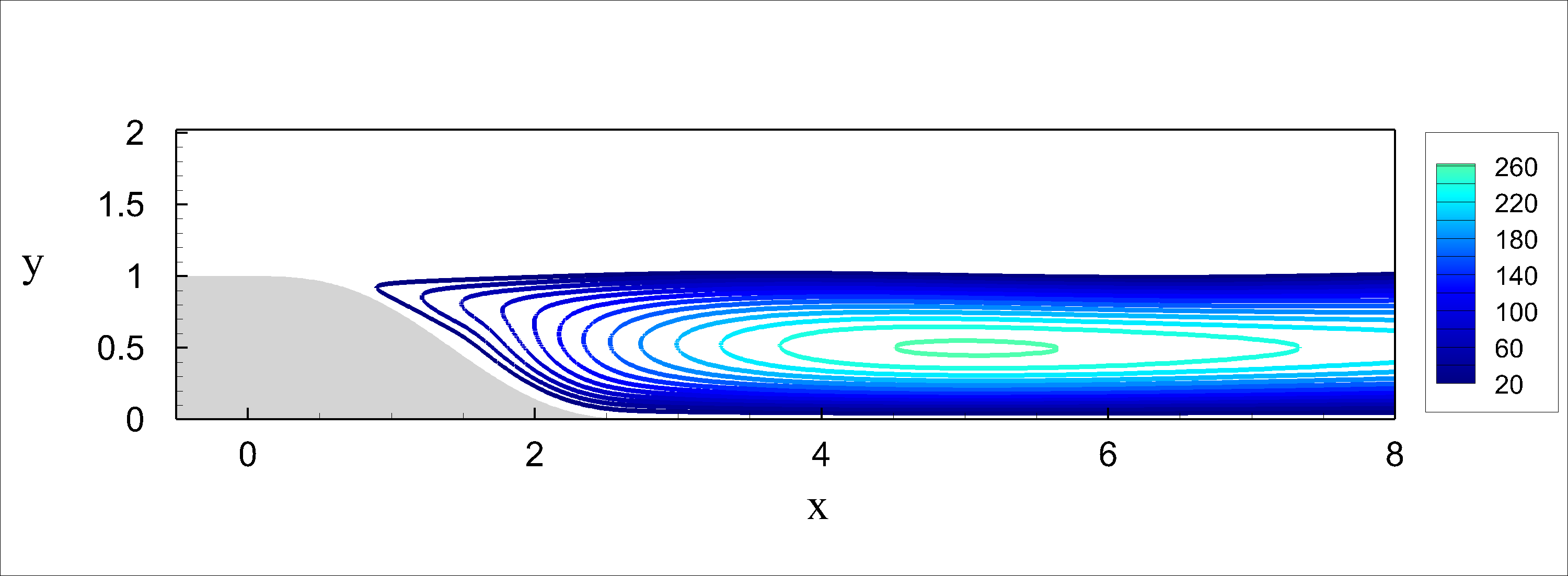} \\
(e) & \includegraphics[trim={0.5cm 0.5cm 0.5cm 0.5cm},clip,width=7cm]{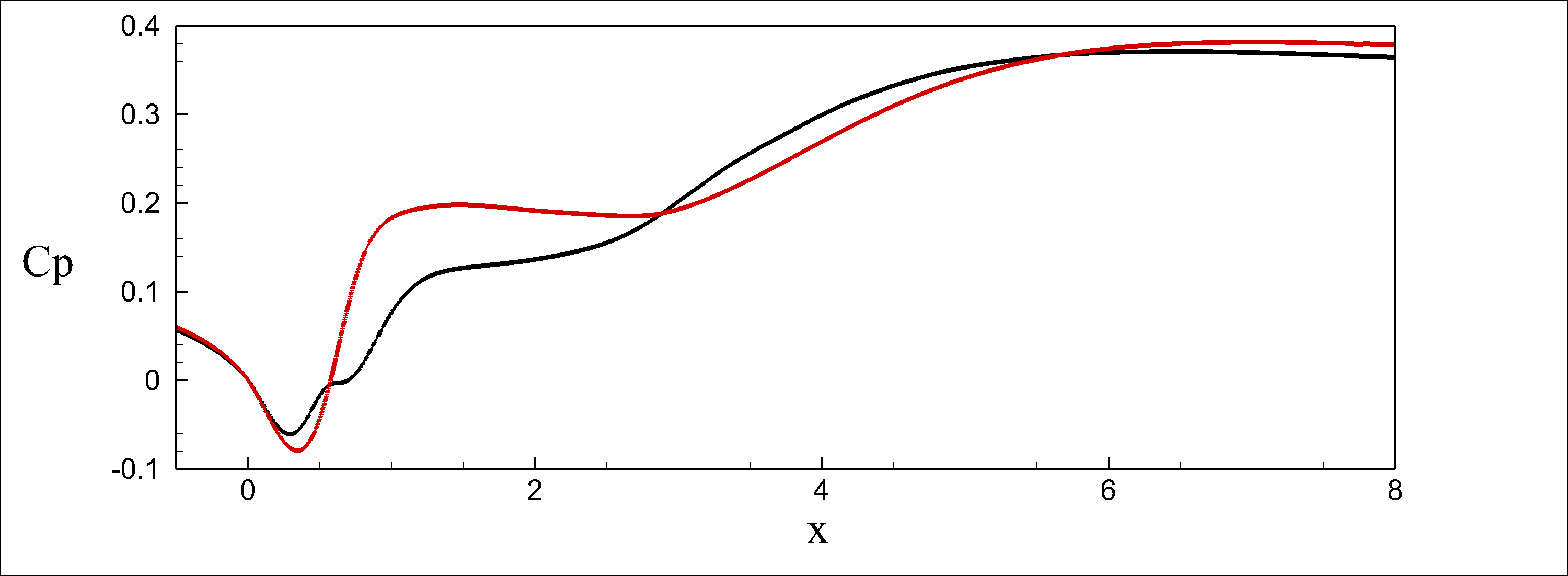} & (f) & \includegraphics[trim={0.5cm 0.5cm 0.5cm 0.5cm},clip,width=7cm]{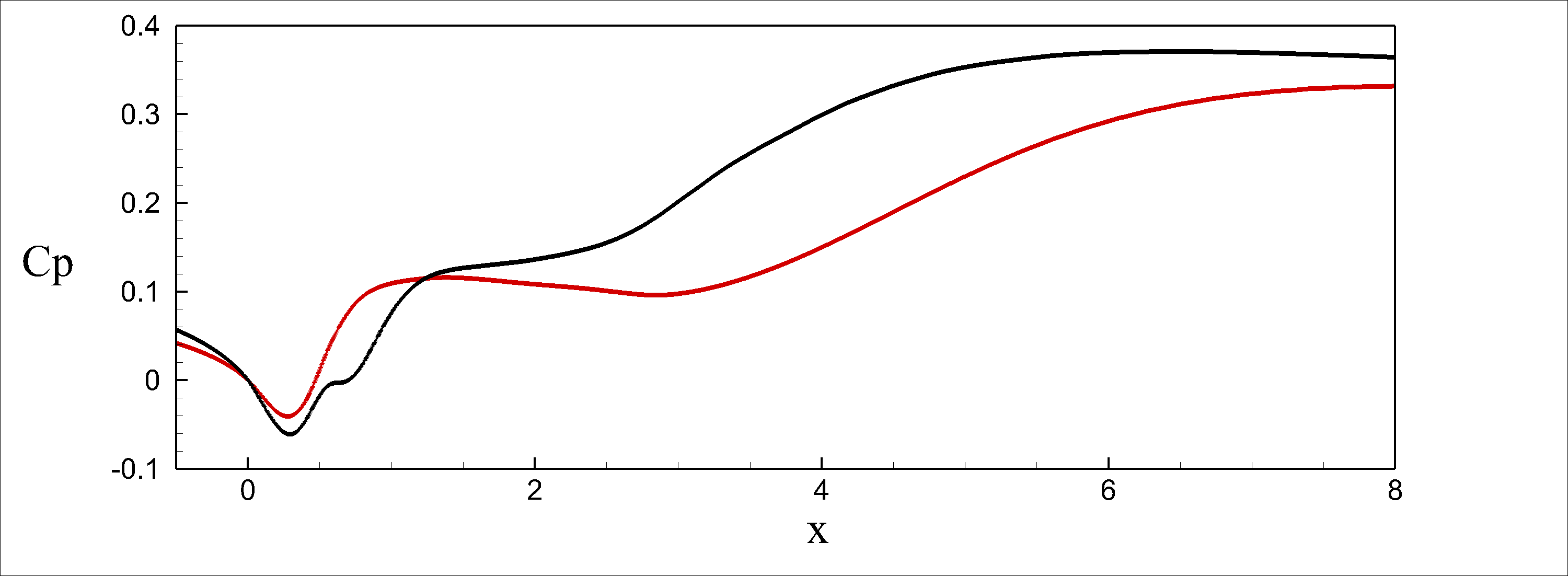} \\
(g) & \includegraphics[trim={0.5cm 0.5cm 0.5cm 0.5cm},clip,width=7cm]{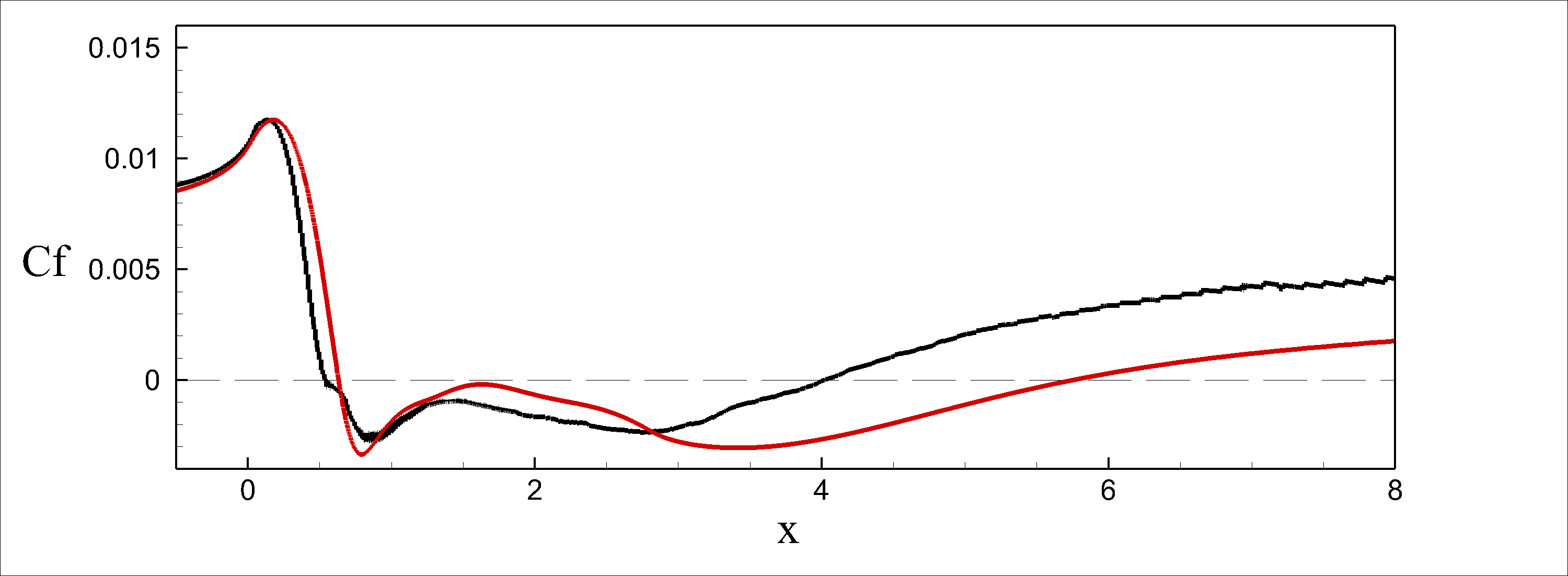} & (h) & \includegraphics[trim={0.5cm 0.5cm 0.5cm 0.5cm},clip,width=7cm]{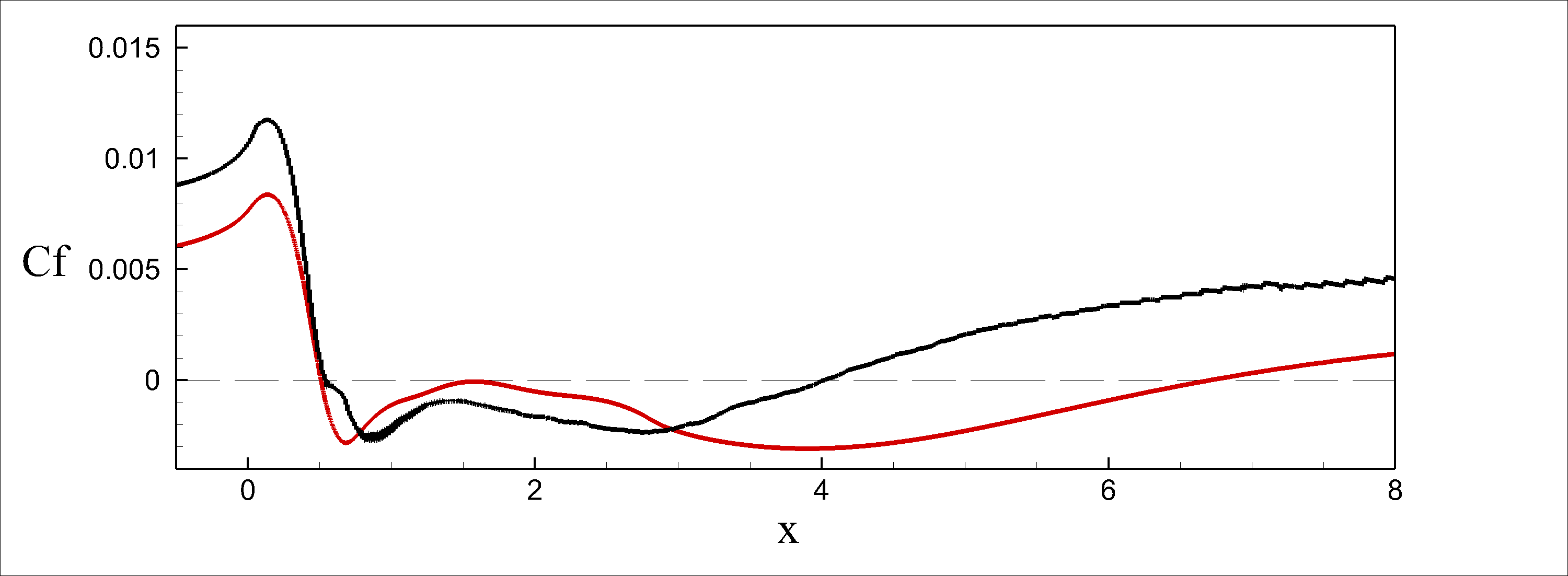} 
\end{tabular}
\caption{Solutions of the (uncorrected) baseline RANS-SA model with two different eddy-viscosity profiles at the inlet.  
Left panels (a,c,e,g) refer to the case of DNS-based eddy-viscosity profile (see figure \ref{fig:DNS}-c) and right panels (b,d,f,h) to the case of constant eddy-viscosity $\nu_t/\nu=3$.
Isocontours in (a,b) depict streamwise velocities $ \bar{u} $ and (c,d) eddy viscosity fields $\nu_t/\nu$.  Panels (e,f) depict pressure $C_p$ and (g,h) friction $C_f$ coefficients along the wall. In (e,f,g,h), the black solid line refers to the reference solution of the DNS, the red solid line to the baseline RANS-SA model.}\label{fig:RANS-SA}
\end{figure}

At the inlet boundary ($x=-2.5$), we impose the turbulent mean velocity profile of the DNS, displayed with the solid line in figure \ref{fig:DNS}(b), and we consider two cases for the profile of the eddy-viscosity-related variable $\tnu$. Firstly, the constant value $\tnu_\nu=(\tnu/\nu)_{\infty} = 3$, recommended in \citet{Allmaras12}, is imposed, with a fast-decay to zero very close to the wall. Secondly, the actual eddy-viscosity of the DNS, displayed with the dashed line in figure \ref{fig:DNS}(c), 
is imposed. It is computed from the DNS statistics as suggested by \citet{mettot2014quasi}. In that case, the maximum value of $ \tnu/\nu $ is close to 40, which should result in filling the streamwise velocity profiles close to the wall and therefore increasing skin friction.  The solutions of the RANS-SA model obtained with those two inflow eddy-viscosity profiles are compared in figure \ref{fig:RANS-SA}. We can see that, with the constant eddy-viscosity profile, the RANS solution presents a very large recirculation region ($x_r\approx6.6$) whereas, for the true eddy-viscosity profile (exhibiting much stronger values), the bubble is (accordingly much) shorter ($x_r \approx 5.8$), but still much longer than the reference value $ x_r=3.93 $ of the DNS. This discrepancy is made more quantitative by examining the streamwise evolution of the wall-pressure and friction coefficients shown in figures \ref{fig:RANS-SA} (e,f), respectively. It is seen that the skin-friction coefficient in the attached boundary layer is well predicted in the case of the reference eddy-viscosity profile, and underestimated in the other case.  With the reference eddy-viscosity profile, the incoming eddy-viscosity values at the step location are higher than with the constant eddy-viscosity profile (see figures \ref{fig:RANS-SA} (c,d)). Hence, the shear-layer undergoes stronger diffusion and the bubble shortens. 
The discrepancies between the mean-flow results obtained with the baseline RANS-SA model and the reference solution motivate the use of data-assimilation techniques to recover the reference solution by tuning a small correction term in the baseline model. In this study, we will consider the time-averaged DNS as the reference solution from which measurements will be extracted to feed the assimilation process. The RANS-SA model with the two different inlet boundary conditions will be used as baseline models in the data-assimilation procedure, and we  will solely consider turbulence modeling defects as an uncertain parameter. The inflow profiles, that could be viewed as uncertain parameters, will not be inferred in the present study.

\section{Flow models and Data assimilation procedure}
\label{sec:theory}

In this section, after introducing and discussing the two source term corrections in the baseline model (\S \ref{sec:correc}), we present in some details the data-assimilation procedure (\S \ref{sec:procedure}). With respect to \citet{Foures14}, we use an efficient low-memory Broyden–Fletcher–Goldfarb–Shanno (BFGS) algorithm to find the minimum of the cost-functional from sole knowledge of the cost function and its gradient.

\subsection{Corrections of the RANS-SA baseline model} \label{sec:correc}

The baseline RANS-SA models with the two correction terms that will be considered in the following may be written as
\begin{eqnarray}
       \tuu \cdot \nabla \tuu + \nabla \tp & = & \nabla \cdot \left( (\nu+\nu_t(\tnu)) \nabla_s \tuu \right) + \tffx, \;\;\;\nabla \cdot \tuu  = 0    \label{eqn:corrected-RANS-SAb} \\
        \tuu \cdot \nabla \tnu - \nabla \cdot ( \eta(\tnu) \nabla \tnu ) &=& s(\tnu,\nabla \tnu, \nabla \tuu) + \tffnu,      \label{eqn:corrected-SA-tnu}
\end{eqnarray}
with the boundary conditions discussed above. $\tffx$ and $\tffnu$ are the two spatially-dependent source terms introduced to  correct the RANS-SA model. In the following we either consider $\tffx \neq 0$ and $\tffnu=0$ or vice-versa. The volume-force $\tffx$ acts on the momentum equations without directly modifying the production of eddy-viscosity. Such a volume-force allows the Reynolds stresses to escape the Boussinesq constraint.
The volume-force $\tffnu$ acts on the equation governing the turbulent variable, without modifying the momentum equations.
By modifying the balance between the production, destruction and cross-diffusion terms in the Spalart--Allmaras, this term modifies the eddy-viscosity $ \nu_t$ that appears in the momentum equations \eqref{eqn:corrected-RANS-SAb}, and has thus only an indirect effect (via $ \nu_t $) on the velocity $\tuu$ and pressure $\tp$ fields.
The Reynolds stress tensor therefore remains constrained by the Boussinesq assumption, which is valid in shear-dominated flows. This model is sufficiently flexible to allow for example for correction of a wrong free shear-layer expansion ratio, as is the case in figure \ref{fig:RANS-SA}.

As discussed below, the $\tffnu-$ correction is more constrained than the $ \tffx-$one. Indeed, with the $ \tffx-$correction, we are able to encompass all corrections obtained with the $\tffnu-$ correction, while the reverse is not true. {We will see that the model associated with the $\tffnu-$ correction is much more rigid (or less flexible) than the $\tffx-$ correction: for example, in the ideal full-state measurement case, the $\tffnu-$correction is able to reconstruct the full flow-field in a far less accurate way than the $ \tffx-$correction.}
In the following, we will show that rigidity and flexibility properties of the models may both be an advantage or a drawback, depending on the sparsity of the available measurements. 
One has to keep in mind that the inverse problem that aims at determining the tuning function that yields the best measurement match is more or less severely underdetermined: for example, in case of sparse (resp. many) measurements, it might be better to pick the more rigid (resp. flexible) $\tffnu-$correction (resp. $\tffx-$correction). 

\subsection{Data Assimilation procedure} \label{sec:procedure}

In this section, we describe the data-assimilation procedure used here. Let $\omm$ be a set of higher-fidelity or experimental measurements that correspond to information extracted from the flow and $\mathcal{M}(\cdot)$ the measurement operator that allows to extract the corresponding measure from our simulation result $ (\tuu,\tp) $. In this work, since we are dealing only with velocity measurements, this operator will act on the velocity field, $\tuu$, yielding $\tmm = \mathcal{M}(\tuu) \in M$, where $M$ is the measurement space, whose norm is given, generically, by $|| \cdot ||_M$. The data-assimilation problem can now be recast into an optimization one, for which the forcing terms (either $\tffx$ or $\tffnu$) are tuned such that the cost functional:
\begin{equation}\label{eqn:cost-functional}
    J(\tuu) = \frac{1}{2} || \mathcal{M}(\tuu) - \mathbf{m} ||_M^2
\end{equation} 
is minimal, with the velocity field $\tuu$ satisfying the corrected RANS-SA equations \eqref{eqn:corrected-RANS-SAb} and \eqref{eqn:corrected-SA-tnu}. {We remark that, for now, no extra penalization of the correction term (that may be related to a modelling of its uncertainties) is considered}. Following \citet{Foures14}, \citet{Parish16}, \citet{Singh16} or \citet{Mons14}, this optimization problem
may be solved with an iterative gradient-based algorithm. It requires in particular the computation of the cost functional gradient with respect to the correction fields, $\nabla_{\tffx} J$ or $\nabla_{\tffnu} J$. To obtain an expression of the gradient, we resort to a Lagrangian formalism, that allows rewriting the constrained optimization problem into an unconstrained optimization problem. To that aim, the state is augmented with a set of Lagrange multipliers (or adjoint variables) $(\tuua,\tpa,\tnua)$ and we look for critical points of the Lagrangian functional:
\begin{equation}
    \begin{split}
        L ([\tuu,\tp,\tnu],[\tuua,\tpa,\tnua],[\tffx,\tffnu]) = & J (\tuu) \\
        + & \left( \tuua , \tuu \cdot \nabla \tuu + \nabla p - \nabla \cdot \left( (\nu+\nu_t(\tnu)) \nabla_s \tuu \right) - \tffx \right)_{\Omega} \\
        + & \left( \tpa, \nabla \cdot \tuu \right)_{\Omega} \\
        + & \left( \tnua , \tuu \cdot \nabla \tnu - \nabla \cdot ( \eta(\tnu{}) \nabla \tnu ) - s(\tnu, \nabla \tnu, \nabla \tuu) - \tffnu \right)_{\Omega},
    \end{split}
\end{equation}
where $\left(\qq_1,\qq_2\right)_{\Omega} = \int_{\Omega} \qq_1 \cdot \qq_2 \; d\Omega $ represents the inner product related to the classical $ \mathcal{L}_2$ norm. Setting to zero the variation of this Lagrangian with respect to the adjoint variables $[\tuua,\tpa,\tnua]$ yields the governing equations \eqref{eqn:corrected-RANS-SAb} and \eqref{eqn:corrected-SA-tnu}.
Setting to zero its variation with respect to the direct variables $[\tuu,\tp,\tnu]$ provides the adjoint equations of the RANS-SA model:
\begin{eqnarray}
    \nabla \cdot \tuua & =& 0, \label{eqn:RANS-SA-div-adjoint} \\
        \tuua \cdot ( \nabla \tuu)^T - \tuu \cdot \nabla \tuua - \nabla \cdot ( (\nu + \nu_t) \nabla_s \tuua ) - \nabla \tpa& &  \label{eqn:RANS-SA-adjoint} \\
        + \tnua \nabla \tnu + \nabla \cdot (\tnua \partial_{\nabla \tuu} s) & =& - \left( \frac{\partial \mathcal{M}}{ \partial \tuu} \right)^{\dagger} (\mathcal{M}(\tuu) - \omm), \nonumber \\
        - \tuu \cdot \nabla \tnua - \nabla \cdot ( \eta \nabla \tnua ) + (\partial_{\tnu} \eta) \nabla \tnua \cdot \nabla \tnu + ( \partial_{\tnu} \nu_t ) \nabla \tuua : \nabla_s \tuu & & \nonumber \\
        - (\partial_{\tnu} s) \tnua + \nabla \cdot ( \tnua \partial_{\nabla \tnu} s ) & =& 0, \label{eqn:RANS-SA-SA-adjoint}
\end{eqnarray}

We remark that, although we present the continuous formalism, in practise, we solve the discrete adjoint matrix, consisting in the transpose of the Jacobian matrix (which is necessary as well for the Newton method used to obtain the RANS-SA solutions). Indeed, one can show that this implementation of the adjoint represents a valid discretization of the above equations (\ref{eqn:RANS-SA-div-adjoint}, \ref{eqn:RANS-SA-adjoint} and \ref{eqn:RANS-SA-SA-adjoint}) in a Finite-Element Method framework {(see, for example, \citep{houston1999posteriori} for the stabilized Finite-Element on a linear advection equation and references therein)}. Taking now the variation of the Lagrangian with respect to the forcing term ($\tffx$ or $\tffnu$), we obtain the expressions of the gradients as a function of the adjoint variables:
\begin{eqnarray}\label{eqn:gradients}
         \nabla_{\tffx} J = - \tuua \;\;,\;\; \nabla_{\tffnu} J = - \tnua.
\end{eqnarray}

With this gradient information, we follow with the description of the optimization method employed.

\subsection{Optimization Method}

\begin{figure}
\centering
	\includegraphics[trim={0cm 0cm 0cm 0cm},clip,width=10cm]{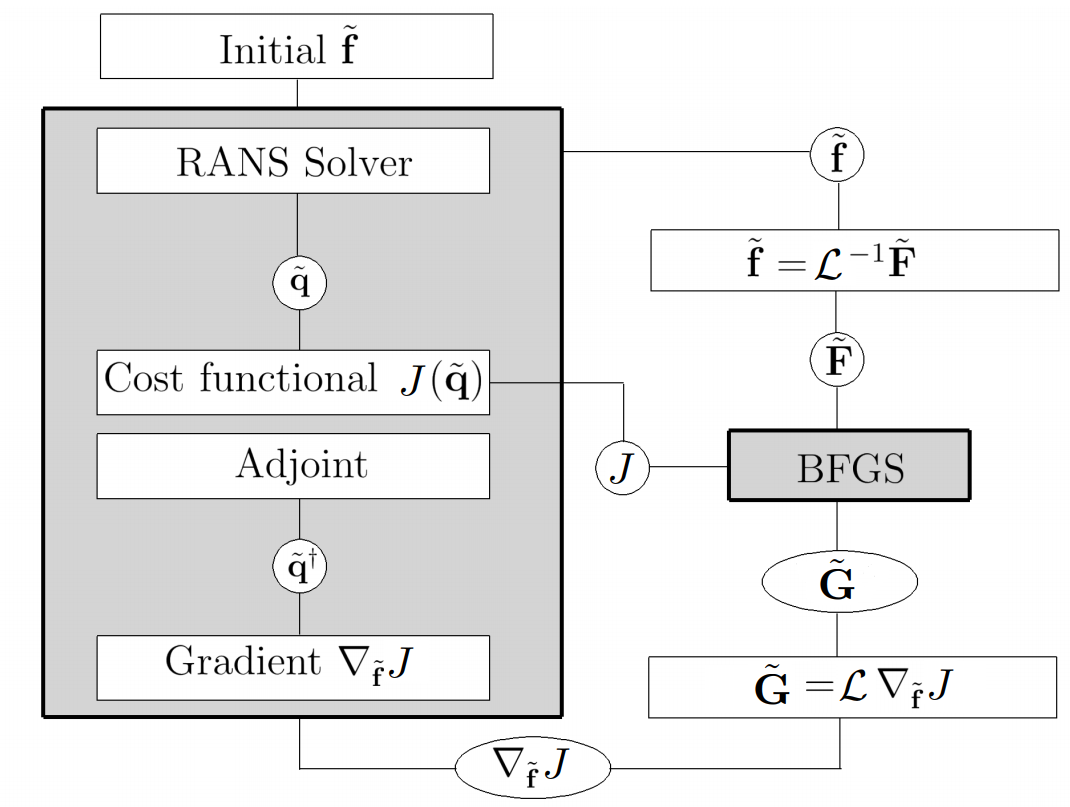}
\caption{Sketch of coupling between the optimization algorithm (BFGS) and the fluid solver.}\label{fig:BFGS}
\end{figure}

As mentioned above, for the optimization algorithm, we choose the (low memory) BFGS (see, for example \citep{bonnans2000}) since it provides a second-order convergence, outperforming, in general, simple gradient descent methods. This higher-order convergence is achieved through an approximation of the Hessian $\mathcal{H} = \nabla_{\tFF} \nabla_{\tFF} J$, which contains the second-order derivatives of the cost functional $J$ with respect to a generic forcing vector $\tFF$. This approximation is then used to find the descent direction by solving $\mathcal{H}_n^{-1} \tGG_n$, where $\tGG_n$ is the numerical gradient at iteration $n$. This matrix is approximated through:
\begin{equation}
	\mathcal{H}_{n+1} = \mathcal{H}_n + \frac{\mathbf{y}_n \mathbf{y}_n^T}{\mathbf{y}_n^T \mathbf{s}_n} - \frac{\mathcal{H}_n \mathbf{s}_n \mathbf{s}_n^T \mathcal{H}_n}{\mathbf{s}_n^T \mathcal{H}_n \mathbf{s}_n},
\end{equation}
with $\mathcal{H}_0=I$,
$\mathbf{y}_n = \tGG_{n+1} - \tGG_{n}$ the difference of the gradient between two successive iterations and $\mathbf{s}_n = \tFF_{n+1} - \tFF_n$ the difference in forcing vectors. From those relations, we can see that all inner products used in this algorithm (and as well in the line-search method, see \citet{Nocedal2000}) correspond to the Euclidean inner product $\tilde{\mathbf{F}}^T \tilde{\mathbf{G}}$, which may be inconsistent with the physical inner product $\left( \tilde{f} , \tilde{g} \right)_{\Omega} = \int_{\Omega} \tilde{f} \tilde{g} d \Omega= 
\tilde{\mathbf{f}}^T \mathcal{B} \tilde{\mathbf{g}}$, where {the symmetric positive matrix $\mathcal{B}$ accounts for the metric corresponding to the spatial discretization. We propose to}  perform the change of variables $\tFF = \mathcal{L} \tff$, where $\mathcal{L}$ is the Cholesky decomposition of $\mathcal{B} = \mathcal{L} \mathcal{L}^T$,  $\tFF$ is the {vector} used in the implemented BFGS algorithm and $\tff$ is the correction vector (the discrete counterpart of either $\tffx$ or $\tffnu$ in our case). By doing so, we take into account the physical inner-product without changing the implementation of the BFGS algorithm, since $\tFF^T \tGG = (\mathcal{L} \tff)^T (\mathcal{L} \tilde{\mathbf{g}}) = \tff^T \mathcal{B} \tilde{\mathbf{g}}$. A mass-lumping technique is used to perform efficiently the Cholesky decomposition, and for consistency, the gradient used in the BFGS algorithm is $\tGG = \mathcal{L}^T \nabla_{\tff} J$.  A sketch of the coupling of the BFGS method with our finite-element flow solver is shown in figure \ref{fig:BFGS}. 

\begin{figure}
    \centering
    \begin{tabular}{lclc}
            (a) &
        \includegraphics[trim={0.5cm 0.5cm 0.5cm 0.5cm},clip,width=6cm]{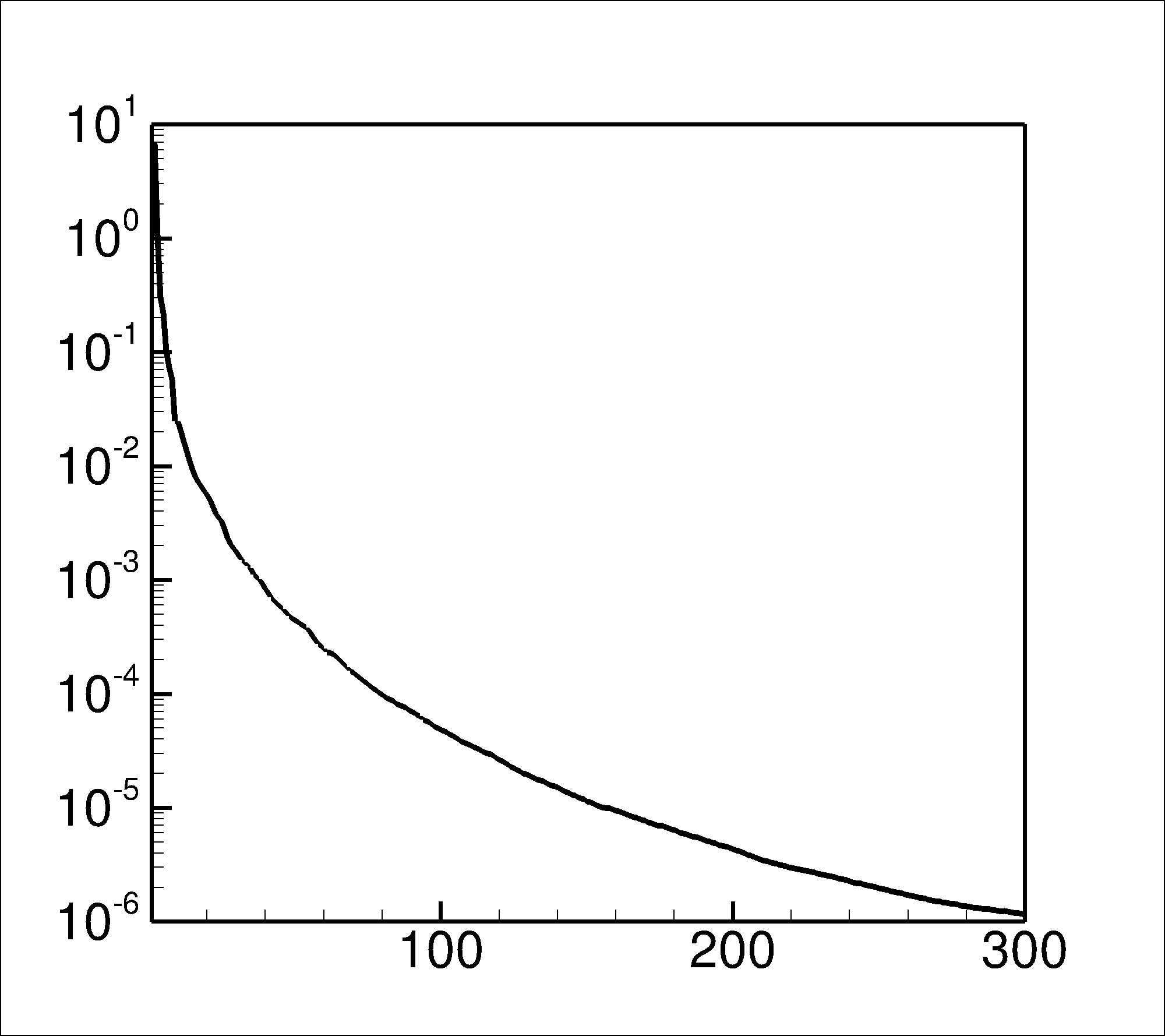}
        & (b) & \includegraphics[trim={1cm 1cm 1cm 1cm},clip,width=5.5cm]{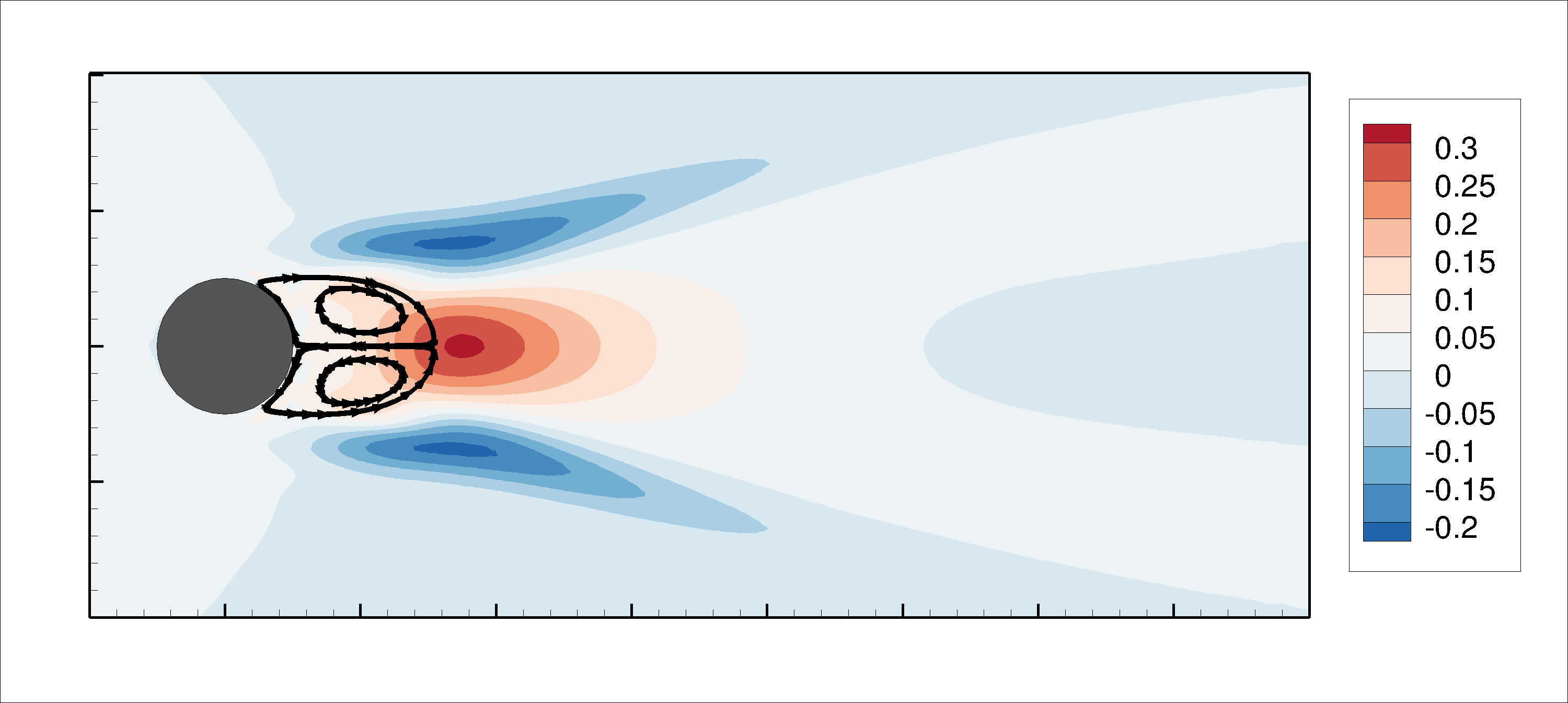} \\
        (c) & \includegraphics[trim={1cm 1cm 1cm 1cm},clip,width=5.5cm]{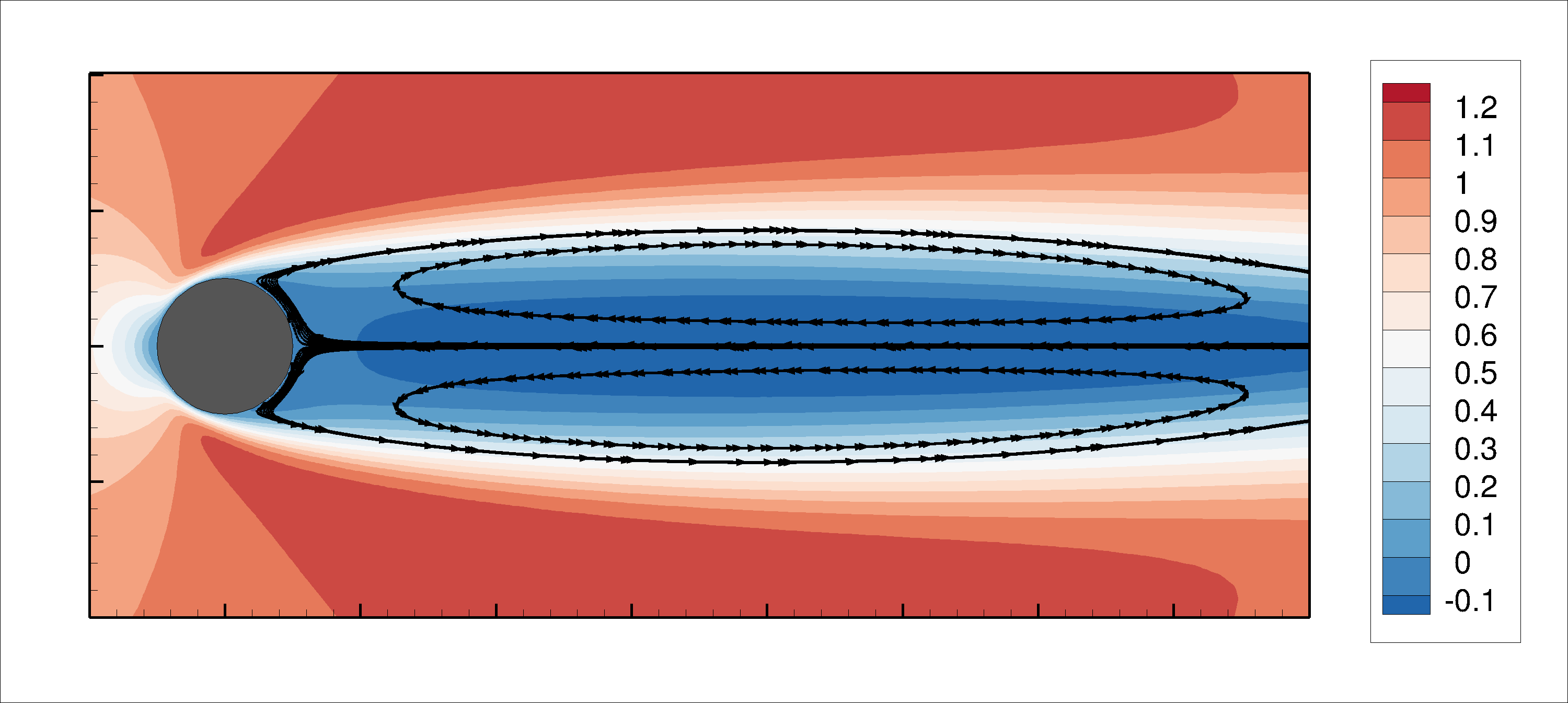} & (d) & \includegraphics[trim={1cm 1cm 1cm 1cm},clip,width=5.5cm]{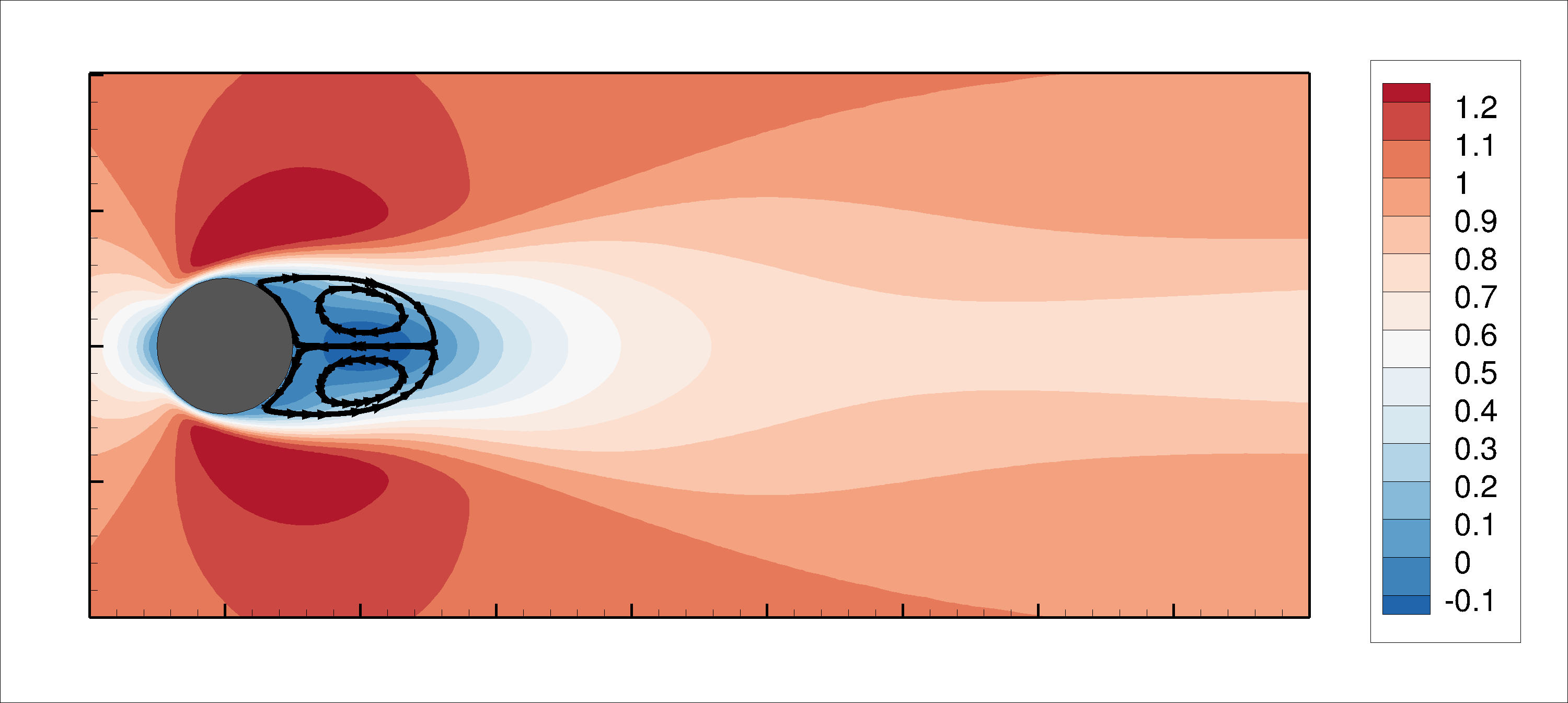}
    \end{tabular}
    \caption{Optimization results for flow over round cylinder at $Re=150$ with entire velocity field as measurement \citep{Foures14}. Cost functional convergence $J_n/J_0$ (a), assimilated forcing term $\tffx$ (b), steady solution given as initial guess to the optimization algorithm (c), assimilated solution $\tuu$ (d).}
    \label{fig:optimcylinder}
\end{figure}

We have validated in figure \ref{fig:optimcylinder} this procedure against the simple laminar case studied in \citet{Foures14}, that is, flow around a circular cylinder at $Re=150$ with the whole velocity field as measurement.
As in \citet{Foures14}, the algorithm was initialized with the base-flow (figure (c)).
In figure (a), we can see that the cost functional converges to a precision of $J_n/J_0 = 10^{-6}$ within $300$ iterations of the BFGS algorithm, which is considerably faster than the $\approx 2000$ iterations taken by the simpler optimization algorithm used in \citet{Foures14}. We also recover the same assimilated field (figure (d)) and forcing (figure (b)).
Note that, although the descent direction used in the BFGS algorithm is not directly the gradient due to the inversion with the Hessian matrix, it nevertheless keeps the soleno\"idal property of the initial gradient, as proven in appendix \ref{apd:BFGS}. The present algorithm therefore provides the same results as the one in \citet{Foures14} but with far less iterations. {It is worth-mentioning that the optimization algorithm remains stuck in the vicinity of the initial condition (the baseline laminar solution) if the BFGS
method is used without correcting the gradient direction (setting $\mathcal{L}=\mathcal{I}$ in figure \ref{fig:BFGS}). This shows that informing BFGS algorithm the metric of the mesh is crucial for the proper convergence of BFGS.}

\section{Results}
\label{sec:results}

We will now analyse reconstruction results of the backward-facing step flow for two different sets of measurements. The quality of the flow reconstruction will be monitored by scrutinizing the $L_2-$norm of the reconstruction error:
\begin{equation}
    e_{\Omega} = \sqrt{ \left. \left( \int_{\Omega} |\tuu - \ouu|^2 d \Omega \right) \middle/ \left( \int_{\Omega} 1 \; d\Omega \right) \right. }.
\end{equation}
{where $\Omega$ is the
domain plotted in all figures from now on, $\Omega=(-0.5,8) \times (0,2)$. This region is considered as the domain of interest here and is smaller than the computational domain sketched in figure \ref{fig:SketchDomain}.}

Firstly (\S \ref{sec:dense}), we will consider the complete mean velocity field in order to assess the capacity of the two corrected-models to accurately recover the velocity, pressure and Reynolds stress forcing. {This is an important step since we will assess the capability of a given correction term to reproduce the various features of the reference solution. }
Secondly (\S \ref{sec:sparse}), we will use only few point-wise velocity measurements and analyse the behaviour of the two-models when the inverse problem underlying the reconstruction procedure is strongly underdetermined. {The evaluation of the performance of a given correction term if only sparse information is available is a complementary and more realistic test for the assessment of the data-assimilation procedure. Due to the rigidity / flexibility properties, a given correction term may indeed perform differently if full or sparse measurements are considered.}

\subsection{Dense velocity field measurements} \label{sec:dense}

We consider the measure $\overline{\mathbf{m}} = \ouu$, where  $\ouu$ refers to the mean velocity field of the DNS interpolated on the mesh used for the optimization procedure (see figure \ref{fig:SketchDomain}). Since that mesh is coarser than the DNS-one, we consider that the error interpolation is negligible. The measure is thus defined on the same mesh as the solution in the optimization procedure. For this reason, the measurement space $M$ ends up being the velocity space itself. The norm associated to the measurement space is chosen to be:  $|| \cdot ||_M = || \cdot ||_{\Omega}$, the $L_2$ norm of the two velocity-components. This cost-functional is directly linked to the reconstruction error $ e_\Omega $. In eq. (\ref{eqn:RANS-SA-adjoint}), we then have:
\begin{equation}
    - \left( \frac{\partial \mathcal{M}}{\partial \tuu } \right)^{\dagger} (\mathcal{M}(\tuu) - \omm) = - (\tuu - \ouu).
\end{equation}

In the following, we first (\S \ref{sec:two}) analyse the favorable case where, in the optimization problem, both the streamwise velocity and the eddy-viscosity profiles at the inlet correspond to those provided by the DNS (see figure \ref{fig:DNS} (c)). The reconstruction error related to this (uncorrected) baseline solution is $ e_\Omega=0.06$.
Then (\S \ref{sec:one}), we consider the case where only the mean-velocity profile stems from the DNS and a constant eddy-viscosity profile is enforced at the inlet ($ \nu_t(y)/\nu=3 $). The reconstruction error associated to the uncorrected solution is then slightly higher, $ e_\Omega=0.094$. In the latter case, we aim at assessing how volume source term corrections manage to compensate boundary condition defects in the model. In all cases, the optimization procedures are initialized with the baseline uncorrected RANS-SA model.  
\subsubsection{Reference eddy-viscosity profile at the inlet}
\label{sec:two}

\begin{figure}
\centering
\begin{tabular}{cccc}
& $\tffx-$corr. DENSE REF & & $\tffnu-$corr. DENSE REF \\
(a) & \includegraphics[trim={0.5cm 0.5cm 0.5cm 0.5cm},clip,width=6cm]{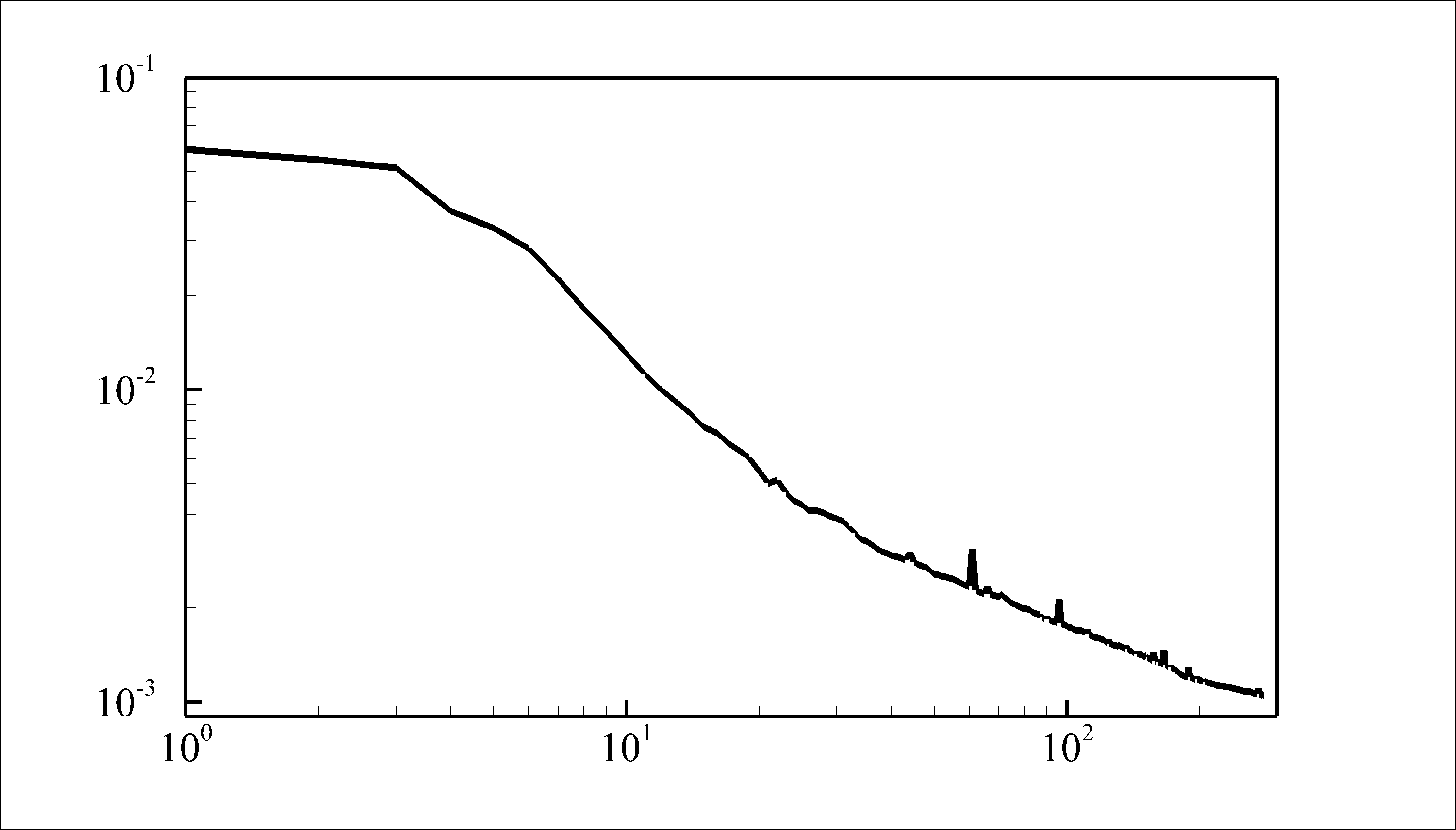} & (b) & \includegraphics[trim={0.5cm 0.5cm 0.5cm 0.5cm},clip,width=6cm]{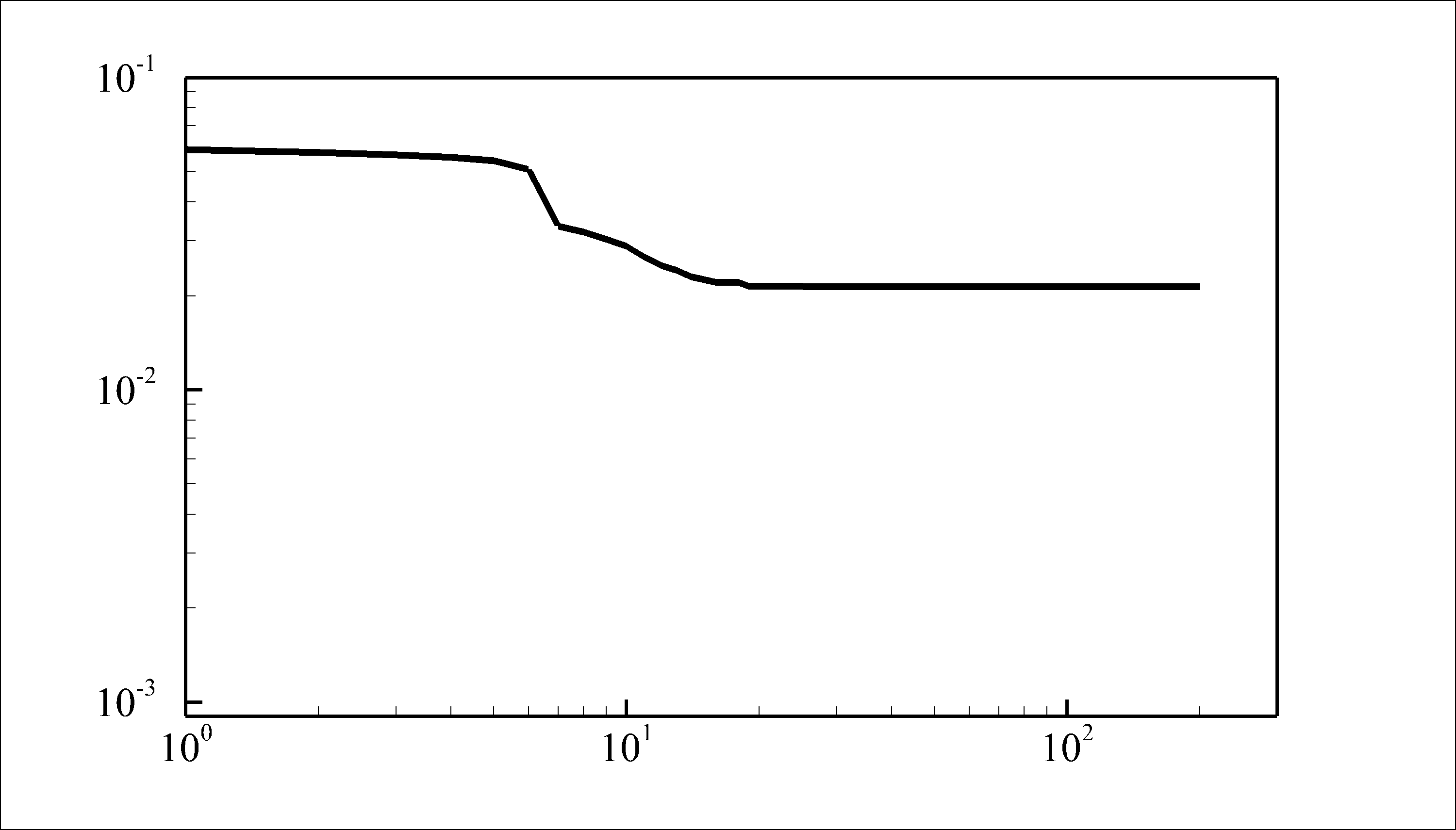} \\
(c)  & \includegraphics[trim={0.5cm 0.5cm 0.5cm 0.5cm},clip,width=7cm]{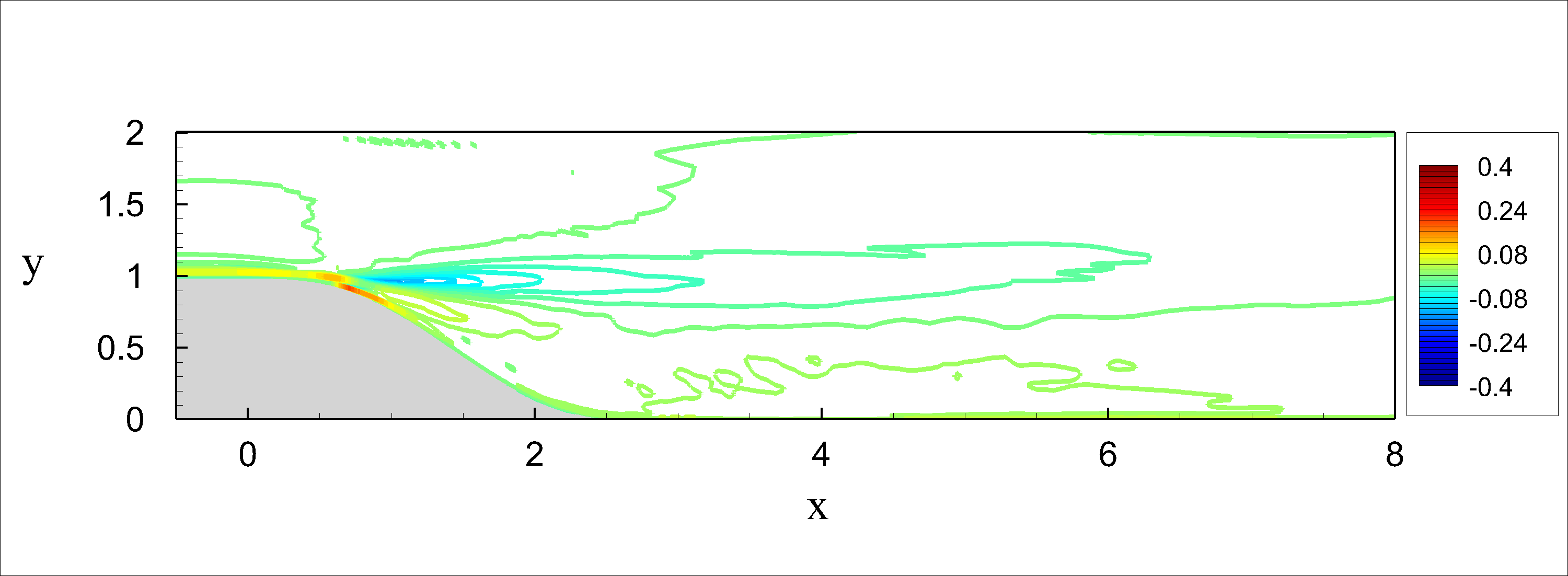} & (d) &   \includegraphics[trim={0.5cm 0.5cm 0.5cm 0.5cm},clip,width=7cm]{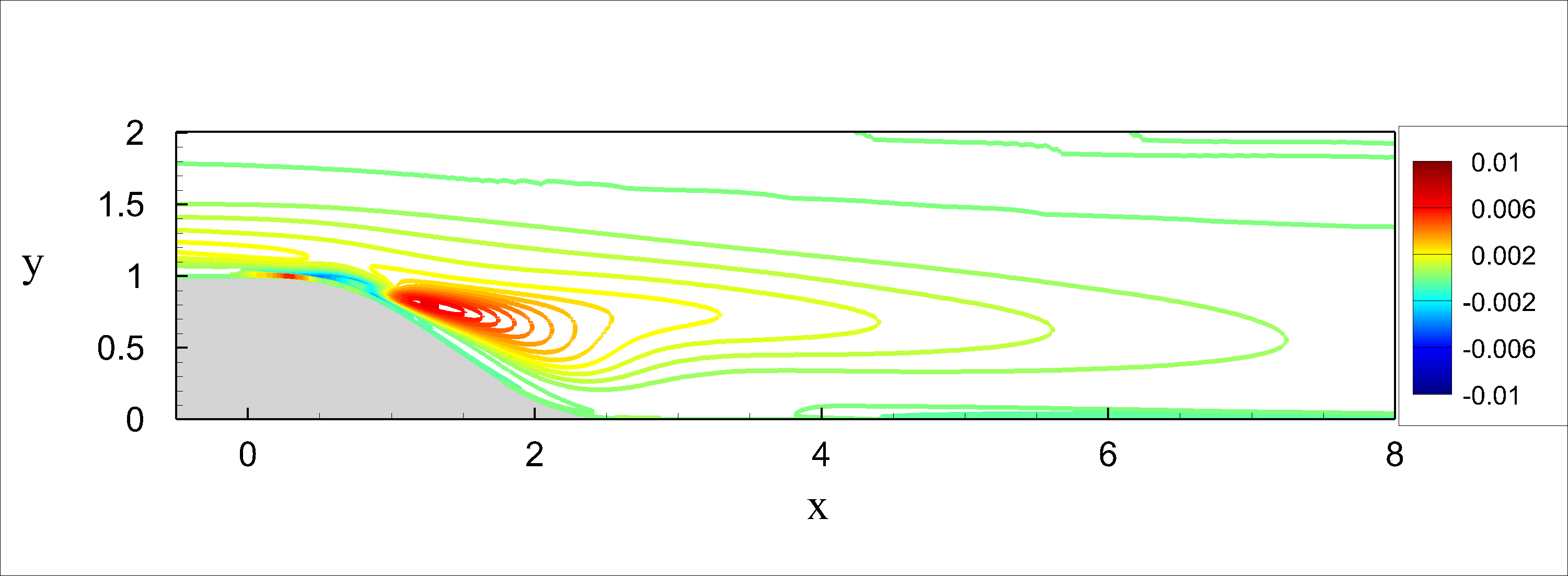} \\
(e) & \includegraphics[trim={0.5cm 0.5cm 0.5cm 0.5cm},clip,width=7cm]{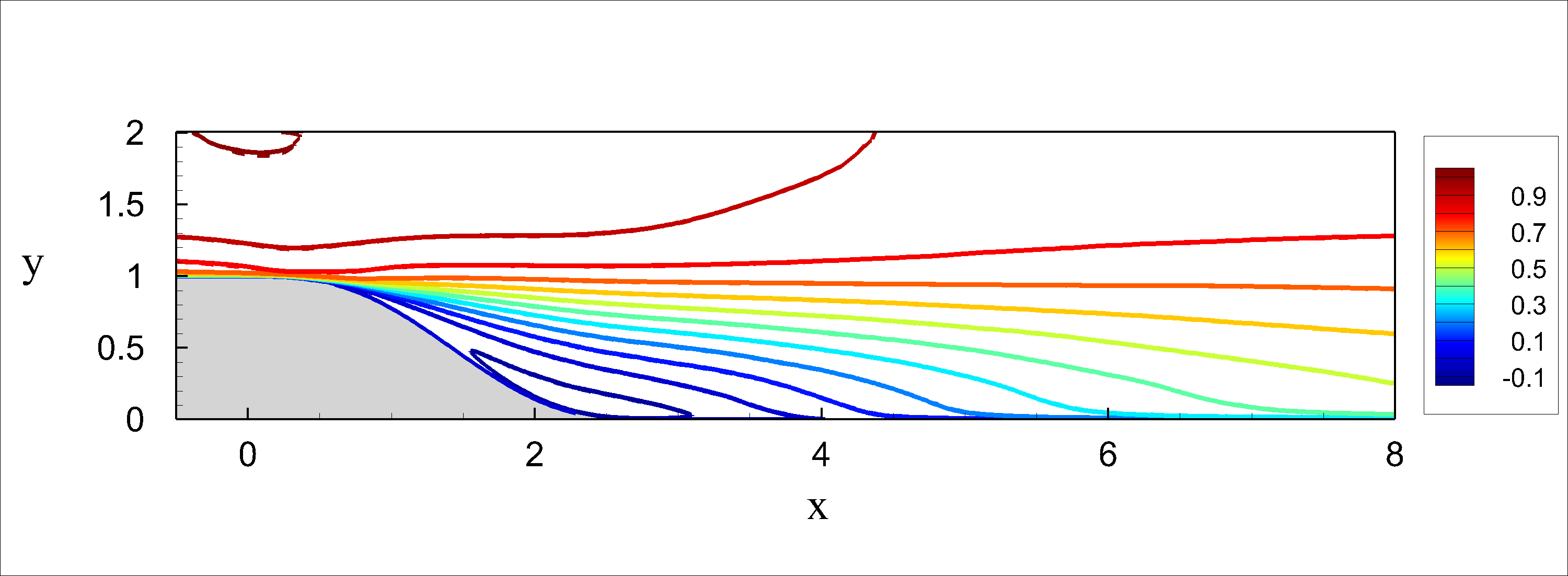} & (f) & \includegraphics[trim={0.5cm 0.5cm 0.5cm 0.5cm},clip,width=7cm]{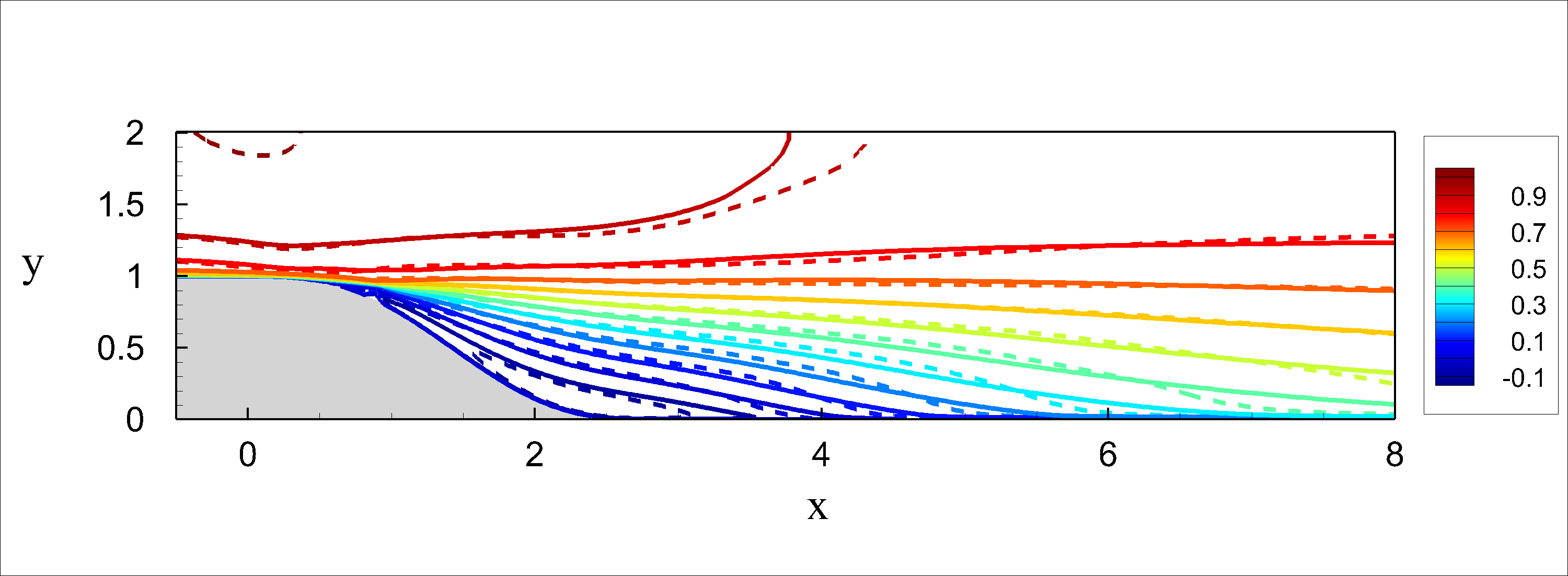} \\
(g) & \includegraphics[trim={0.5cm 0.5cm 0.5cm 0.5cm},clip,width=7cm]{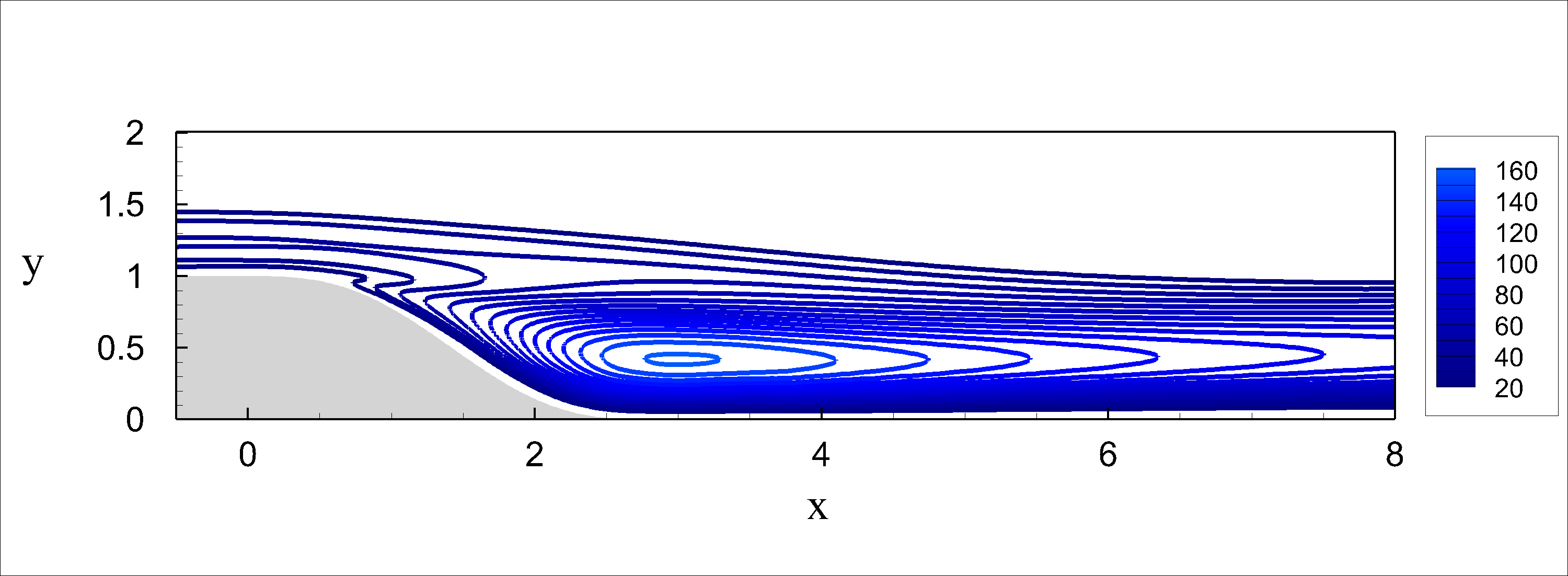} & (h) & \includegraphics[trim={0.5cm 0.5cm 0.5cm 0.5cm},clip,width=7cm]{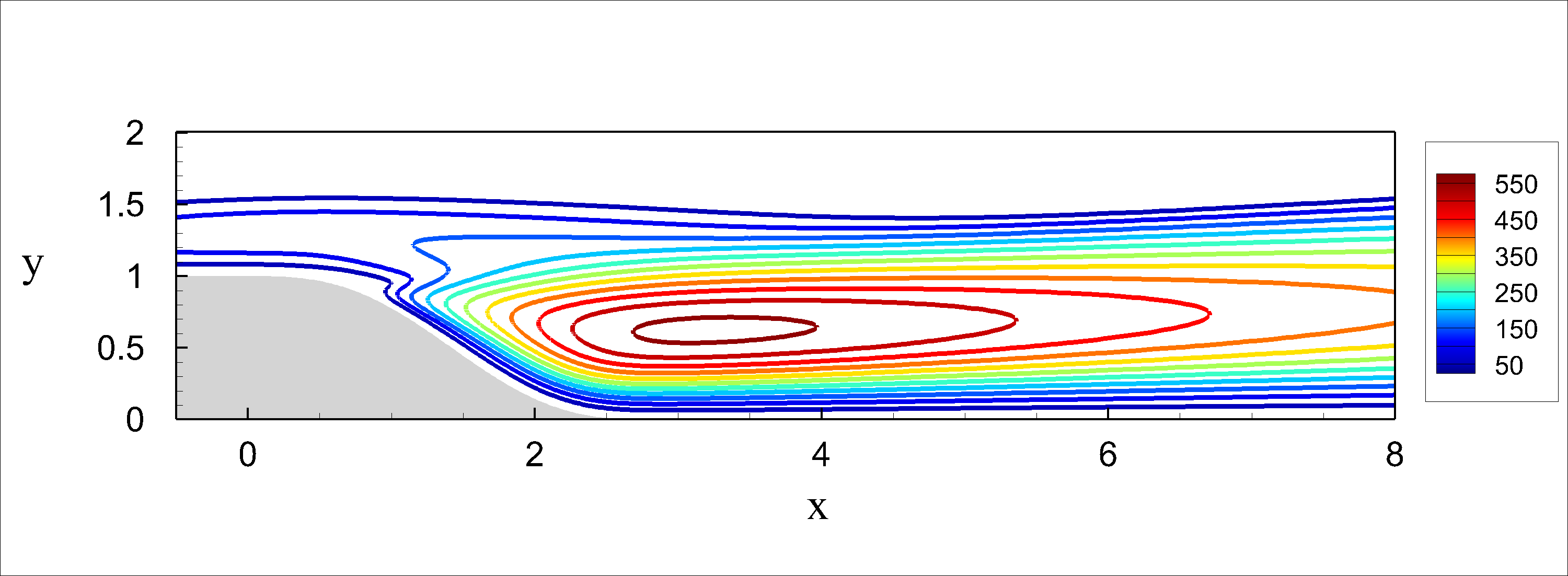} \\
(i) & \includegraphics[trim={0.5cm 0.5cm 0.5cm 0.5cm},clip,width=7cm]{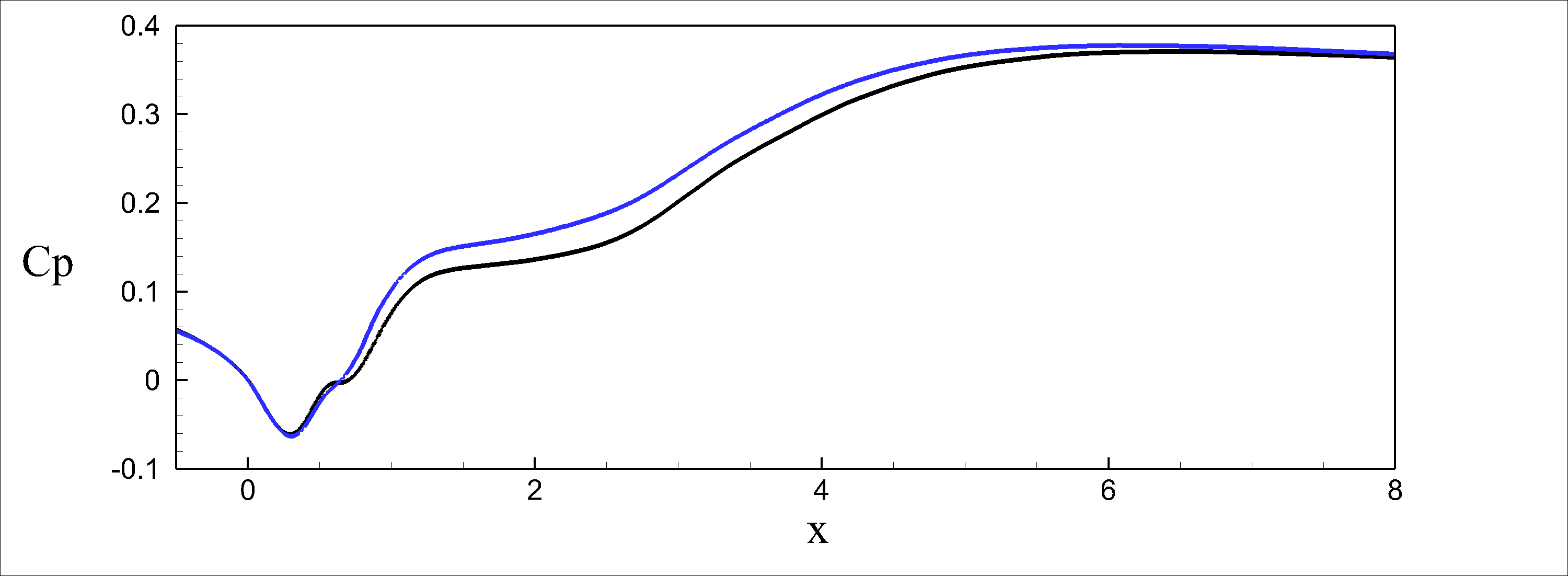} & (j) & \includegraphics[trim={0.5cm 0.5cm 0.5cm 0.5cm},clip,width=7cm]{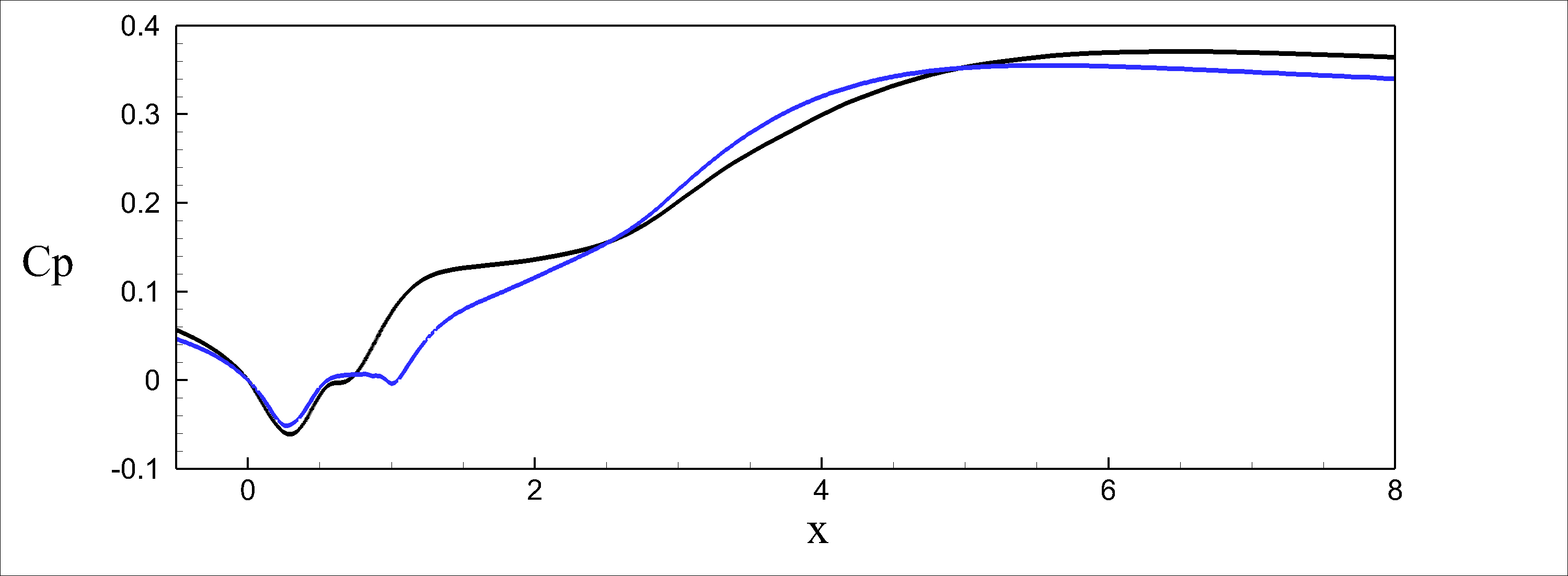} \\
(k) & 
\includegraphics[trim={0.5cm 0.5cm 0.5cm 0.5cm},clip,width=7cm]{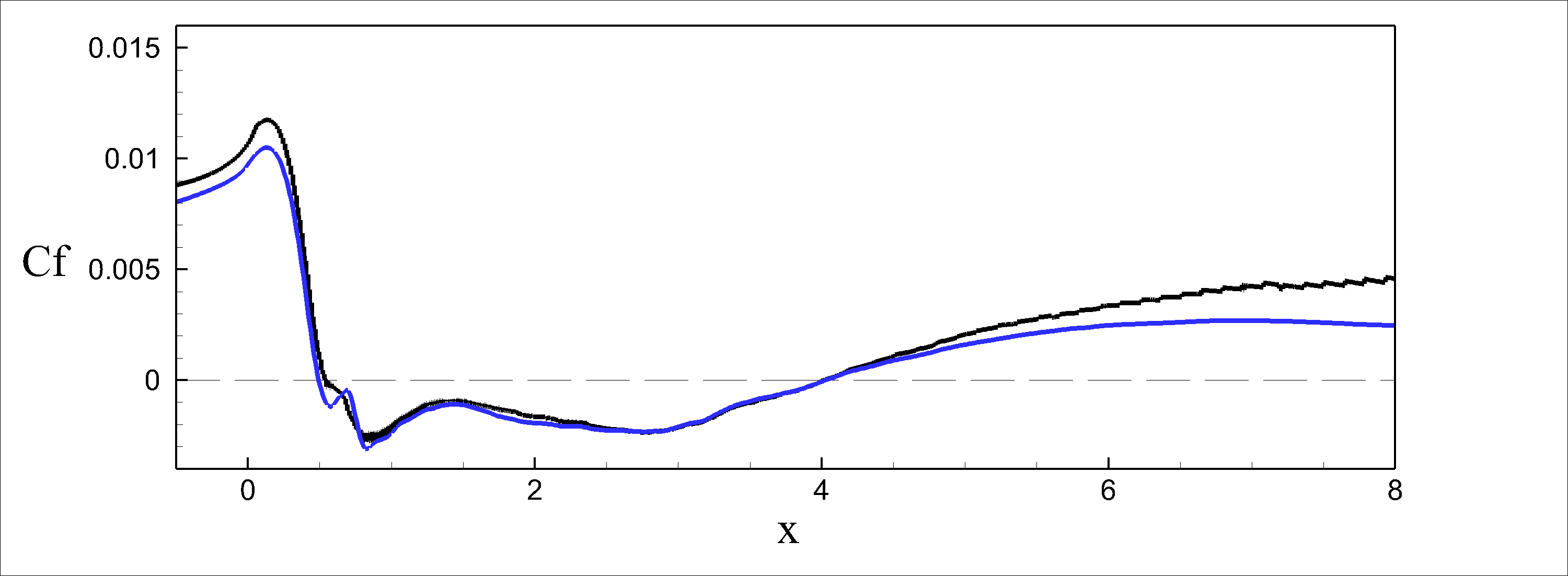} & (l) & \includegraphics[trim={0.5cm 0.5cm 0.5cm 0.5cm},clip,width=7cm]{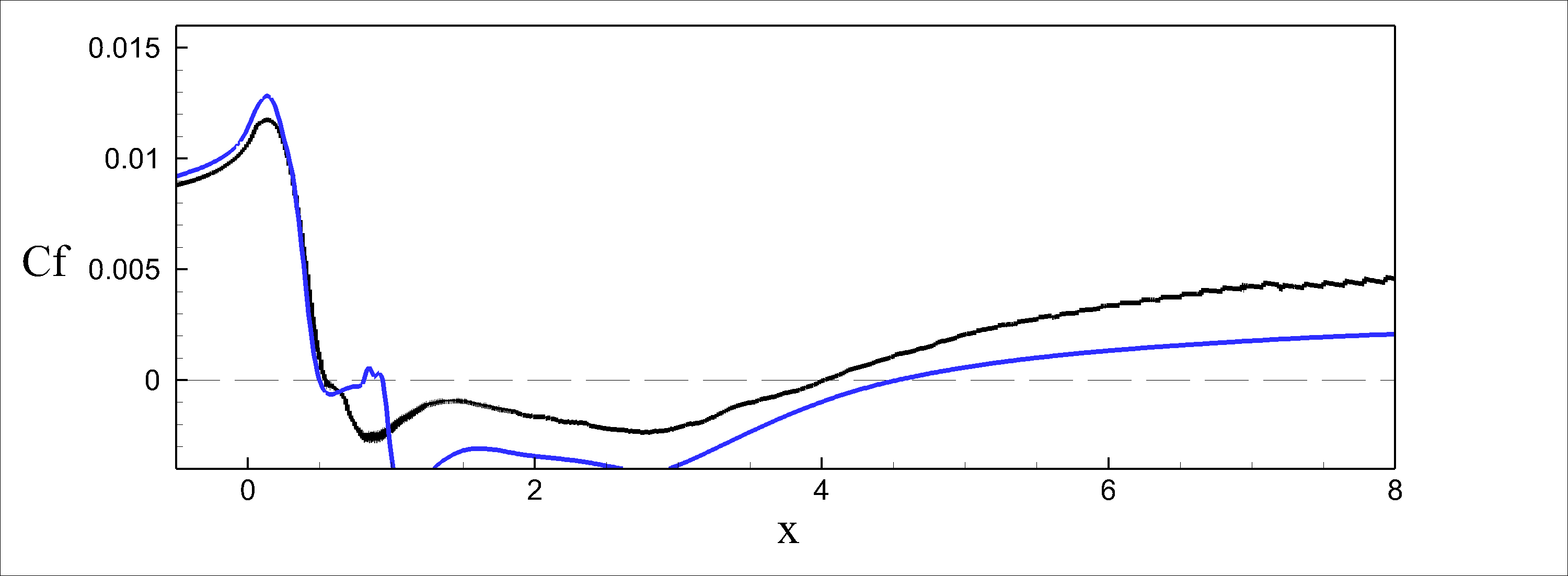}
\end{tabular}
\caption{Assimilation of {\it dense} velocity measurements with {\it reference eddy-viscosity profile} at the inlet. Left panels (a,c,e,g,i,k) concern optimization with $ \tffx-$correction and
right panels (b,d,f,h,j,l) to $\tffnu-$correction.
(a,b): error $e_{\Omega}$ as a function of optimization iteration $n$.
(c): streamwise component of $\tffx-$correction. (d): $\tffnu-$correction.
(e,f): streamwise component of assimilated velocity field. The solid and dashed lines correspond to the assimilated and reference results, respectively (in figure (e) the contours are indistinguishable).
(g,h): eddy-viscosity field $\nu_{t}/\nu$. (i,j): wall-pressure along lower wall.
(k,l); friction coefficient along lower wall.
In (i,j,k,l), the black solid line corresponds to the reference solution and the blue solid line to the reconstructed solution.}
\label{fig:J-u}
\end{figure}

\begin{figure}
\centering
\begin{tabular}{lclc}
(a) & & (b) &  \\
	& \includegraphics[trim={0.5cm 0.5cm 0.5cm 0.5cm},clip,width=7cm]{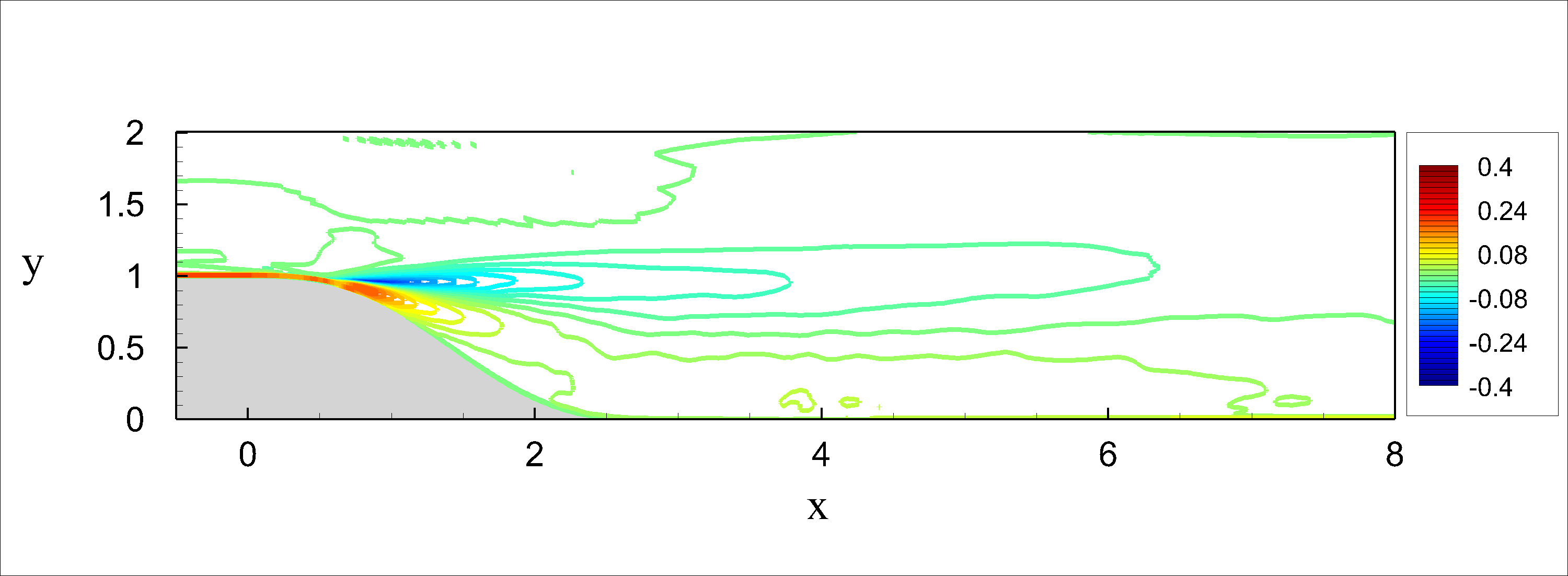} & & \includegraphics[trim={0.5cm 0.5cm 0.5cm 0.5cm},clip,width=7cm]{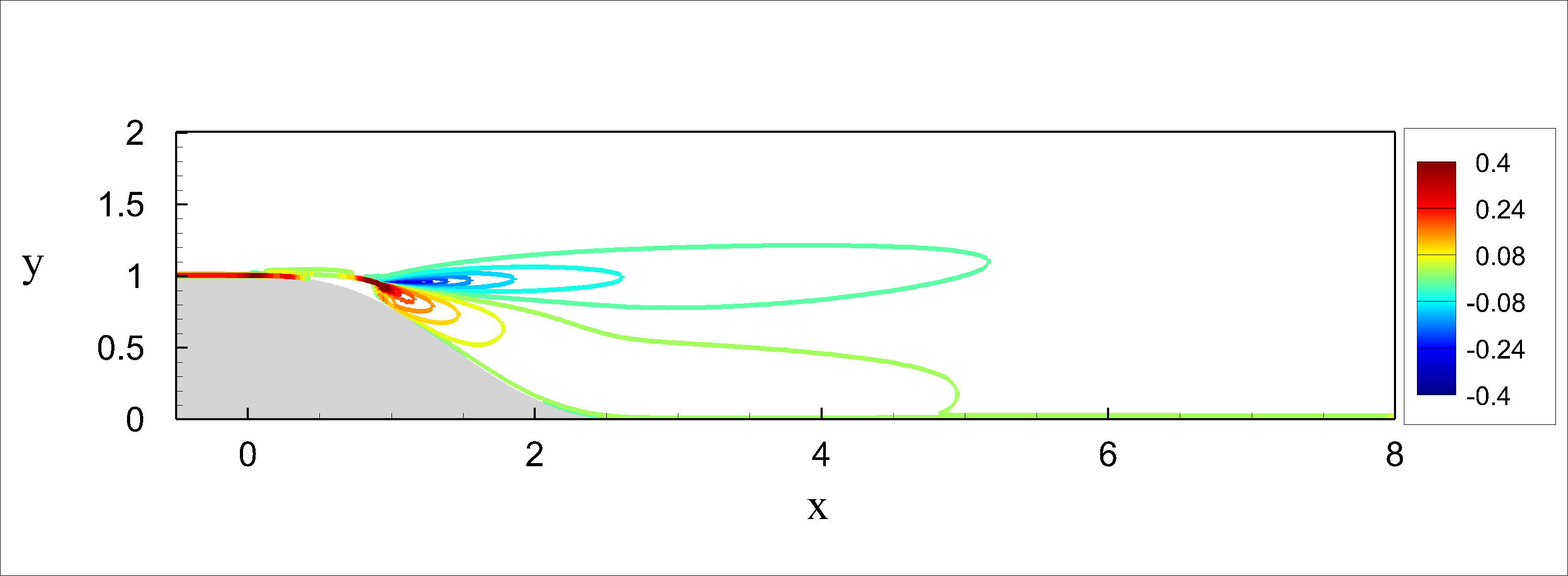} 
	\end{tabular}
\caption{Assimilation of {\it dense} velocity measurements with {\it reference eddy-viscosity profile} at the inlet. Resulting volume force $\nabla \cdot (\nu_t \nabla_s \tuu)+\tffx$). (a): optimization with $ \tffx-$correction. (b): with $\tffnu-$correction.}\label{fig:J-u2}
\end{figure}

The optimization results, for both correction terms (left column figures for $\tffx-$correction, right-column for $\tffnu-$correction), are provided in figure \ref{fig:J-u} in the case of the reference eddy-viscosity profile.
Figures (a,b) show the reconstruction error $ e_\Omega$ as a function of the optimization iterations. 
The $ \tffx-$correction manages
to reconstruct the reference DNS solution extremely accurately due to the high flexibility of the model: the reconstruction error reaches very small values of the order of $10^{-3}$.
This decrease is achieved in less than 200 iterations, similarly to the case of laminar flow over a cylinder. When using the $ \tffnu-$correction, the optimization procedure reaches a plateau in $n \approx 10$ iterations, for which the reconstruction error is about $e_\Omega=0.02$ (to be compared with the baseline value that was $ e_\Omega=0.06$).
The convergence process is much faster but also less accurate than with the $ \tffx-$correction, due to the fact that the $\tffnu-$correction is far more constrained than the $\tffx-$correction.
As for the reconstructed velocity fields (figs (e,f)), we can accordingly see that the $\tffx-$correction manages to reconstruct the streamlines of the reference solution represented by the dashed lines (barely visible in figure (e)), whereas the $\tffnu-$correction only approximately achieves this goal (dashed-lines can be seen distinct from the solid-lines). In figures (k,l), the skin friction is accurately reconstructed in the separated flow region (reattachment point at $x_r=4.02$) for the $ \tffx-$correction, while the agreement is more approximate for the $ \tffnu$-correction (reattachment point around $x_r=4.64$, the DNS reference and uncorrected RANS-SA values being respectively $x_r=3.93$ and $ x_r=5.8 $).
It is interesting to note that with the $ \tffx-$correction, the final eddy-viscosity values (figure (g)) have decreased with respect
to the baseline values (the maximum is now around $\nu_t/\nu \approx 160$, compared to $220$ for the uncorrected RANS-SA solution). This indicates that the baseline SA turbulence model induces errors that may be compensated most efficiently by replacing part of the Reynolds-stresses modeled by a Boussinesq approximation by a general unconstrained forcing. In contrast, with the $ \tffnu-$ correction, the final eddy-viscosity values $ \nu_t $ have strongly increased for the bubble to become shorter, its maximum value being here $\nu_t / \nu \approx 550$ (figure (h)). Yet, the resulting reconstruction error remains much higher than with the $\tffx-$correction. Both optimized correction terms $\tffx$ and $\tffnu$ (figs. (c,d)) are located in the vicinity of the baseline separation point. In figs \ref{fig:J-u2}(a,b), we can also see that the resulting overall momentum forcing terms (eddy-viscosity forcing together possibly with $\tffx$) are similar for both corrections and that they are qualitatively close to the reference Reynolds-stress force $\off = - \nabla \cdot \underline{\tau}$ from the DNS (see figure \ref{fig:DNS}(b)). 

\subsubsection{Constant eddy-viscosity profile at inlet}
\label{sec:one}

\begin{figure}
\centering
\begin{tabular}{cccc}
& $\tffx-$corr. DENSE CST & & $\tffnu-$corr. DENSE CST \\
(a) & \includegraphics[trim={0.5cm 0.5cm 0.5cm 0.5cm},clip,width=6cm]{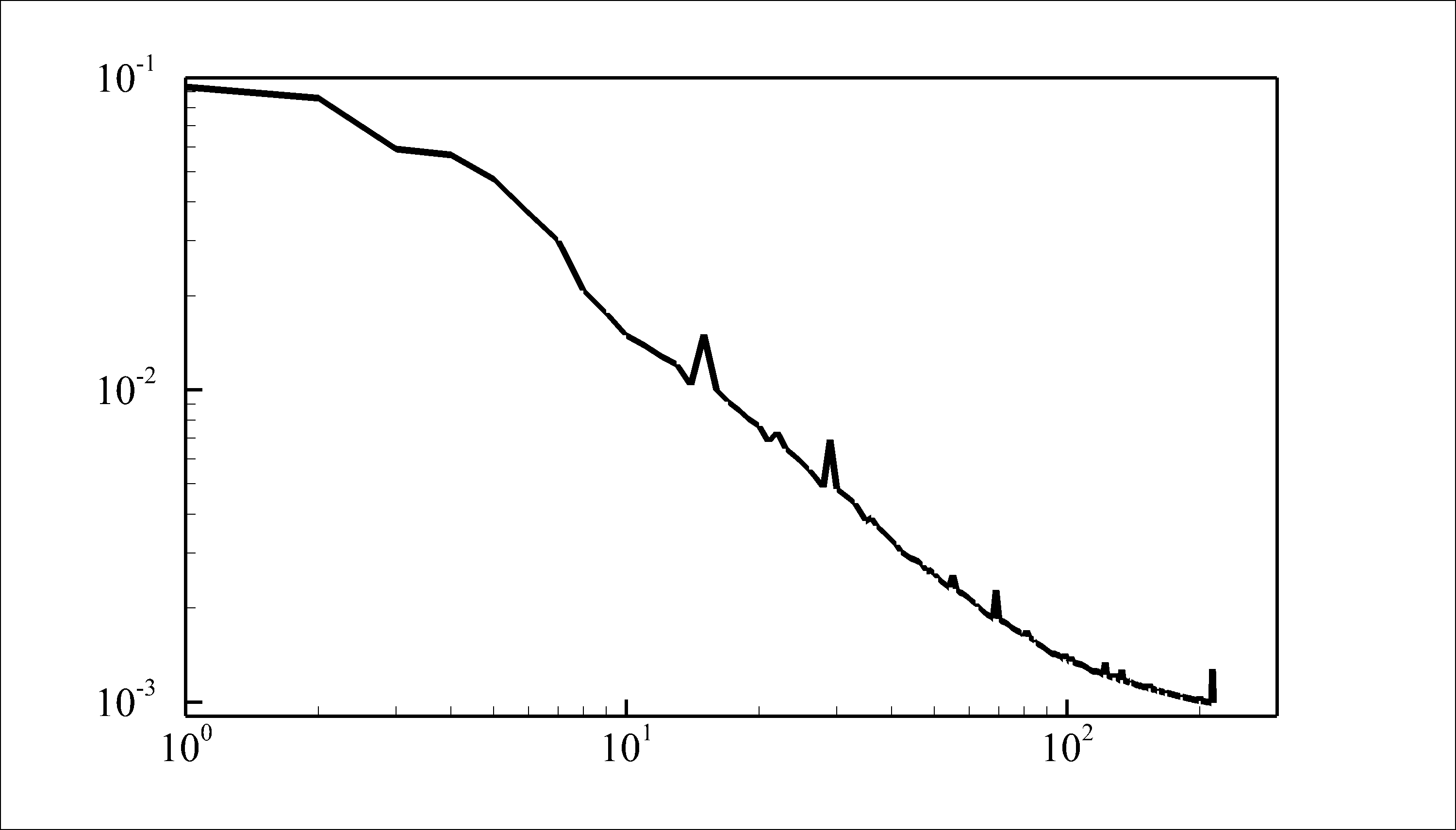} & (b) & \includegraphics[trim={0.5cm 0.5cm 0.5cm 0.5cm},clip,width=6cm]{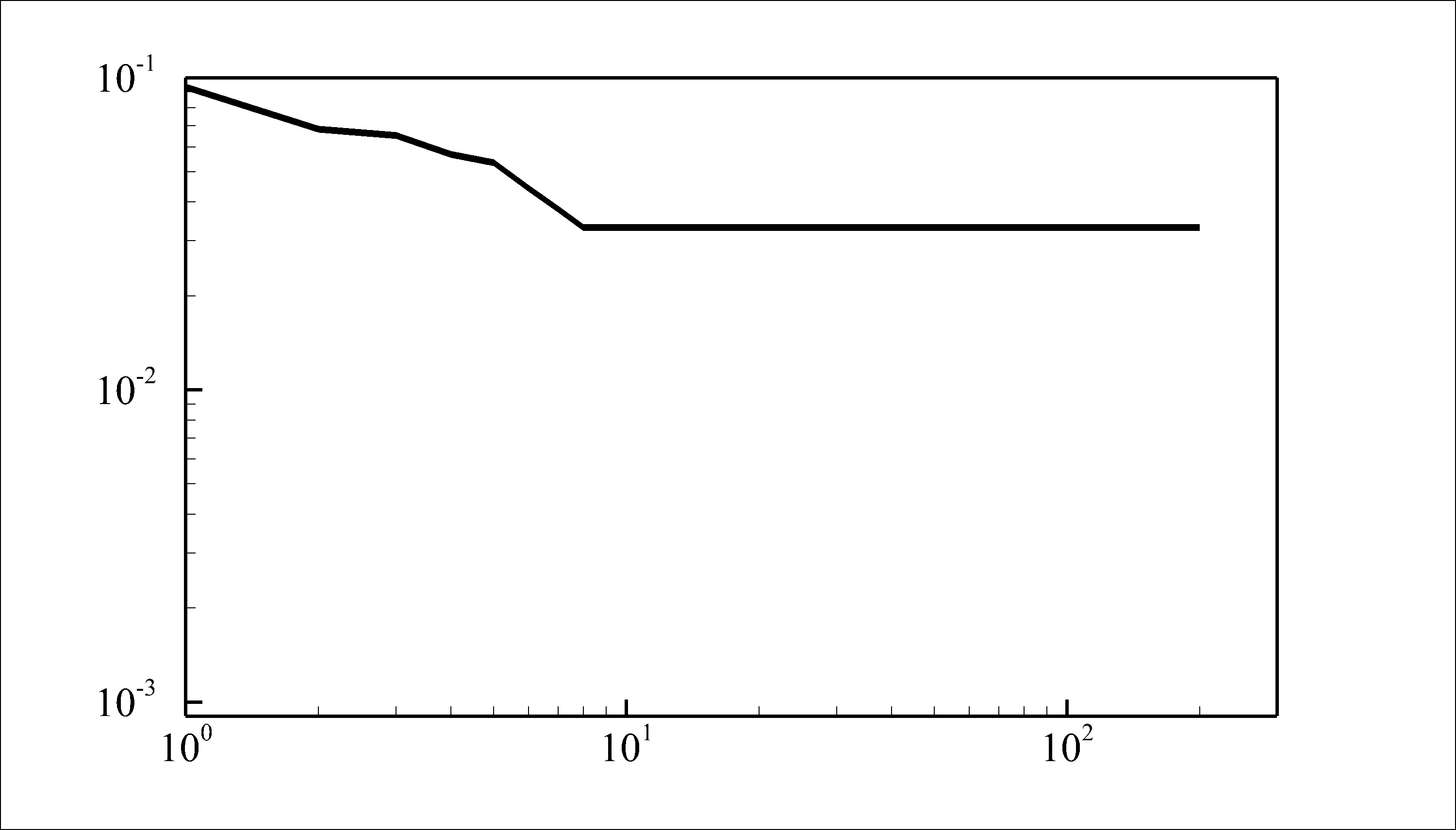} \\
(c)   & \includegraphics[trim={0.5cm 0.5cm 0.5cm 0.5cm},clip,width=7cm]{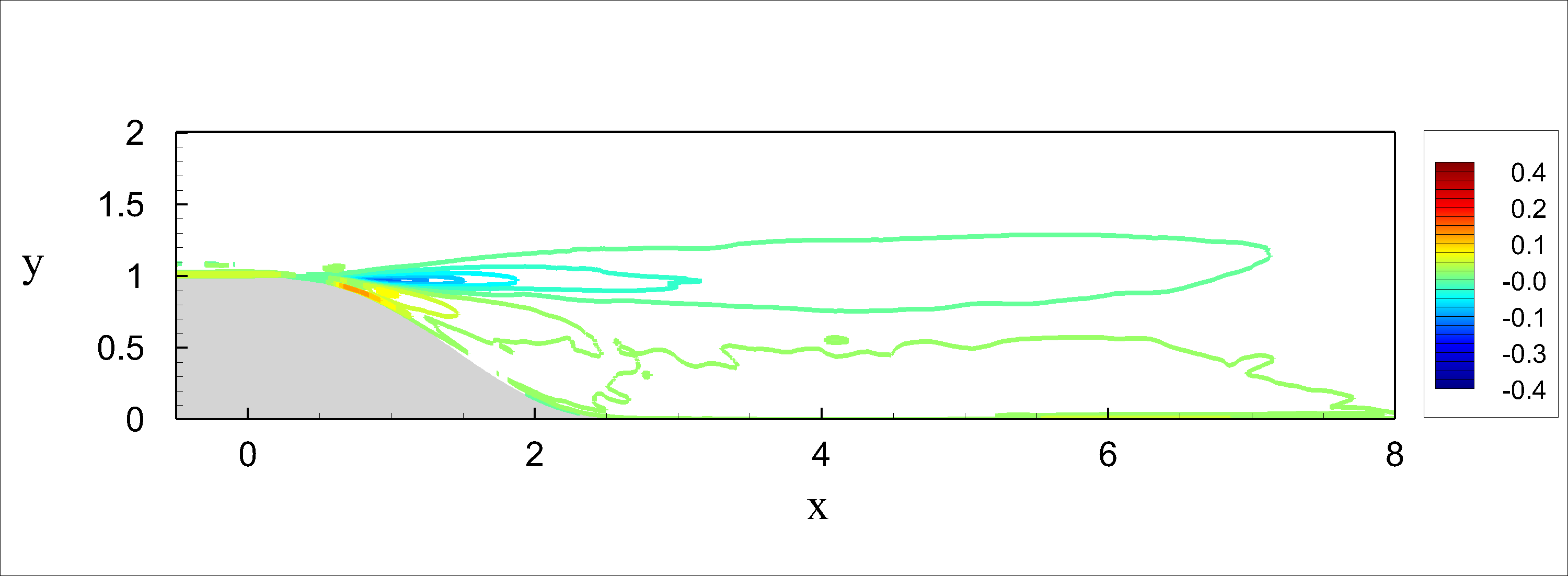} & (d) & \includegraphics[trim={0.5cm 0.5cm 0.5cm 0.5cm},clip,width=7cm]{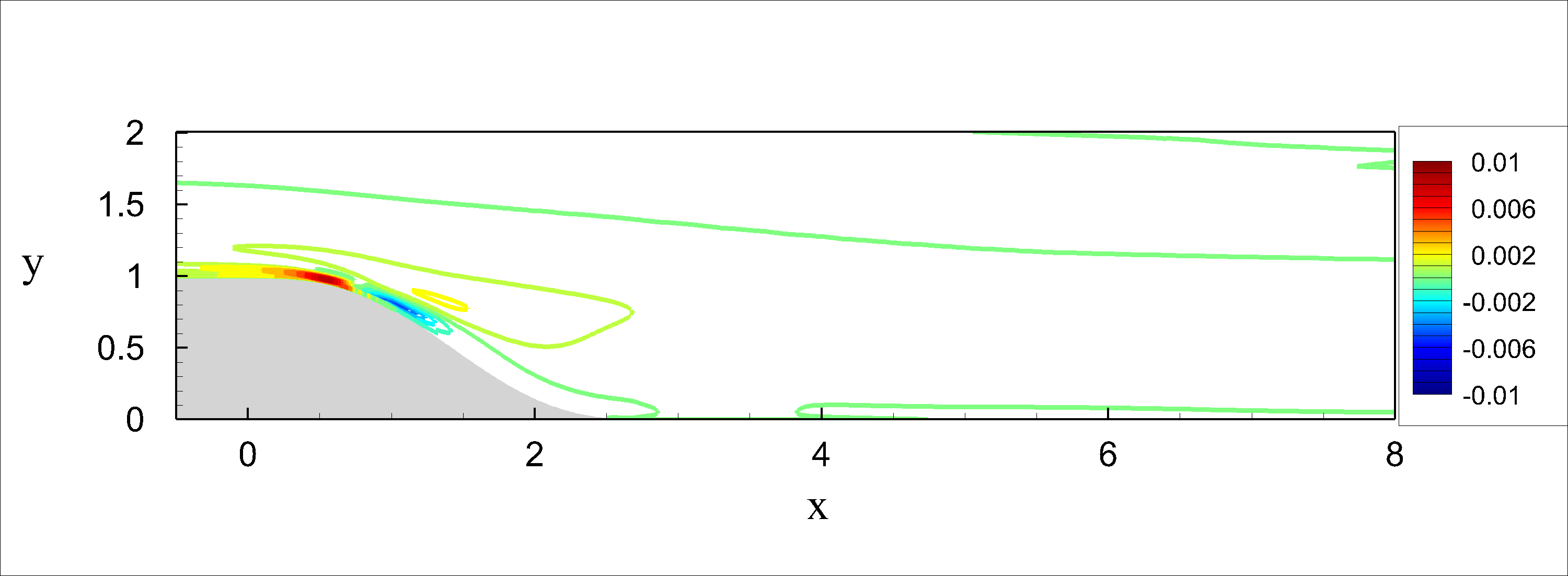} \\
(e) & \includegraphics[trim={0.5cm 0.5cm 0.5cm 0.5cm},clip,width=7cm]{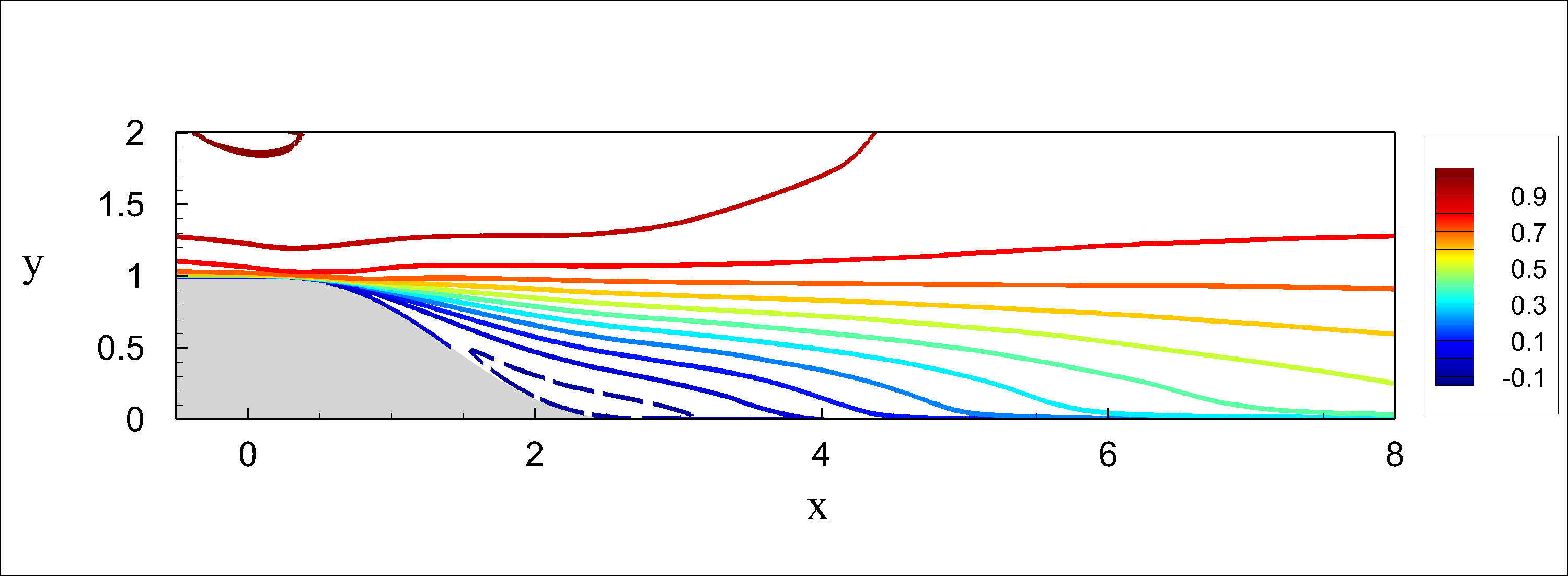} & (f) & \includegraphics[trim={0.5cm 0.5cm 0.5cm 0.5cm},clip,width=7cm]{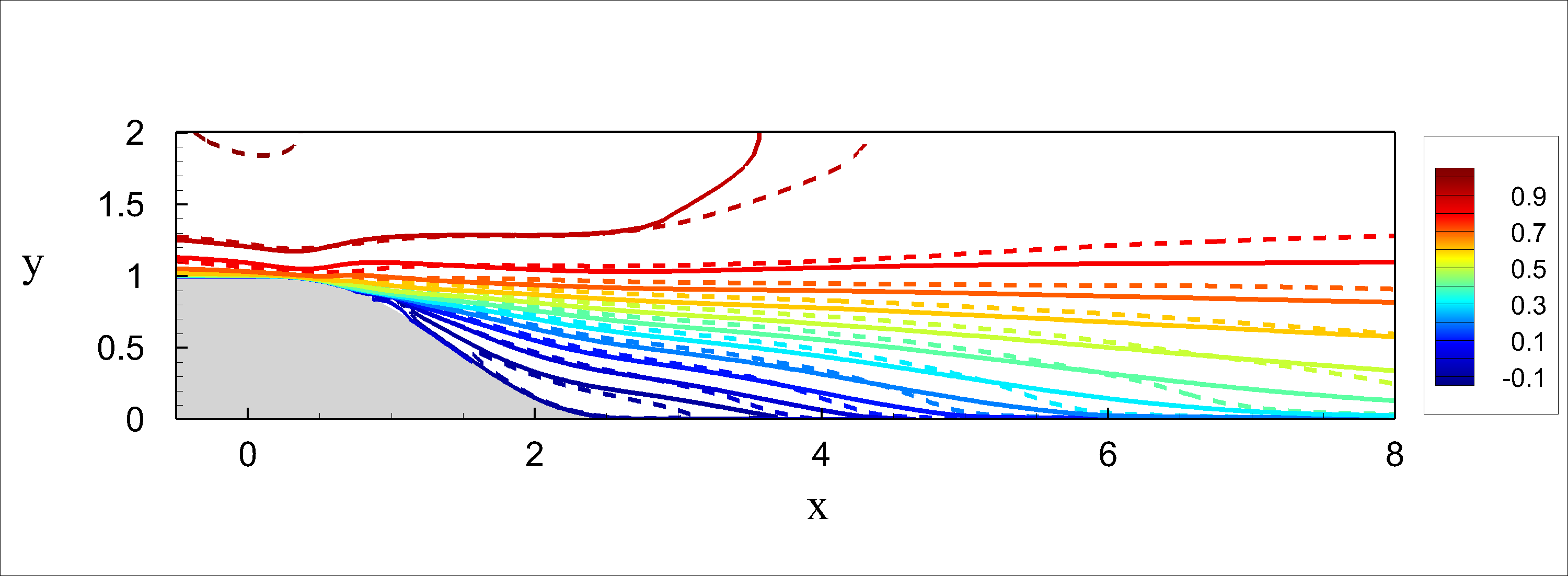} \\
(g) & \includegraphics[trim={0.5cm 0.5cm 0.5cm 0.5cm},clip,width=7cm]{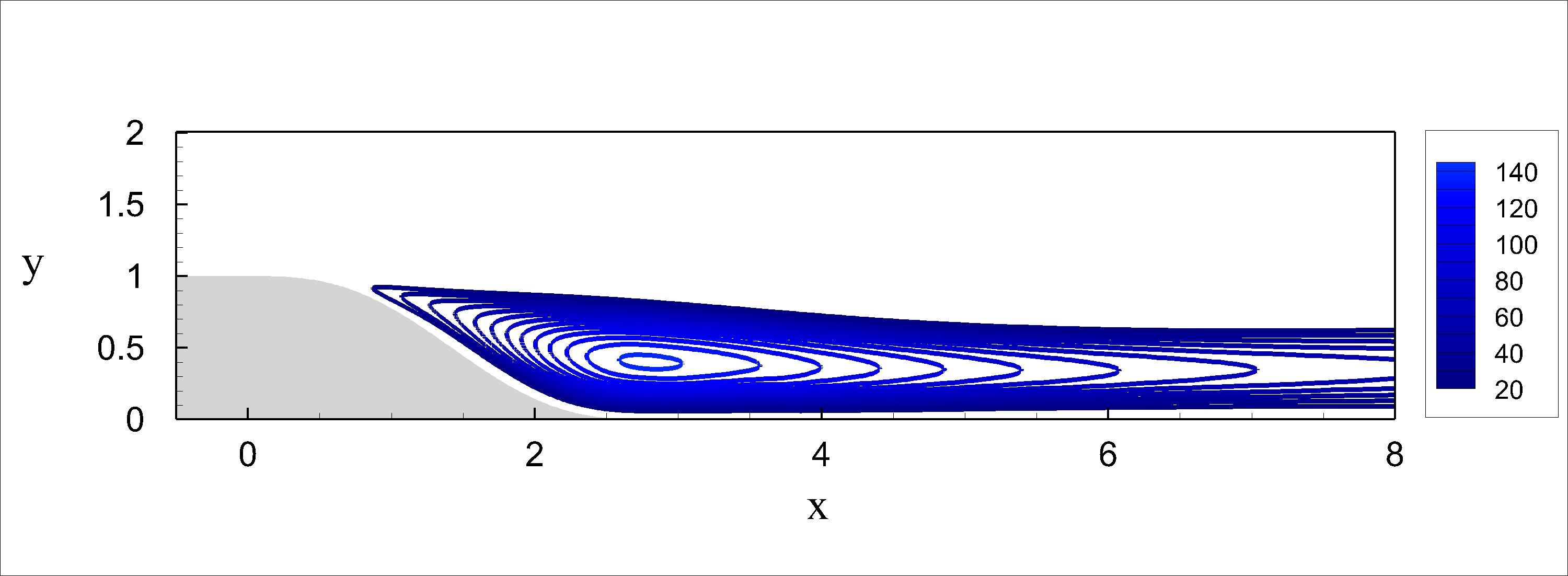} & (h) & \includegraphics[trim={0.5cm 0.5cm 0.5cm 0.5cm},clip,width=7cm]{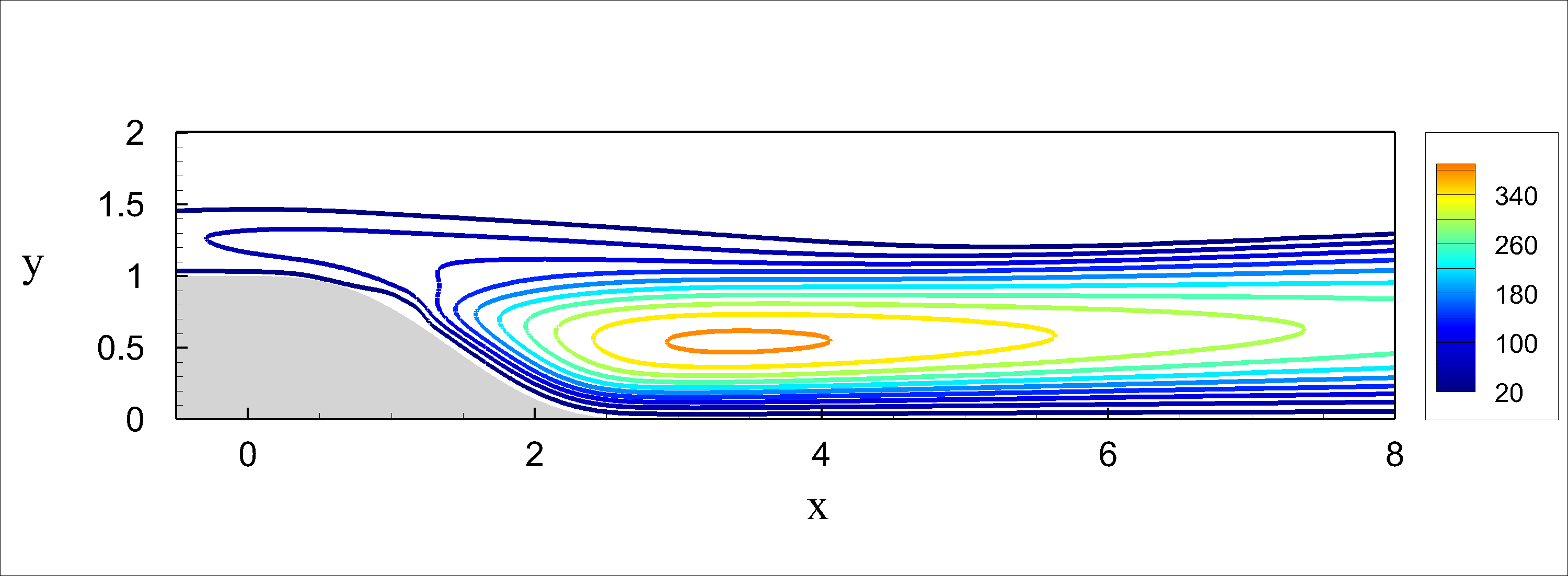} \\
(i) & \includegraphics[trim={0.5cm 0.5cm 0.5cm 0.5cm},clip,width=7cm]{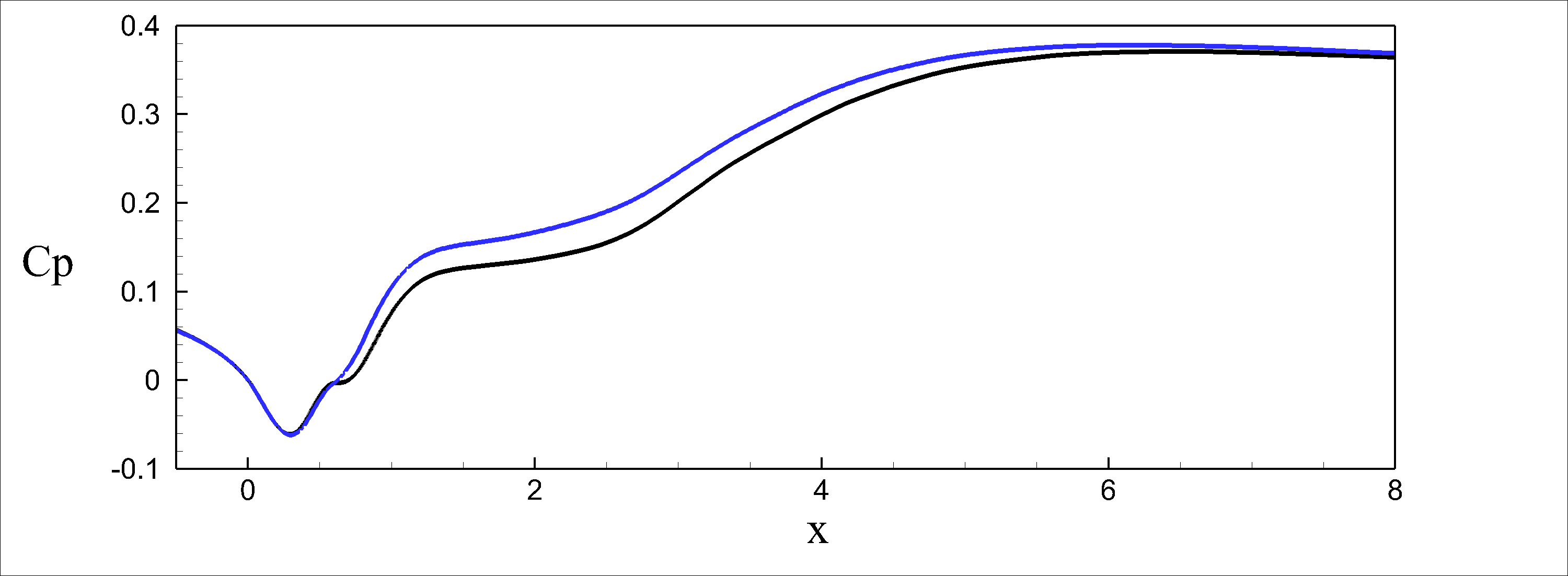} & (j) & \includegraphics[trim={0.5cm 0.5cm 0.5cm 0.5cm},clip,width=7cm]{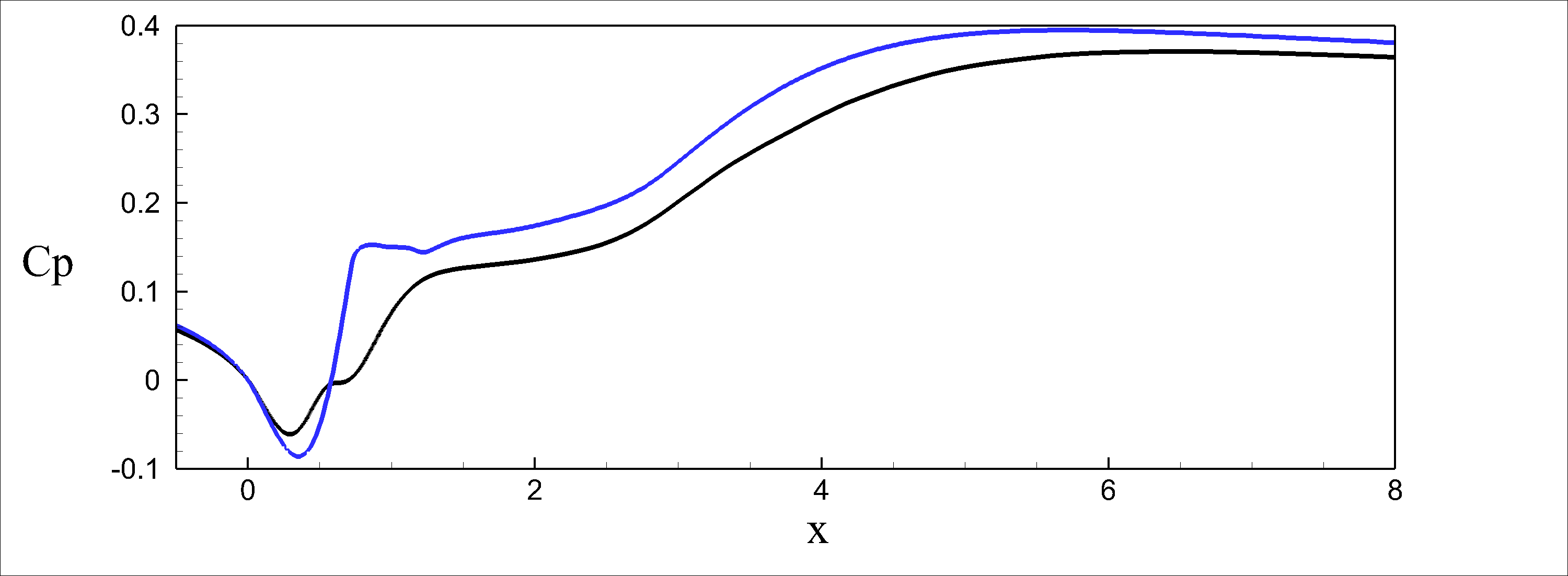} \\
(k) & \includegraphics[trim={0.5cm 0.5cm 0.5cm 0.5cm},clip,width=7cm]{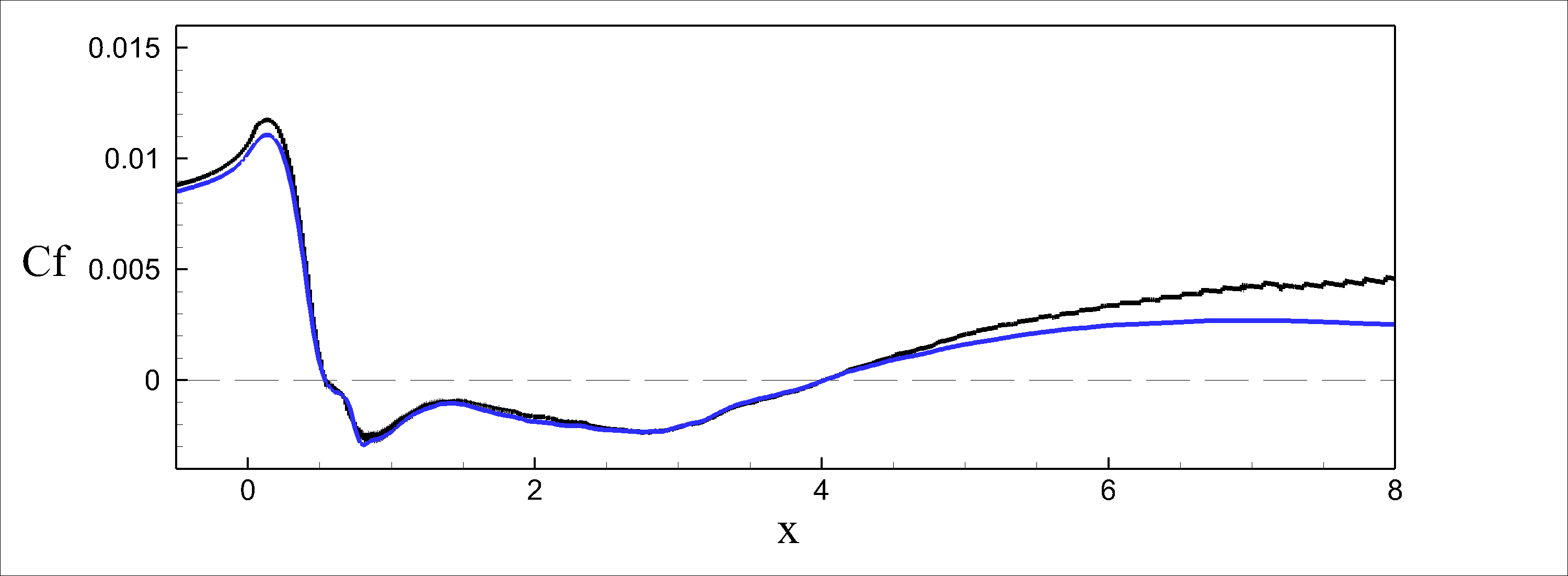} & (l) & \includegraphics[trim={0.5cm 0.5cm 0.5cm 0.5cm},clip,width=7cm]{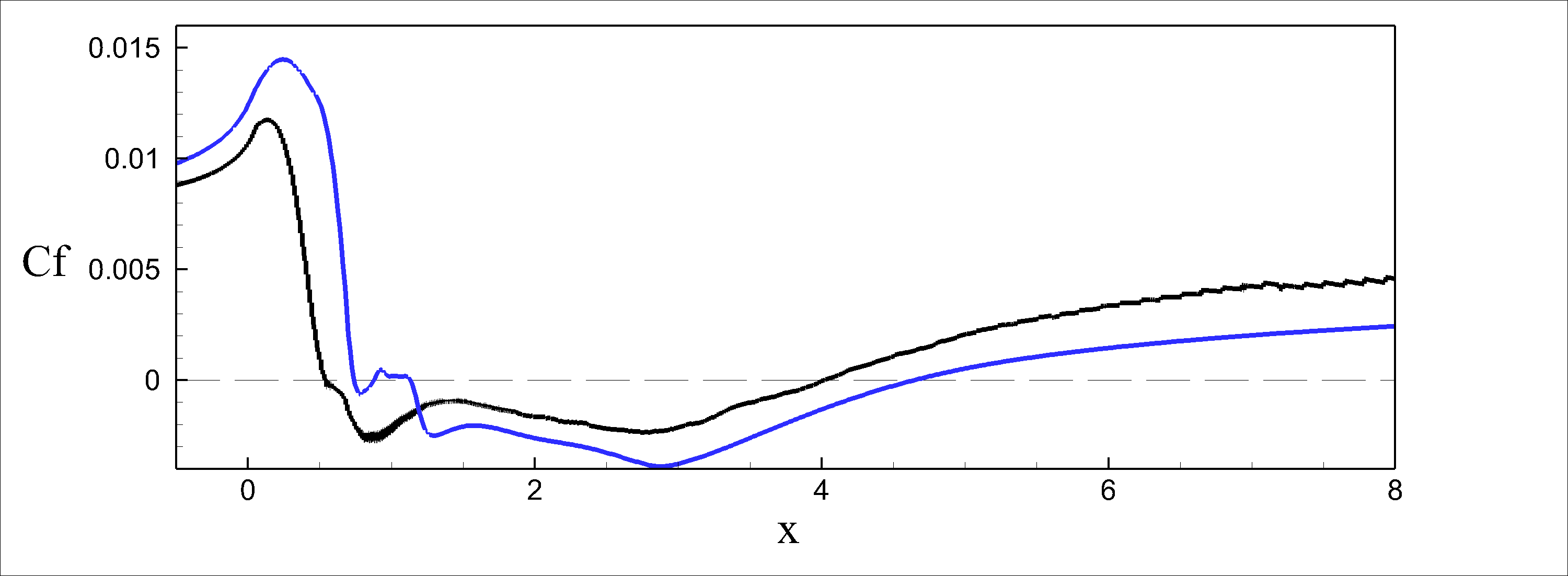}
\end{tabular}
\caption{Assimilation of {\it dense} velocity measurements with {\it constant eddy-viscosity profile} at the inlet. For more details see caption of figure \ref{fig:J-u}.}
\label{fig:J-u-cste}
\end{figure}

In the case of a defect in the boundary conditions of the baseline model, here an erroneous constant eddy-viscosity profile, results remain similar, as shown in figure \ref{fig:J-u-cste}.
We just stress the main differences. The uncorrected baseline solution is slightly further away from the reference solution since it presents a larger recirculation region ($e_\Omega=0.094$ versus $ e_\Omega=0.06$ before). This can be seen in figs (a,b). The $\tffx-$ correction manages again to reach reconstruction errors of $ 10^{-3}$, while in the case of the $\tffnu-$correction we obtain $0.035$ compared to $ 0.02 $ in the reference case.
In figure (g), we observe that, contrary to before, there is no eddy-viscosity seen in the upstream boundary layer with the $\tffx-$correction: the latter forcing directly reconstructs the Reynolds-stress forcing and does not proceed through the Boussinesq term by triggering eddy-viscosity. The maximum value of $ \nu_t/\nu$ is now $ 160 $.  
On the contrary (figure (h)), the $\tffnu-$correction induces higher values of eddy-viscosity both in the upstream boundary layer (to compensate the boundary condition defect at the entrance) and in the recirculation region, where $ \nu_t/\nu=360$. An overshoot of the friction coefficient is even observed in the upstream boundary layer in order to manage to reduce as much as possible the size of the bubble (reconstruction errors over the whole bubble region dominate those in the upstream boundary layer). The resulting reattachment point in this case is $x_r \approx 4.03$, slightly above the value obtained in the previous section.  
 
\subsection{Sparse velocity  measurements} \label{sec:sparse}

In this section, we keep a baseline model where a constant eddy-viscosity profile is enforced at the inlet (see \S \ref{sec:one}).
We consider the case where fewer velocity measurements are considered in the assimilation process.
More precisely, the measurement operator is:
\begin{equation}
    \mathcal{M}(\tuu) = \{ \tuu(\xx_m^i) \}_{i=1, \cdots, N}.
\end{equation}
Here $\xx_m^i$ are $N$ points where velocity measurements are available. Typically, these $N$ measurements may correspond to cross-stream profiles of $ N_y $ points at $ N_x $ streamwise stations, such that $N=N_x \times N_y$. This way, the cost functional reads:
\begin{equation}
    J = \frac{1}{2} \sum_{i=1}^{N} | \tuu(\xx_m^i) - \omm_i |^2,
\end{equation}
where $\omm_i = \ouu(\xx_m^i)$. With this definition of the measurement operator, the right-hand-side of eq.(\ref{eqn:RANS-SA-adjoint}) becomes: 
\begin{equation}\label{eqn:adj-J-u-dots}
    - \left( \frac{\partial \mathcal{M}}{\partial \tuu } \right)^{\dagger} (\mathcal{M}(\tuu) - \mathbf{m}) = - \sum_{i=1}^N (\tuu(\xx_m^i) - \omm_i) \delta_{\xx_m^i} 
\end{equation}
where $\delta_{\xx_m}$ is the Dirac-mass centered at $\xx_m$.

\subsubsection{Results for the $\tffnu-$correction}

\begin{figure}
\centering
\begin{tabular}{cc}
& $\tffnu-$corr. SPARSE CST \\
(a) & \includegraphics[trim={0.5cm 0.5cm 0.5cm 0.5cm},clip,width=6cm]{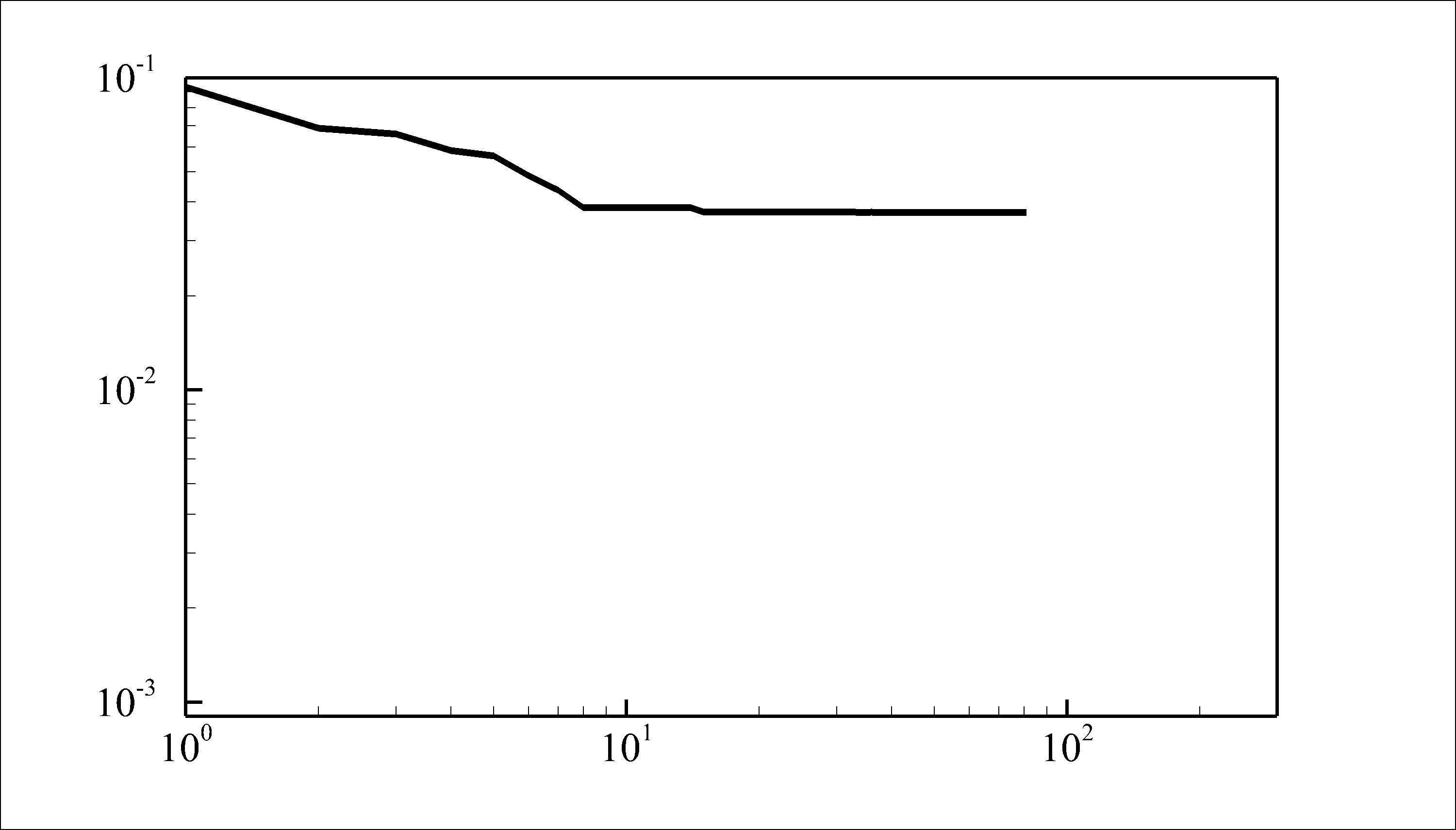} \\
(b) & \includegraphics[trim={0.5cm 0.5cm 0.5cm 0.5cm},clip,width=7cm]{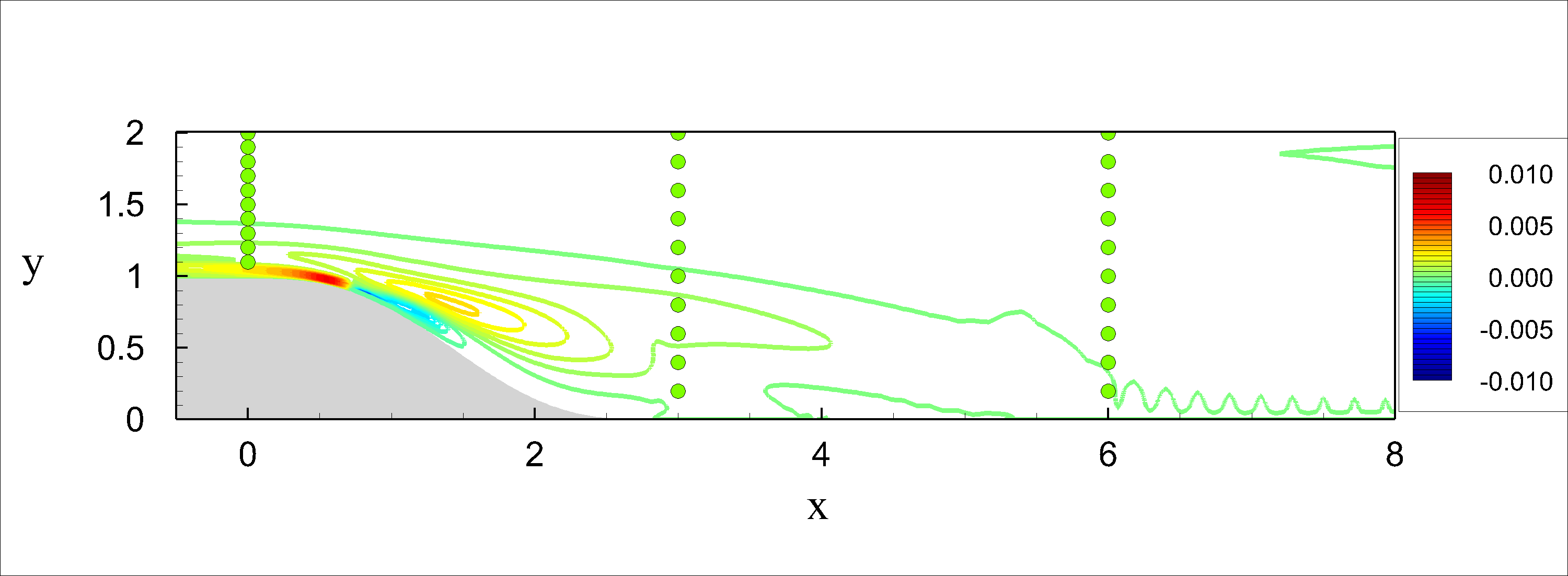} \\
(c) & \includegraphics[trim={0.5cm 0.5cm 0.5cm 0.5cm},clip,width=7cm]{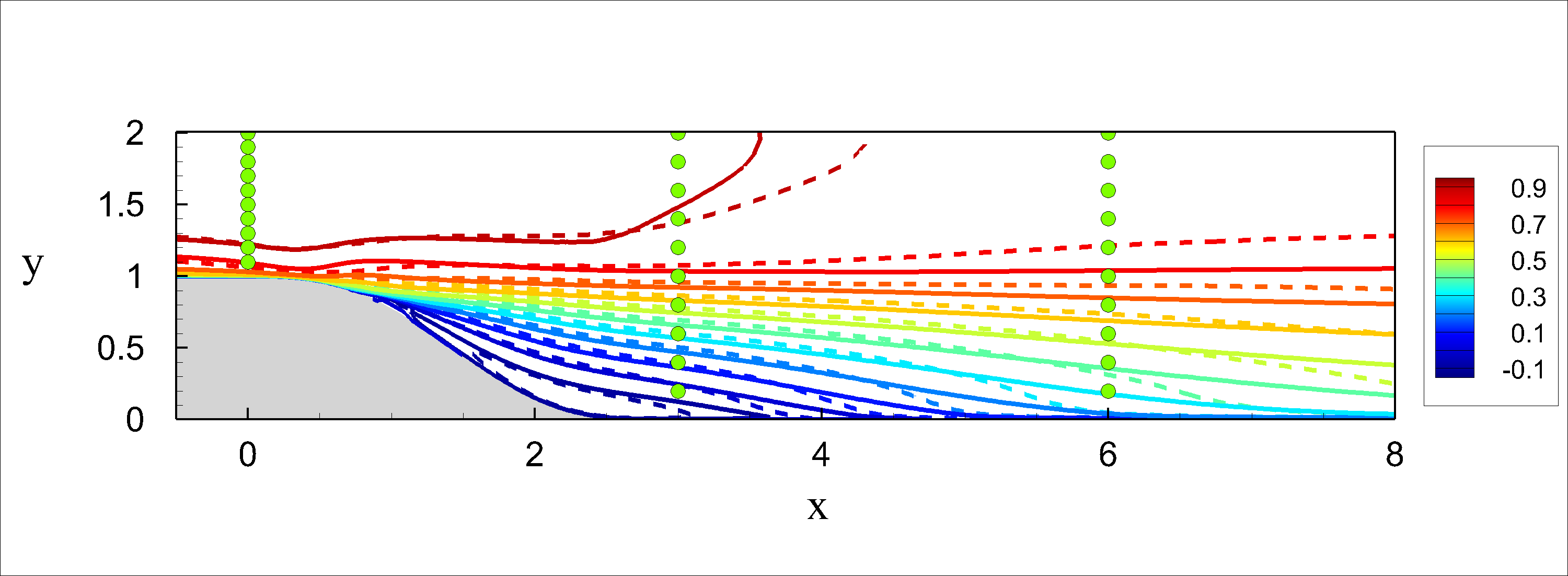} \\
(d) & \includegraphics[trim={0.5cm 0.5cm 0.5cm 0.5cm},clip,width=7cm]{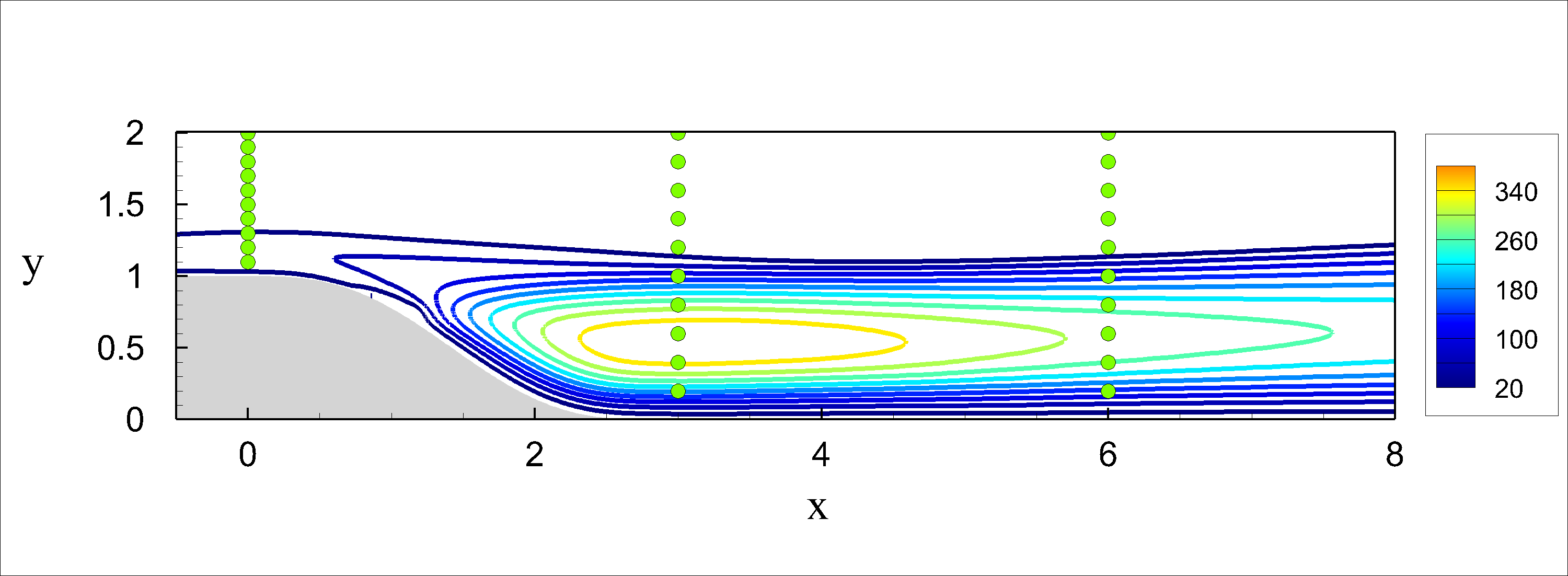} \\
(e) & \includegraphics[trim={0.5cm 0.5cm 0.5cm 0.5cm},clip,width=7cm]{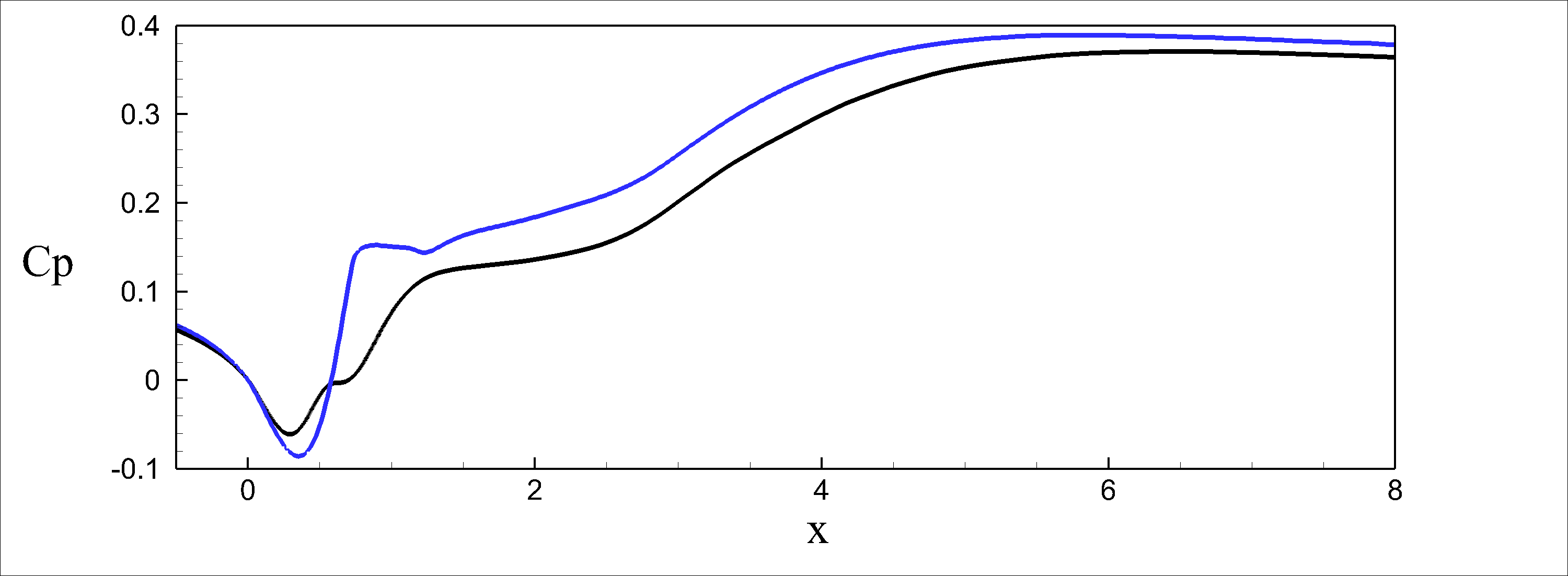} \\
(f) & \includegraphics[trim={0.5cm 0.5cm 0.5cm 0.5cm},clip,width=7cm]{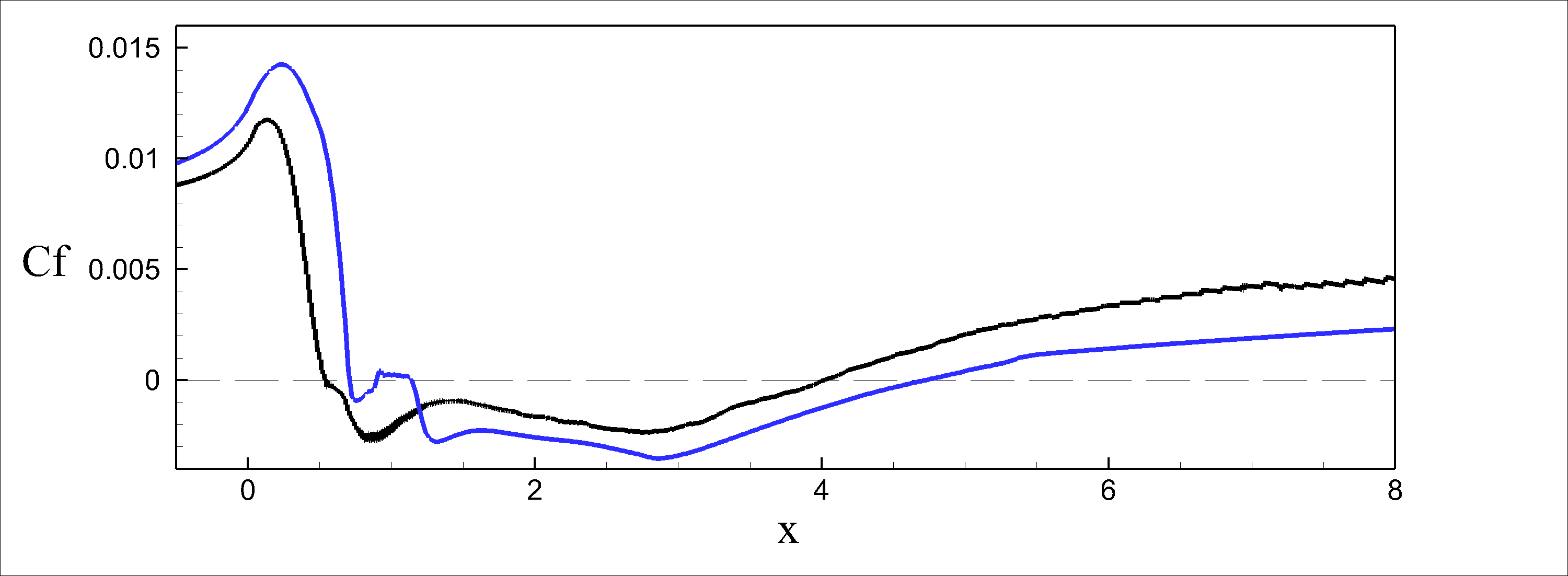}
\end{tabular}
\caption{Assimilation of \textit{sparse} velocity measurements with {\it constant eddy-viscosity profile} at the inlet. $N_x=3$ and $N_y=10$ measurement have been used.  For more details see caption of figure \ref{fig:J-u}.}
\label{fig:J-u-dots2}
\end{figure}

Results of the optimization with $N_x=3$ and $N_y=10$ are reported in 
figure \ref{fig:J-u-dots2} with the $\tffnu-$correction. These plots should be compared to the optimization results with dense data shown in the right plots of figure \ref{fig:J-u-cste}.
We can see that we reach $ e_{\Omega}=0.037$, which is very close to the dense value ($e_{\Omega}=0.035$). The $ \tffnu-$correction in figure (b) is seen to be less peaked at the separation point. All other plots are close.
Other optimization results for $ N_x$ and $N_y$
are reported in table \ref{tab:J-u-dots}: it is seen that the reconstruction error $e_\Omega$ is not very sensitive to the number of available measurements.
Hence, in the case of sparse measurement data, the $ \tffnu-$ correction provides robust reconstruction results that are close to the dense case. We say thus that this model presented itself as `rigid', or almost insensitive to the amount of data to be assimilated.

\subsubsection{Results for the $\tffx-$correction}

\begin{figure}
\centering
\begin{tabular}{cccc}
& $\tffx-$corr. SPARSE CST & &  Penal. $\tffx-$corr. SPARSE CST \\
(a) & \includegraphics[trim={0.5cm 0.5cm 0.5cm 0.5cm},clip,width=6cm]{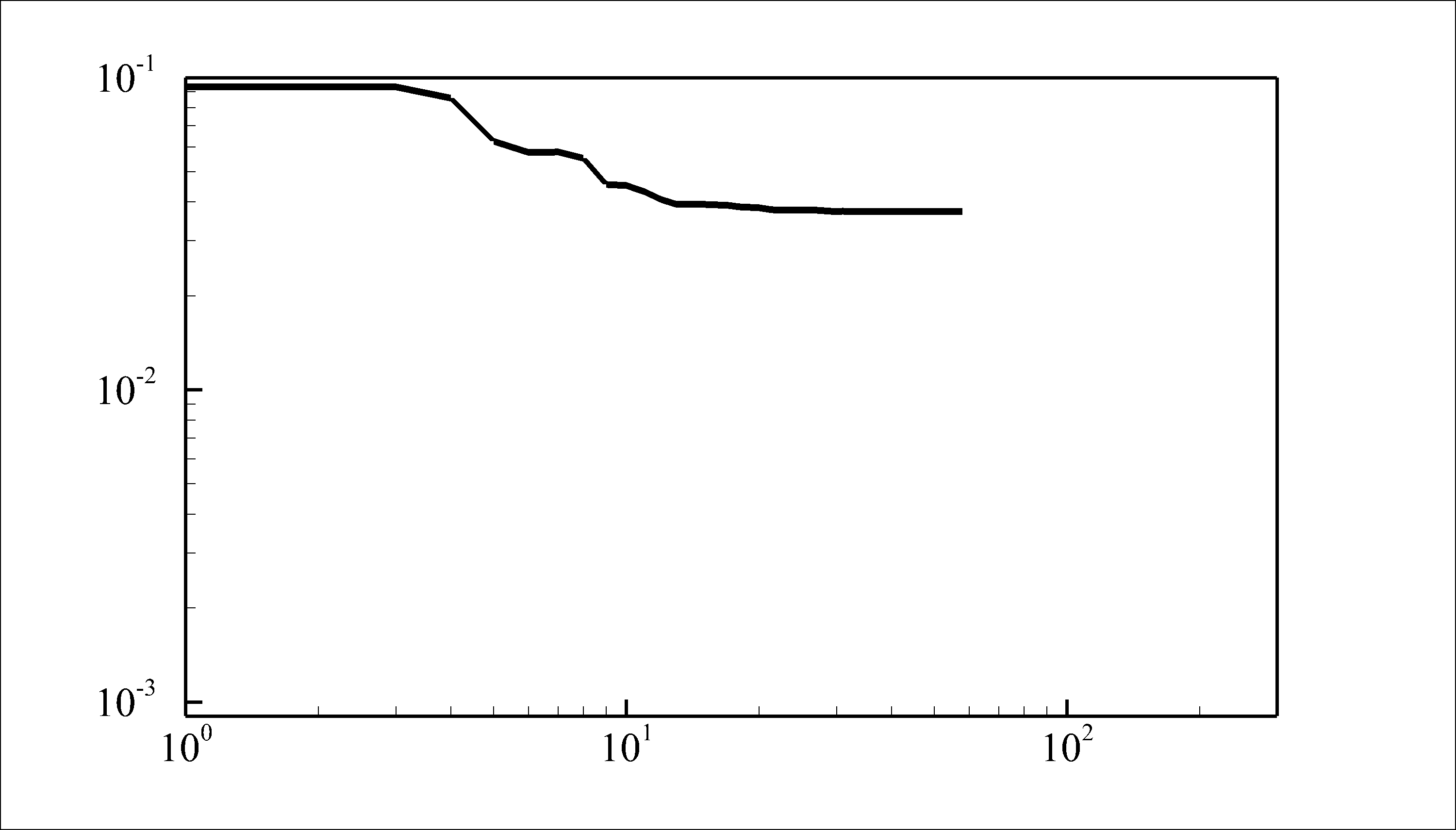} & (b) & \includegraphics[trim={0.5cm 0.5cm 0.5cm 0.5cm},clip,width=6cm]{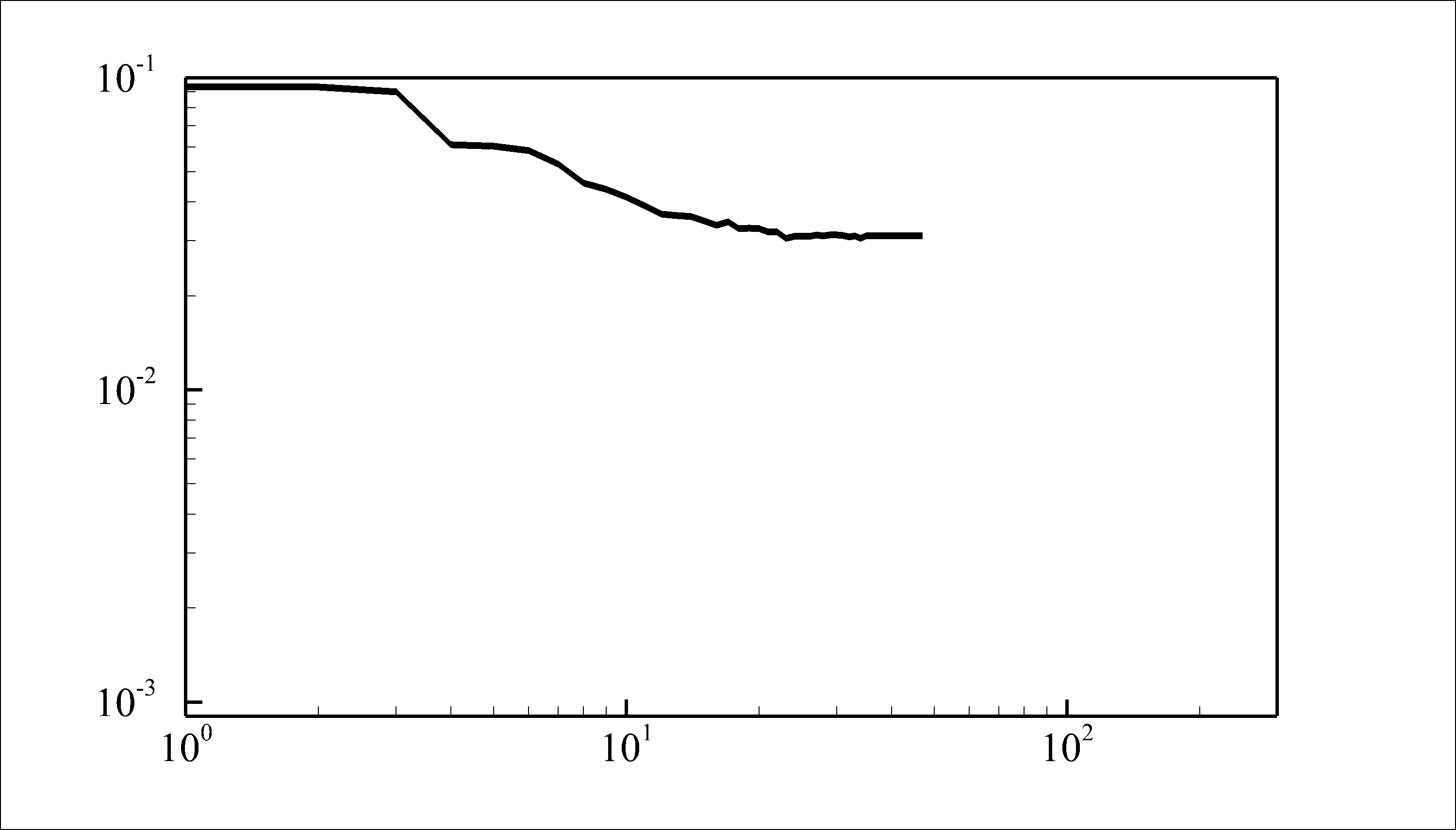} \\
(c)   & \includegraphics[trim={0.5cm 0.5cm 0.5cm 0.5cm},clip,width=7cm]{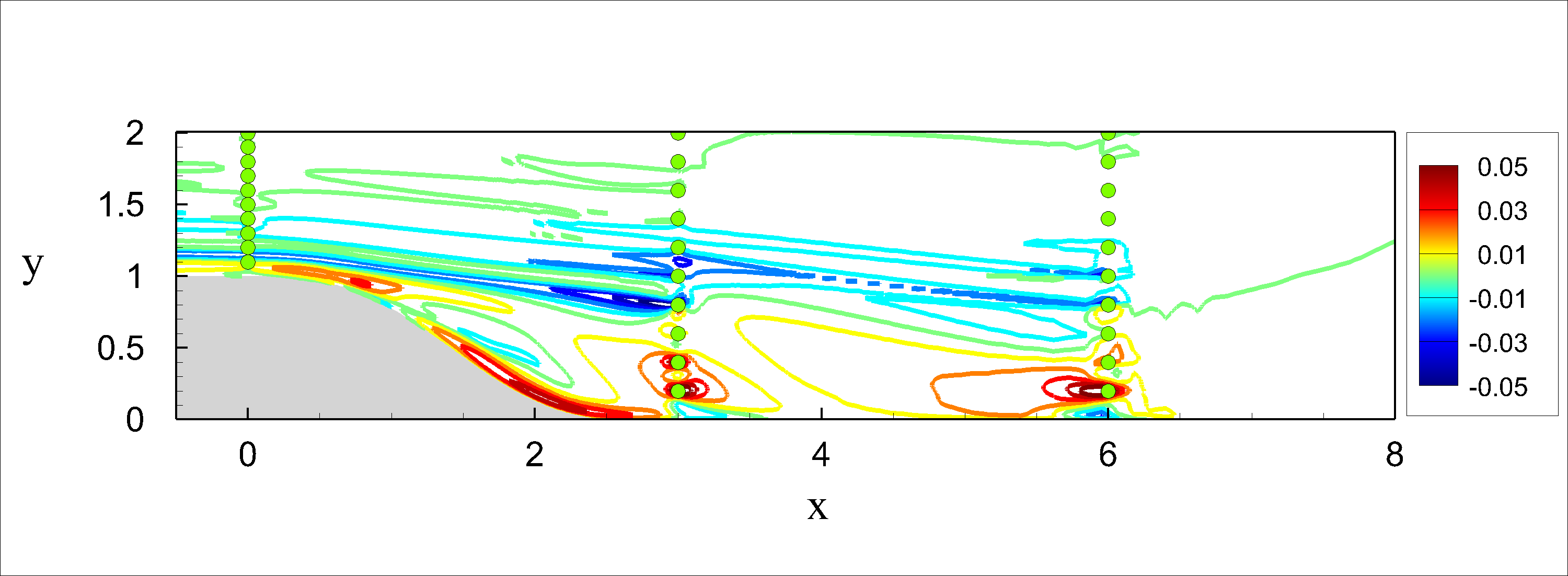} & (d) & \includegraphics[trim={0.5cm 0.5cm 0.5cm 0.5cm},clip,width=7cm]{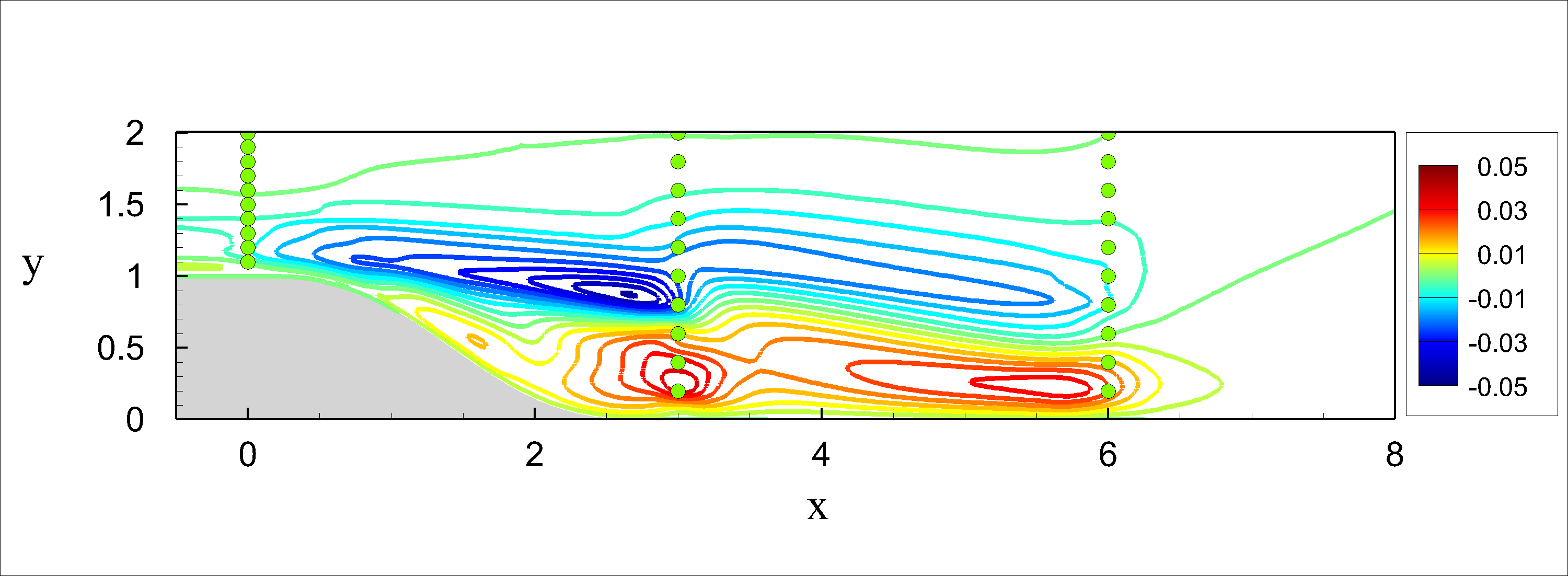} \\
(e) & \includegraphics[trim={0.5cm 0.5cm 0.5cm 0.5cm},clip,width=7cm]{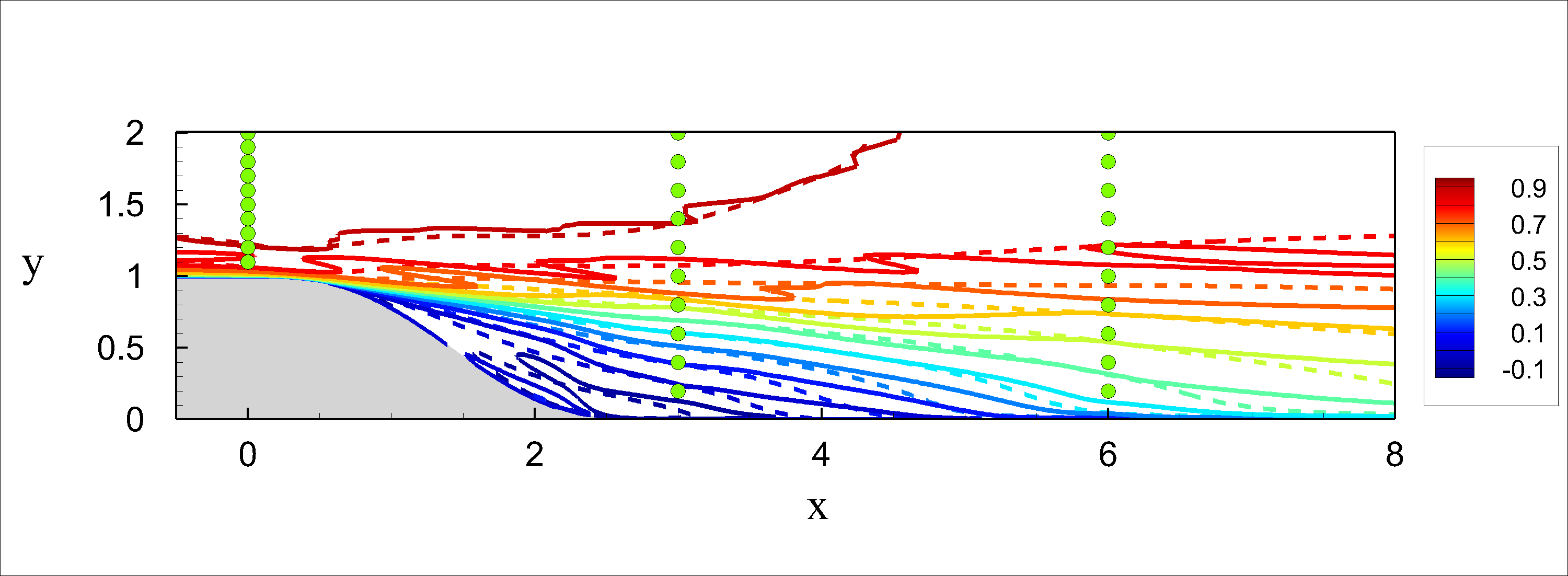} & (f) & \includegraphics[trim={0.5cm 0.5cm 0.5cm 0.5cm},clip,width=7cm]{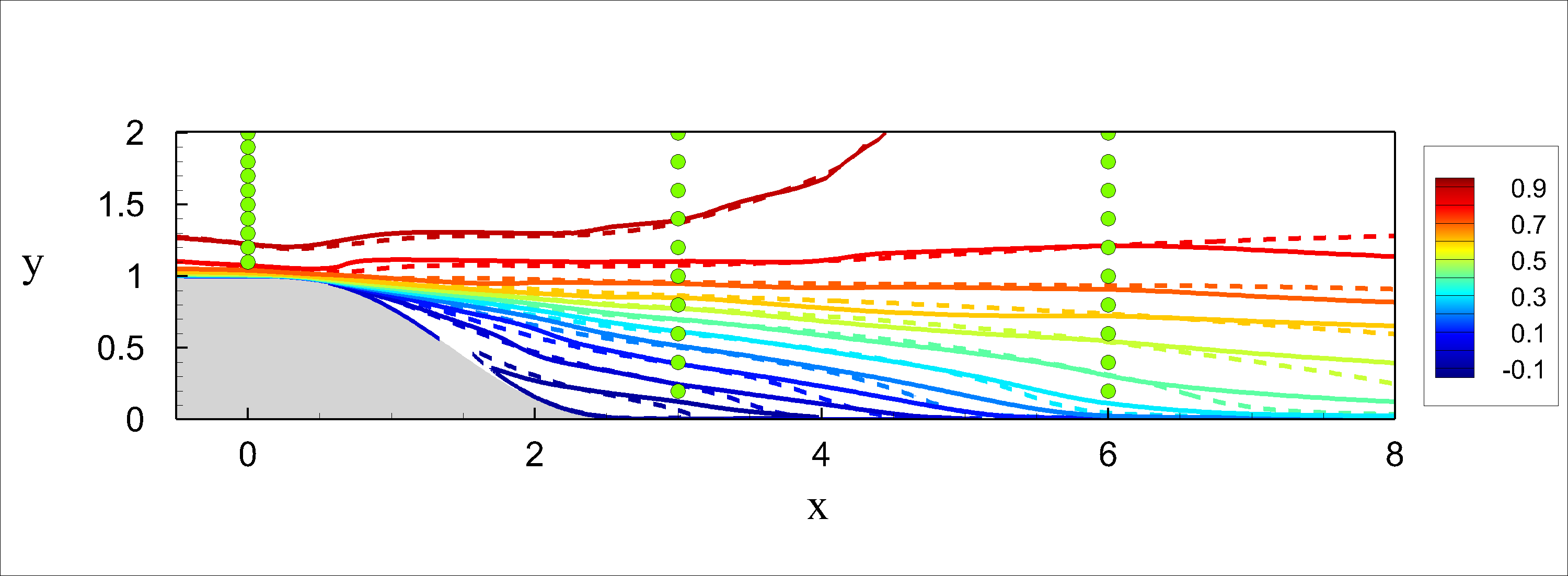} \\
(g) & \includegraphics[trim={0.5cm 0.5cm 0.5cm 0.5cm},clip,width=7cm]{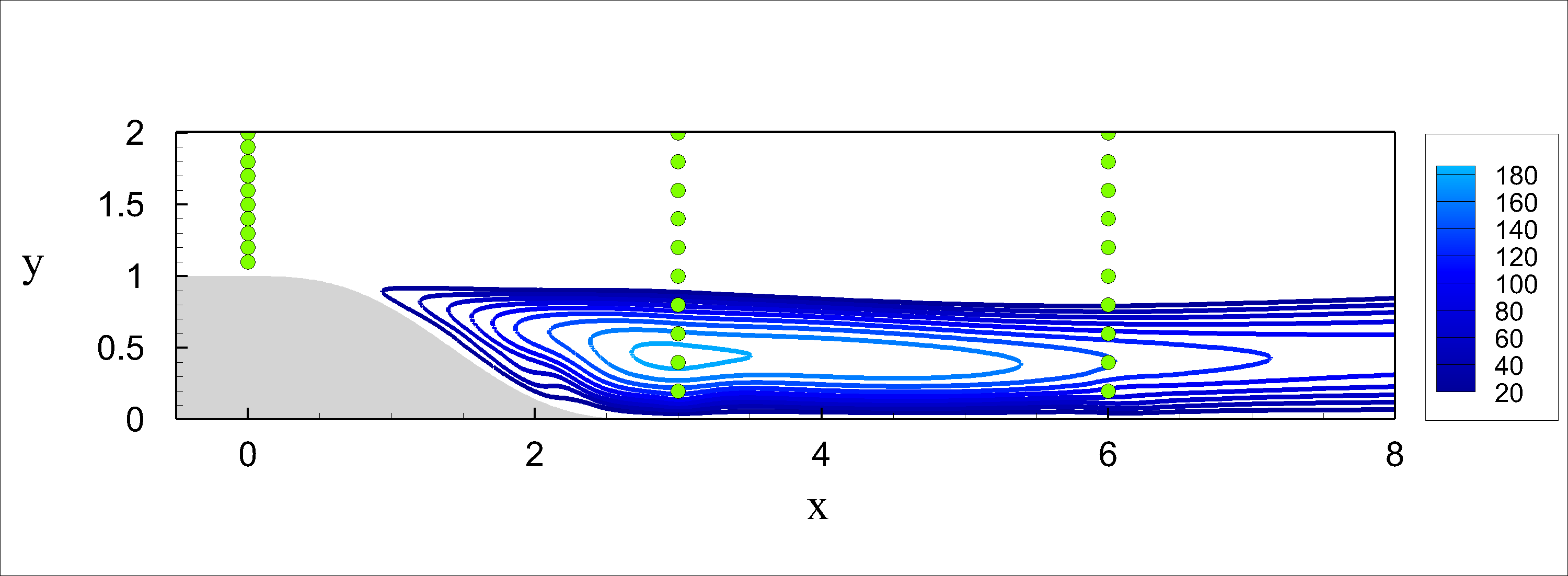} & (h) & \includegraphics[trim={0.5cm 0.5cm 0.5cm 0.5cm},clip,width=7cm]{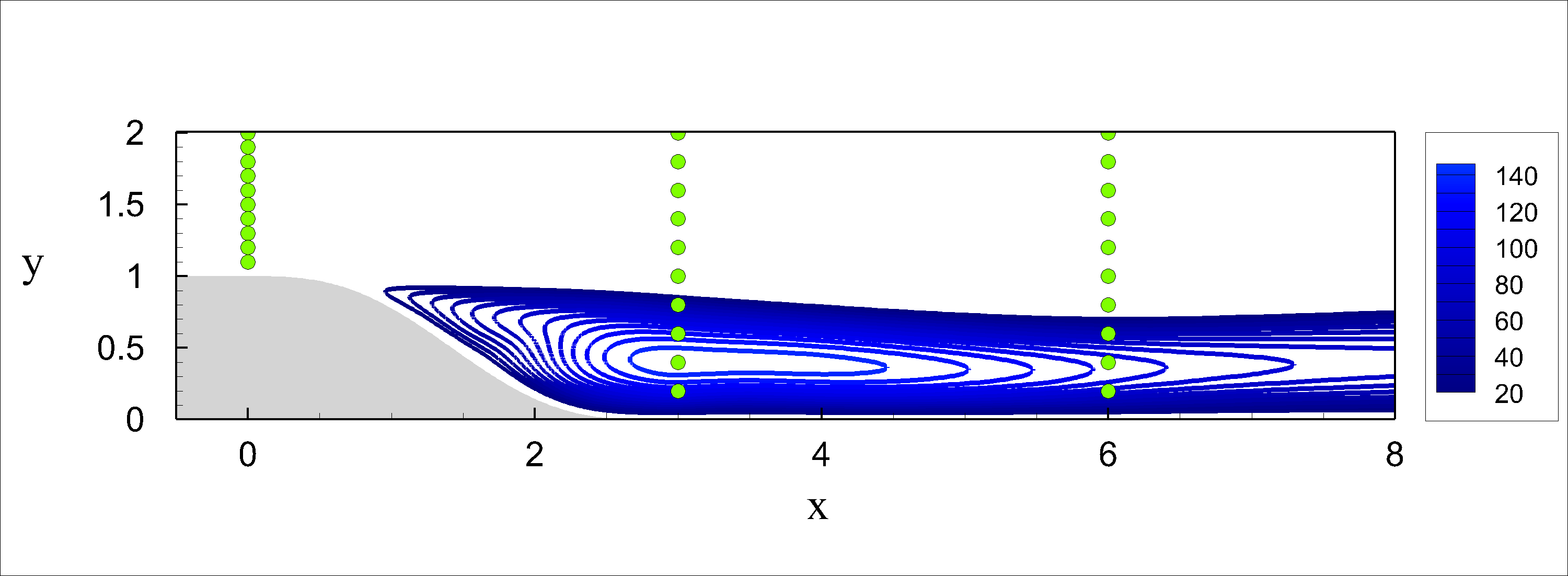} \\
(i) & \includegraphics[trim={0.5cm 0.5cm 0.5cm 0.5cm},clip,width=7cm]{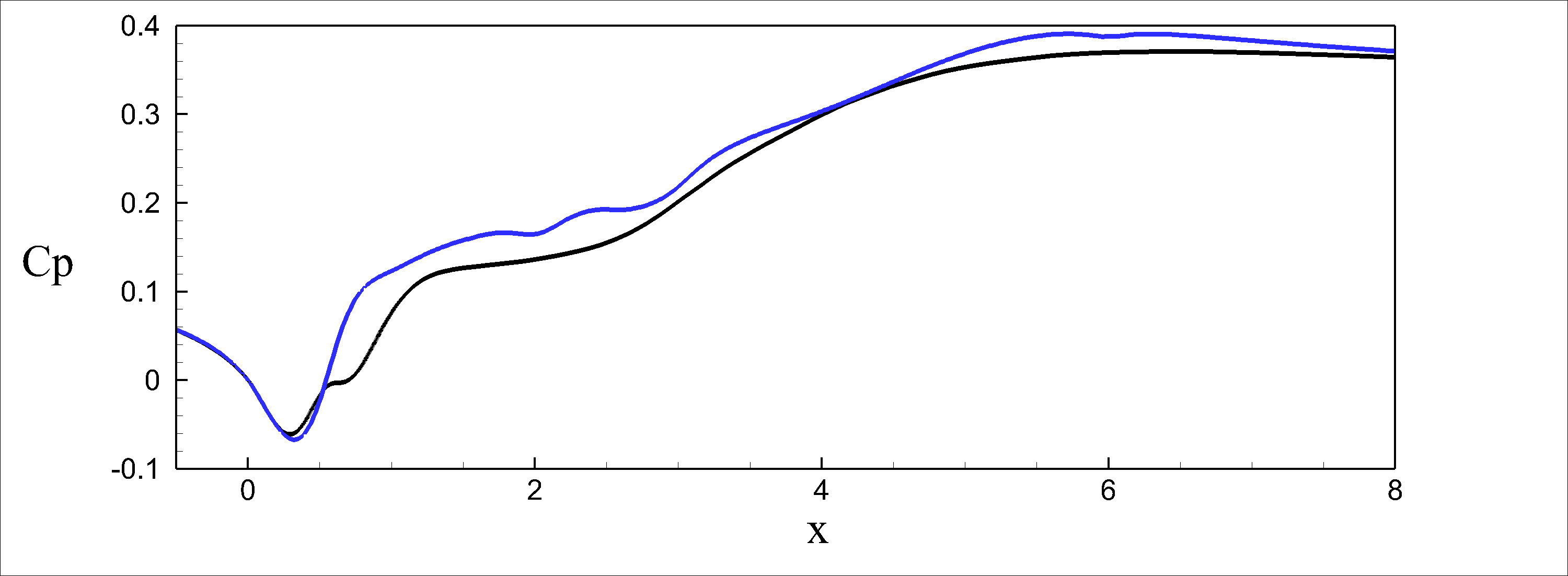} & (j) & \includegraphics[trim={0.5cm 0.5cm 0.5cm 0.5cm},clip,width=7cm]{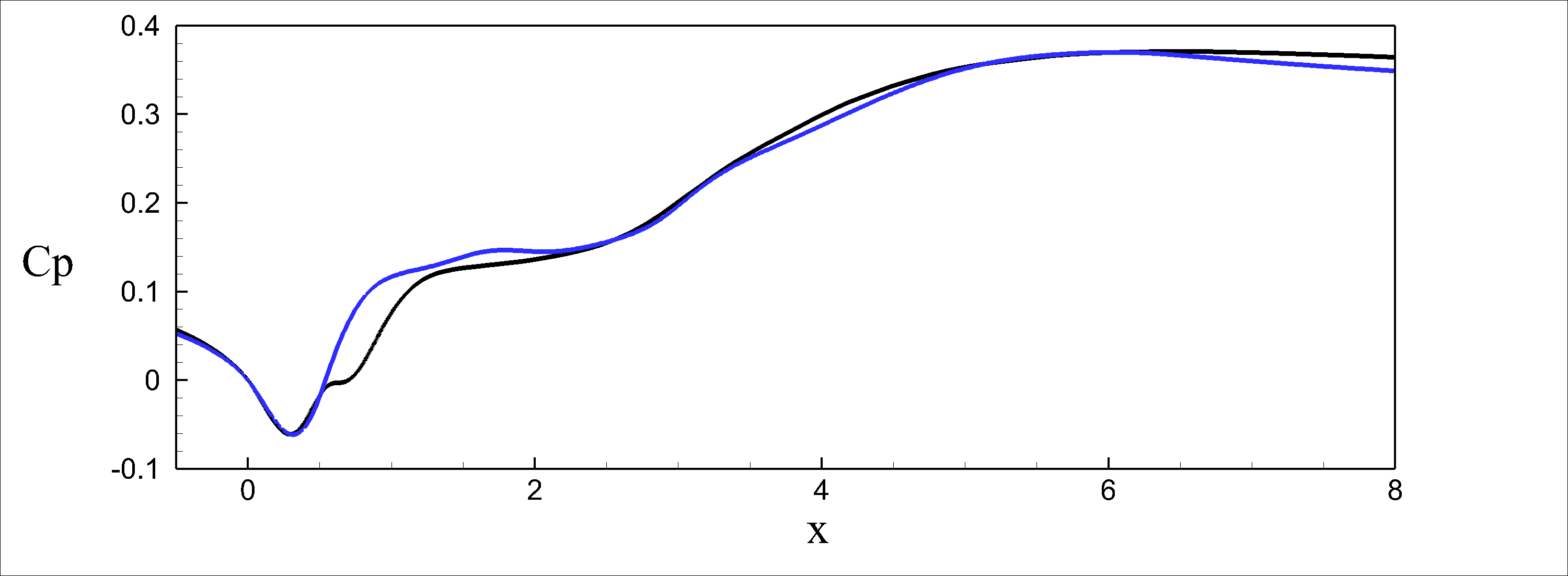} \\
(k) & \includegraphics[trim={0.5cm 0.5cm 0.5cm 0.5cm},clip,width=7cm]{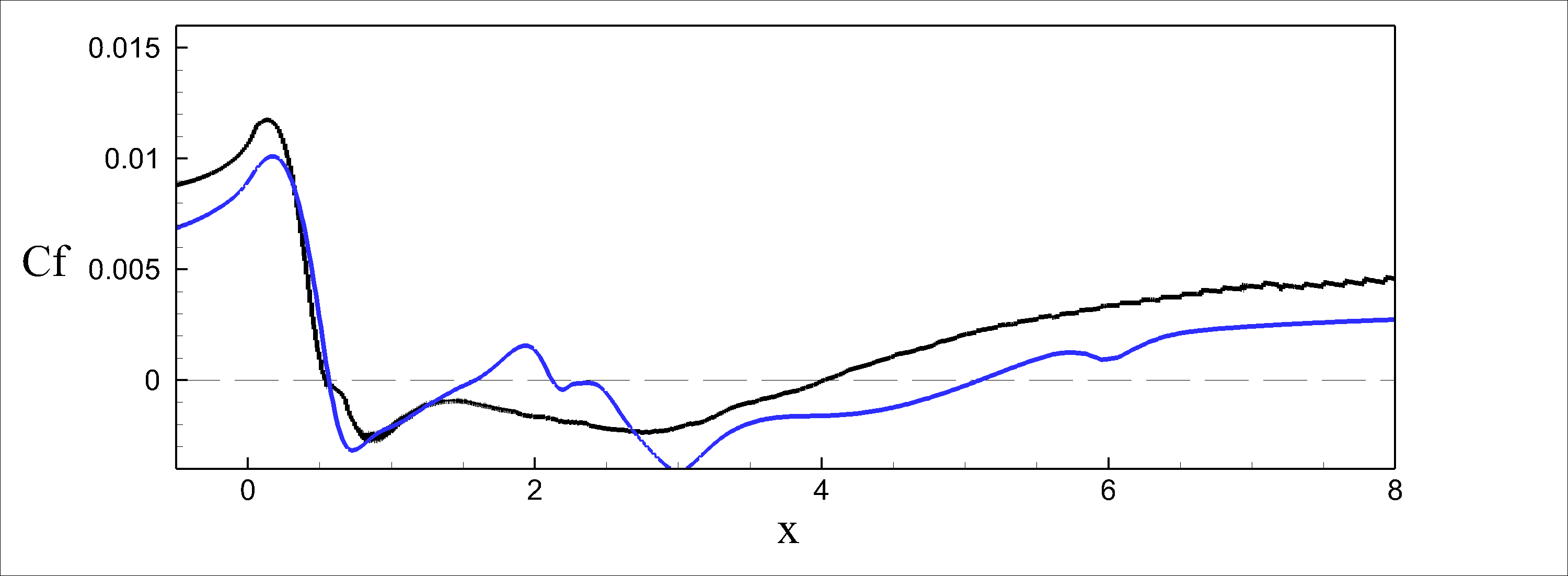} & (l) & \includegraphics[trim={0.5cm 0.5cm 0.5cm 0.5cm},clip,width=7cm]{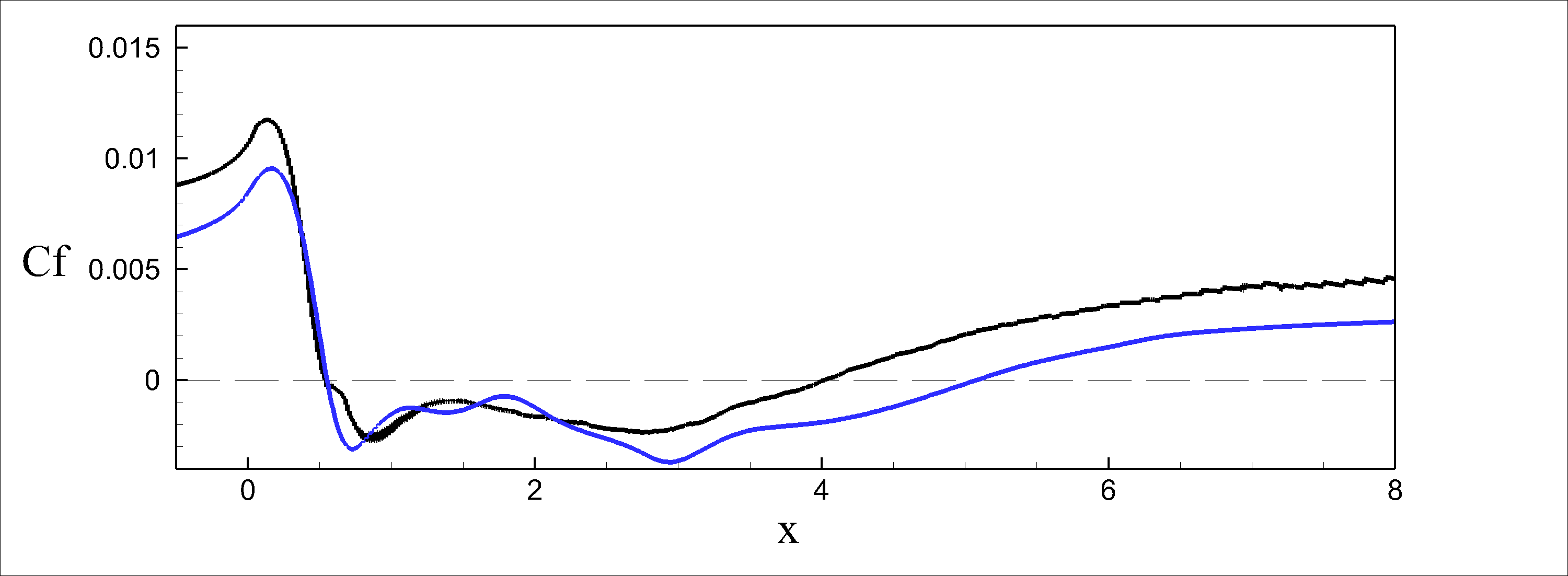}
\end{tabular}
\caption{Assimilation of \textit{sparse} velocity measurements with {\it constant eddy-viscosity profile} at the inlet. $N_x=3$ and $N_y=10$ measurement have been used. For more details see caption of figure \ref{fig:J-u}.}
\label{fig:J-u-dots}
\end{figure}

Optimization results with $N_x=3$ and $N_y=10$ for the $ \tffx-$correction are shown in the left plots of figure \ref{fig:J-u-dots}.
We can see from figure \ref{fig:J-u-dots} (e) that the iso-contours of the reconstructed velocity exhibit non physical wiggles. Also, the wall-friction coefficient shown in figure \ref{fig:J-u-dots} (k) shows two large re-circulation bubbles instead of one. This can be explained by observing that the $\tffx-$correction is given by a linear combination of adjoint velocity fields with right-hand-sides given by \eqref{eqn:adj-J-u-dots}. Since those equations involve a sum of Dirac masses on the momentum equations, the adjoint velocity fields are `peaked' around the measurements, which contaminates the solution, leading to those undesired oscillations. On the other hand, this does not occur with the $\tffnu$-correction, since the $\tnu$-adjoint field is not directly forced by Dirac masses, leaving the $\tnu$-adjoint field smooth. Varying the number of measurements, table \ref{tab:J-u-dots} shows that increasing the number of measurements improves the reconstruction error $ e_{\Omega} $ progressively ($e_\Omega=0.051$ for $ N_x=3$ and $N_y=5$ to $ e_{\Omega}=0.027$ for $ N_x=6$ and $ N_y=20$).
Hence, in the case of a larger number of measurements, the $ \tffx-$correction seems to exhibit more potential.

In the following, we show how to improve the reconstruction associated to the $ \tffx-$correction and in particular how to suppress the non-physical wiggles that appear when sparse measurements are considered. For this, we add some knowledge about the $ \tffx-$field by noting that the Reynolds-stress forcing usually exhibits large-scale structures that vary on the length scales of the mean-flow. Therefore we propose to penalize unphysical small-scale features by considering the following modified cost functional $\tilde{J}$:
\begin{equation}\label{eqn:penalizedfunction}
    \tilde{J} = \frac{J}{J_0} + \frac{\gamma^2}{2} \int_{\Omega} | \nabla \tff |^2 \; d\Omega
\end{equation}
Dividing the original cost functional $J$ by its value at the first iteration, we normalize the term related to the measurement discrepancy to a unity value, making the penalisation term independent of the measurement and only dependent on the parameter $\gamma^2$. This parameter should not be taken too small (the penalisation needs to be effective) and not too large (we still want to decrease the measurement discrepancy). This cost-functional makes the algorithm favour a {spatially} smoother solution, since the tuning term $ \tffx $ in the momentum-equations will exhibit smaller gradient values. This is a physically sound constraint since we know that the Reynolds-stress forcing is smooth. {It is interesting to notice that, in a non-deterministic framework, this extra term is related to the penalization of the objective function with the covariance matrix of the correction term.} 

The results for this modified data-assimilation procedure are shown in the right plots of figure \ref{fig:J-u-dots} for $\gamma^2 = 1$.
We can see that the resulting solution is smoother and still matches well the measurements ($J_n/J_0 \lessapprox 1 \%$, see table \ref{tab:J-u-dots}), suggesting that the penalisation (here  derivatives of $\tffx$) still allows enough freedom for the tuning field to match the measurements, while constraining the solution in a  smooth subspace.
Furthermore, in most of the situations (varying $N_x$ and $N_y$, see table \ref{tab:J-u-dots}), the penalized algorithm provides an overall better reconstruction of the flow (see $e_{\Omega}$ in table \ref{tab:J-u-dots}).
Indeed, the penalized $\tffx-$correction nearly always outperforms both the non-penalized $\tffx-$ and the $\tffnu-$reconstruction.
Only, in the cases where very few measurements are provided (here $N_x=3$, $N_y=5$), the error with $ \tffnu $ is smaller than with the penalized $\tffx-$term. This means that, whenever we have very few measurements, it may be preferable to use the more robust $ \tffnu-$correction than the more flexible penalized $ \tffx-$correction.

\begin{table}
	\centering
    \caption{Various optimization results (dense and sparse measurements) for the case of constant inflow eddy-viscosity profile.}\label{tab:J-u-dots}
    \begin{tabular}{cccccccccccc} \hline
		 & & $\;$ & \multicolumn{3}{c}{$e_{\Omega}$} & $\;$ & \multicolumn{3}{c}{$J_{n}/J_0$} \\ \hline
		    		 baseline &          & $\;$ &       & 0.094 &     & $\;$ &      & -- &    \\ \hline
		    & & $\;$ & $\tffx$ & $\tffx$, $\gamma^2 = 1$ & $\tffnu$ & $\;$ & $\tffx$ & $\tffx$, $\gamma^2 = 1$ & $\tffnu$ \\ \hline 
		\cmidrule{3-5}
		\cmidrule{6-8}
   		 Dense &         & $\;$ & $\sim 10^{-3}$   & --   & 0.035    & $\;$ & $\sim 10^{-4}$     & -- & 12.3 \%   \\ \hline
   		         & $N_y=5$  & $\;$ & 0.051 & 0.042 & 0.036  & $\;$ & $\sim 10^{-6}$ & 0.1 \% & 8.7 \% \\
   		 $N_x=3$ & $N_y=10$ & $\;$ & 0.037 & 0.031 & 0.037  & $\;$ & $\sim 10^{-6}$ & 0.3 \% & 8.7 \% \\
   		         & $N_y=20$ & $\;$ & 0.030 & 0.028 & 0.037  & $\;$ & $\sim 10^{-6}$ & 0.4 \% & 9.0 \% \\  
   		         & $N_y=5$  & $\;$ & 0.042 & 0.035 & 0.037  & $\;$ & $\sim 10^{-6}$ & 0.1 \% & 11.9 \% \\
   		 $N_x=6$ & $N_y=10$ & $\;$ & 0.032 & 0.030 & 0.037  & $\;$ & $\sim 10^{-6}$ & 0.5 \% & 9.4 \% \\
   		         & $N_y=20$ & $\;$ & 0.027 & 0.027 & 0.038  & $\;$ & $\sim 10^{-6}$ & 1.1 \% & 13.4 \% \\   
		\hline
	\end{tabular}
\end{table}

\subsection{Understanding the `rigidity' of the $\tffnu-$correction model through observability Gramian analysis}

This section is devoted to the understanding of the 'rigidity' of the $\tffnu-$correction model, pointed out in the previous section. One possible way to understand this observation is to examine the linearized optimization problem, say, around the RANS-SA solution. In this linear framework, a variation of the control vector $\delta \tff$ (here, either $\tffx$ or $\tffnu$) induces a variation of the state variable $\delta \tqq = (\delta \tuu,\delta \tp,\delta \tnu)$ that satisfies:
\begin{equation}\label{eqn:linearizedeq}
    \delta \tqq = ( \partial_{\tqq} \mathcal{R} )^{-1} P \delta \tff, 
\end{equation}
where {$\partial_{\tqq} \mathcal{R} $ denotes the linearization of the RANS-SA equations with the state variable} and $P$ a linear operator that maps the control vector variations to the actual forcings of the Jacobian. This variation in the state induces a variation on the measurements according to:
\begin{equation}\label{eqn:linearizedmeas}
    \delta \tmm =  ( \partial_{\tqq} \mathcal{M} ) \delta \tqq = \mathcal{A} \, \delta \tff,
\end{equation}
where the operator $\mathcal{A} = ( \partial_{\tqq} \mathcal{M} ) ( \partial_{\tqq} \mathcal{R} )^{-1} P$ is obtained after using  (\ref{eqn:linearizedeq}).

To be able to identify which forcing term induces the most energetic variation on the measure, we may optimize the gain:
\begin{equation}\label{eqn:gain}
    G (\delta \tff) = \frac{|| \delta \tmm ||_{M}^2}{|| \delta \tff ||^2} = \frac{ \left \langle \delta \tmm , \delta \tmm \right \rangle_{M}}{ ( \delta \tff , \delta \tff )_{\Omega}}.
\end{equation}
This can straightforwardly be done by solving the eigenvalue problem:
\begin{equation}
    \mathcal{A}^{\dagger} \mathcal{A} \delta \tff_i = \lambda_i^2 \delta \tff_i, 
\end{equation}
where the positive eigenvalues $\lambda_i^2$ are ranked by decreasing order ($\lambda_i^2\geq\lambda_{i+1}^2$). By taking $\delta \tff=\delta \tff_i$ in (\ref{eqn:gain}), we obtain $ G=\lambda_i^2$. The values $ \lambda_i$ quantifies the measurement variations
along the unit optimal measurement directions $\delta \tmm_i$ =$ \lambda_i^{-1} \mathcal{A} \delta \tff_i$, induced by the unit optimal forcing directions $\delta \tff_i$.
 If all the eigenvalues-values are of the same order, we can state that any measurement $\delta \tmm$ is equally reachable (or, more suitably, `observable') and {the non-linear baseline model can be considered as more flexible}. If we have a strong separation of singular-values, that is $\lambda^2_0 \gg \lambda^2_i$ for some index $i$, we have measurement states that cannot be reached ('observed') with the chosen model, {that is therefore more rigid}.  This can be seen by considering that, for a given $\delta \tff$, the resulting perturbation on the measure is given by:
\begin{equation}
    \delta \tmm = \mathcal{A} \delta \tff = \mathcal{A} \sum_i \alpha_i \delta \tff_i = \sum_i \alpha_i \lambda_i \delta \tmm_i = \lambda_0 \sum_i \alpha_i \frac{\lambda_i}{\lambda_0} \delta \tmm_i.
\end{equation}
Since $\delta \tmm_i$ is unit norm, 
the amplitude $ \alpha_i$ required for the forcing term to achieve a unit measurement variation $\delta \tmm_i$ scales as $\lambda_0/\lambda_i$, which can be very large, and therefore not achievable.

\begin{figure}
\centering
    \includegraphics[trim={1cm 1cm 1cm 1cm},clip,width=8cm]{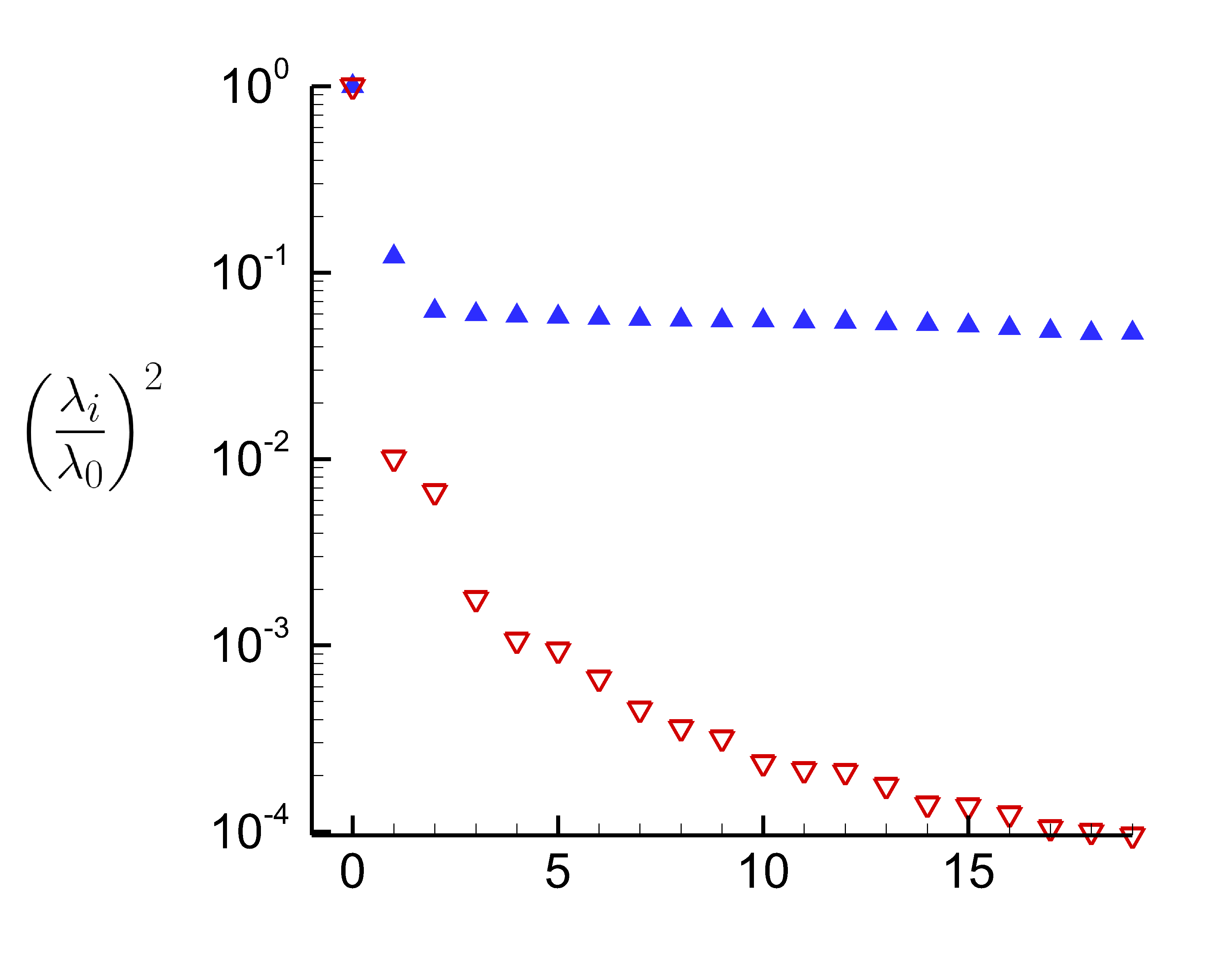}
\caption{Separation of eigenValues $(\lambda_i/\lambda_0)^2$ for the $\tffx-$correction model (blue triangles) and the $\tffnu-$correction model (red triangles). Variations have been performed with the baseline RANS-SA model and with the full-state measurement operator ${\mathcal M}(\tilde{\mathbf{u}})=\tilde{\mathbf{u}}$. }\label{fig:J-u-gains}
\end{figure}

\begin{figure}
\centering
\begin{tabular}{lclc}
 & $\tffx-$corr. & & $ \tffnu-$corr. \\
     (a)  & \includegraphics[trim={0.5cm 0.5cm 0.5cm 8cm},clip,width=7cm]{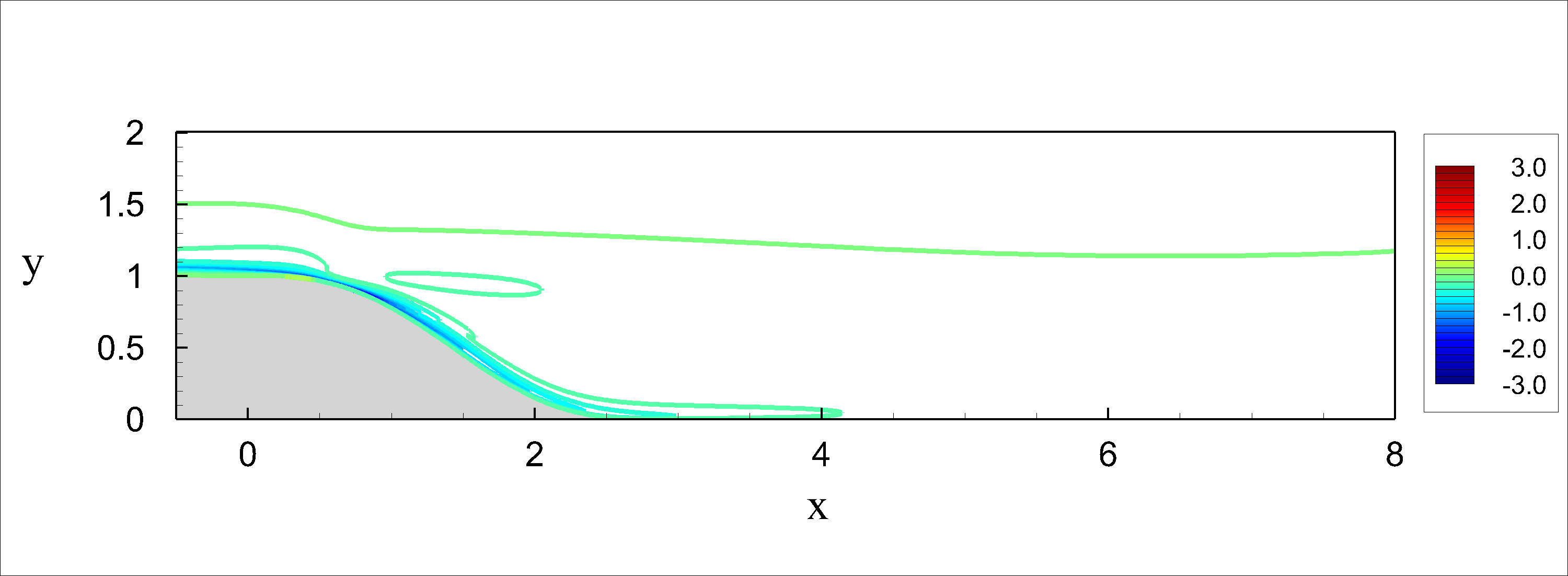} & (b)  
       & \includegraphics[trim={0.5cm 0.5cm 0.5cm 8cm},clip,width=7cm]{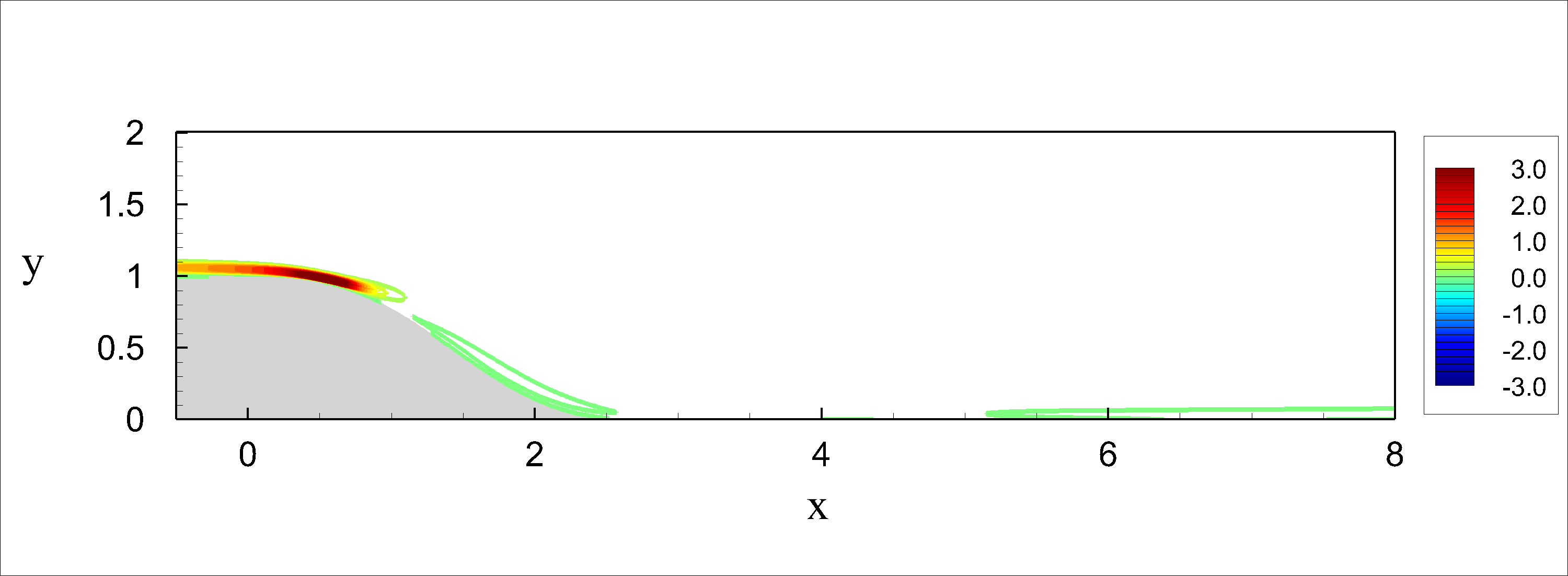} \\
  (c)     & \includegraphics[trim={0.5cm 0.5cm 0.5cm 8cm},clip,width=7cm]{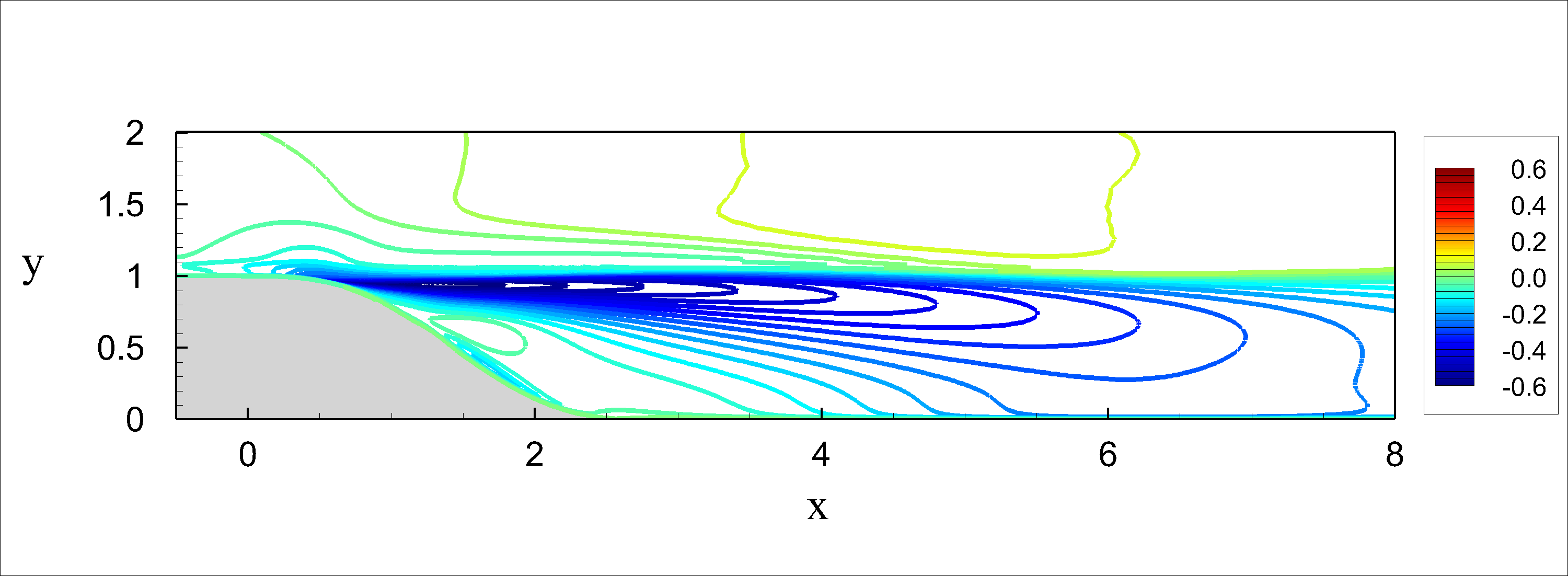} & (d)
       & \includegraphics[trim={0.5cm 0.5cm 0.5cm 8cm},clip,width=7cm]{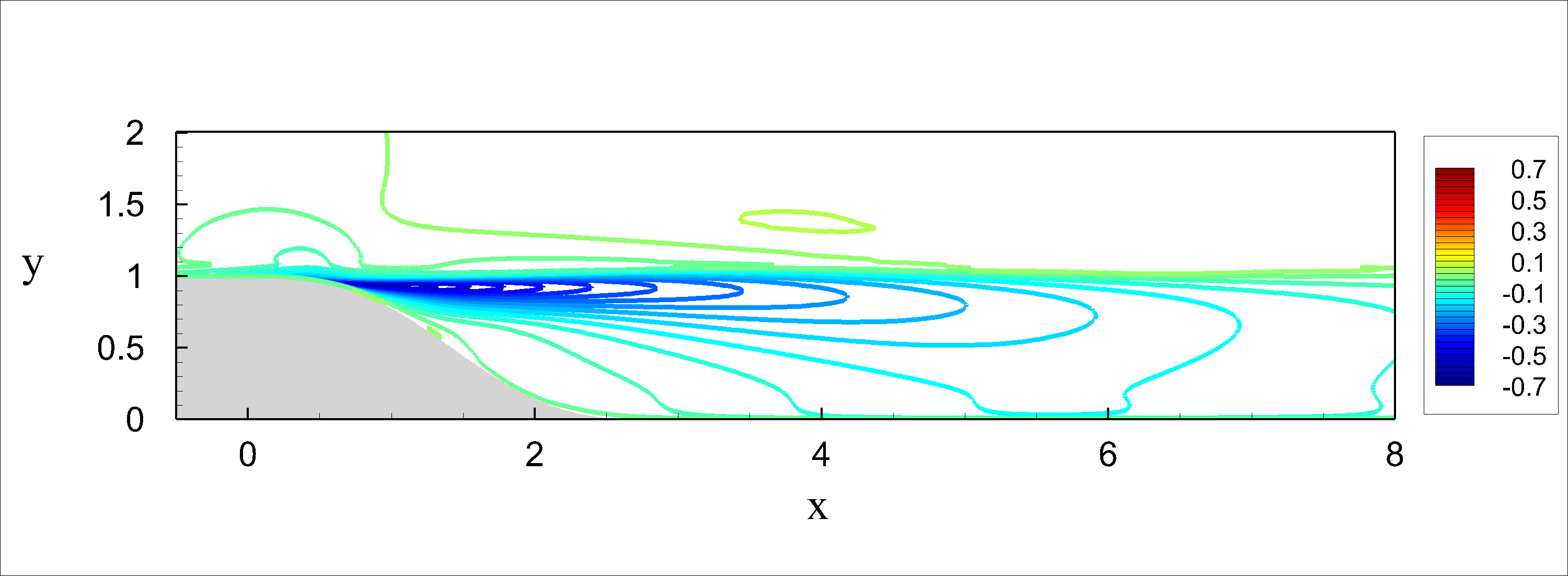} \\
\end{tabular}
\caption{Gramian modes: streamwise component of optimal forcing $\delta \tff_0$ (a,b) and optimal response $\delta \tmm_0$ (c,d).
(a,c): $ \tffx-$correction,
(b,d): $ \tffnu-$ correction.
Variations have been performed with the baseline RANS-SA model and with the full-state measurement operator ${\mathcal M}(\tilde{\mathbf{u}})=\tilde{\mathbf{u}}$.
}
\label{fig:J-u-gramian-modes}
\end{figure}

In figure \ref{fig:J-u-gains}, we display the quantities $(\lambda_i/\lambda_0)^2$ for the $ \tffx-$ and $\tffnu-$correction models, considering the full velocity field $\mathcal{M} (\tqq) = \tuu$ as the measurement operator.
We can see that {more measurement states $\delta \tmm$ can be reached with the $ \tffx-$correction compared to the $ \tffnu-$correction, for which the separation between the eigenvalues is very strong and increases very rapidly for $ i>1$. This shows that the (linearized) $\tffx-$correction model is much more flexible than the $\tffnu-$one, in agreement with results of the previous section.} In figure \ref{fig:J-u-gramian-modes}, we can also see that the leading forcing mode for both models is localized in the vicinity of the separation point and they both produce a similar measurement perturbation. This shows that, for both models, the separation point is the region where the tuning terms are most efficient to correct the reconstructed field and that the associated change in the reconstructed field consists in modifying the length of the separation bubble. Hence, both models are indeed able to correct easily this important feature of backward-facing step flow.

\section{Conclusion}

In this paper, we have introduced a methodology for the reconstruction of mean-flow features from mean-velocity measurements for high-Reynolds number turbulent flows.
The baseline model is in all cases the RANS Spalart--Allmaras model. Two correction terms to account for modelling uncertainties in the Spalart--Allmaras model have been introduced. These uncertainties have been tuned thanks to the knowledge of external measurements by minimizing the measurement discrepancy between the model and the data.
The first correction-term consists in adding a volume-source term $ \tffx$ in the momentum equations, in a similar way as in \citet{Foures14}, the difference being that we have a background eddy-viscosity turbulence model, making the numerical procedure more robust and well-posed for high Reynolds number flows. The second one consists in the correction of the equation governing the eddy-viscosity. We have then considered a smooth turbulent backward-facing step flow to showcase the methodology. If the whole velocity field is known (like in a PIV setup), the $\tffx-$correction model produces a solution that matches exactly the reference, showing the high flexibility of this model.
On the contrary, the $\tffnu-$ correction model is not capable to reproduce exactly the reference state. The model is too `rigid' and lots of measurements are not accessible with this model. If only few point-wise velocity measurements are available, we have shown that the first model, despite its high flexibility to exactly recover the measurements, may lead to noisy / unphysical state reconstructions. For this reason, an additional penalisation of the gradients of the tuning field had to be considered to obtain smooth physical solutions. As for the second model, it turned out that, independently of the quantity of available measurements, the optimal state was approximately the same. {The Boussinesq hypothesis induces a strong constraint on the reconstructed state. The correction term acting only by modification of the diffusion strength $ \nu_t$, it is possible, for example, to correct the expansion rate of the  shear-layer, but not to finely adapt the flowfield in the vicinity of the attachement point}. The flexibility / rigidity of the two considered models has finally been analysed by  looking at the input / output properties of the linear operator between the forcing space and the measurement space. We showed that the controllability/observability of the $\tffnu-$correction model was actually restricted to a rather small subspace. Such a rigidity property may be helpful whenever the number of measurements is very low since the reconstructed state is almost independent of the number of measurements. Hence, the $ \tffnu-$correction should be favoured in the case of very few available measurements while the penalized $\tffx-$correction is more suited in the case of many measurements. 
{As a possible alternative to the scenario presented in this paper, we could point to other kinds of RANS-based turbulence modeling such as Reynolds-Stress Models (RSM), where no Boussinesq hypothesis is assumed but the whole tensor is modeled by additional equations which can be perturbed by some tuning parameter. Since this model is (in theory) fully controlable/observable, it could accommodate an arbitrary number of measurements, while retaining (hopefully) physical features.}.
To conclude with, we note that RANS-based data-assimilation is a way to shed some new light on a field, turbulence modeling, in which progress has been difficult over the last 30 years, {since data-assimilation can be regarded as a first step towards data-driven turbulence modeling}. 

\appendix
\section{Spalart--Allmaras model}\label{apd:SA}

Based on the definition given by equation \ref{eqn:SA-tnu}, we have the production, destruction and cross-diffusion terms, respectively:
\begin{equation}
P(\tilde{\nu},\nabla \overline{\mathbf{u}}) = c_{b1} \tilde{\nu} \tilde{S}, \;\;\; D(\tilde{\nu},\nabla \overline{\mathbf{u}}) = c_{w1} f_{w1} \left( \frac{\tilde{\nu}}{d} \right)^2, \;\;\; C(\nabla \tilde{\nu}) = \sigma^{-1} |\nabla \tilde{\nu}|
\end{equation}
the eddy-viscosity $\nu_t$ and the SA diffusivity $\eta$ given by:
\begin{equation}
\nu_t = \left\{\begin{matrix}
\tilde{\nu} f_{v1}, & \tilde{\nu} \geq 0
\\ 
0, & \tilde{\nu} < 0
\end{matrix}\right.
, \;\;\; \eta(\tilde{\nu}) = \nu \left( 1 + \chi + \frac{\chi^2}{2} \right)
\end{equation}
Those functions are all $\mathcal{C}^1$ with respect to the state, since the auxiliary function $f_{v1}$ has a smooth and null first derivative at $\tilde{\nu} = 0$, leaving the Jacobian of the system continuous. The auxiliary functions that close the model are:
\begin{equation}
\begin{split}
\chi = \frac{\tilde{\nu}}{\nu}, \;\;\;  f_{v1} = \frac{\chi^3}{c_{v1}^3 + \chi^3}, \;\;\; \tilde{S} = | \nabla \times \overline{\mathbf{u}} | + \frac{ \tilde{\nu} f_{v2} }{k^2 d^2}, \;\;\; f_{v2} = 1 - \frac{\chi}{1 + \chi f_{v1}} \\
f_w = g \left[ \frac{1 + c_{w3}^6}{g^6 + c_{w3}^6} \right]^{\frac{1}{6}}, \;\;\; g = r + c_{w2}( r^5 - r ), \;\;\; r' = \frac{\tilde{\nu}}{\tilde{S} k^2 d^2} \\
r = \left\{\begin{matrix}
r', & 0 \leq r' \leq 10
\\ 
10, & r' < 0, r'>10
\end{matrix}\right. , \;\;\; g_n = 1 - 1000 \frac{\chi^2}{1 + \chi^2}
\end{split}
\end{equation}
Although the function $r = r(r')$ is not differentiable (and not even continuous at $r'=0$), the function $r = r(\tilde{\nu},\nabla \overline{\mathbf{u}})$ is (see \citet{Crivellini13} for more details).

\section{SUPG Implementation}\label{apd:SUPG}

The numerical implementation of the RANS-SA equations are based on the Finite Element Method (FEM), available in the FreeFem++ code (see \citet{Hecht12}). Since FEM is naturally numerically unstable at high Reynolds numbers, some stabilization scheme needs to be employed. Here, we choose the Streamline-Upwind Petrov-Galerkin (SUPG) formulation, as proposed by \citet{brooks1982streamline}. In this formulation, the test function is advected with the local velocity field, giving an upwind effect, stabilizing the scheme. Several different formulations have been proposed in the literature (see \citet{franca1992stabilized1}, \citet{franca1992stabilized2}) for various different equations. Here, we employ a simplified version of it, common for unsteady problems (\citet{bao2011numerical}), where only the advection terms are treated. In a simplified notation, we write the nonlinear residual of the RANS-SA equations in the weak form:
\begin{equation}
    \begin{split}
        R([\tuu,\tp,\tnu],[\tvv,\tilde{q},\check{\nu}]) & = \int_{\Omega} \left( \tuu \cdot \bnabla \tuu \right) \cdot \tvv + \int_{\Omega} \left( - p I + (\nu+\nu_t) \nabla_s \tuu \right) : \bnabla \tvv - \int_{\Omega} \left( \bnabla \cdot \tuu \right) \tilde{q} \\
        & + \int_{\Omega} \left( \tuu \cdot \bnabla \tnu - s \right) \check{\nu} + \int_{\Omega} \eta \bnabla \tnu \cdot \bnabla \check{\nu} \\ 
		& + \sum_{\Omega_k} \int_{\Omega_k} \tau_{SUPG} \tuu \cdot \bnabla \tvv \left( \tuu \cdot \bnabla \tuu \right) + \sum_{\Omega_k} \int_{\Omega_k} \tau_{SUPG} \uu \cdot \bnabla \check{\nu} \left( \tuu \cdot \bnabla \tnu \right)
    \end{split}
\end{equation}
where the last two terms correspond to the SUPG formalism and the remaining terms are due to the classical (unstable) Finite-Element formulation. The function $\tau_{SUPG}$ regulates the amount of numerical diffusivity and depends on the local Reynolds ($Re_h$) number as:
\begin{equation}
	\tau_{SUPG} = \frac{\xi(Re_h) h_T}{2 |\uu|}, \;\;\;
	\xi(Re_h) = \left\{\begin{matrix}
    Re_h/3 & Re_h \leq 3 \\ 
    1      & Re_h >    3
    \end{matrix}\right. , \;\;\; 
    Re_h = \frac{|\uu| h_T}{2 \nu}
\end{equation} 
where the function $\xi(Re_h)$ is constant for high Reynolds number, saturating this way the amount of numerical dissipation introduced. The parameter $h_T$ indicates the local element size and is taken here as $h_T = \sqrt{2} A/h_T^{max}$, minimizing the numerical dissipation for higly elongated mesh elements (see \citet{mittal2000performance}).

\section{Incompressibility and BFGS}\label{apd:BFGS}

We notice as well that, even if the gradient has been modified through the Hessian matrix, its divergence (in the case of the volume-force correction $\tffx$) seems to be null. To be able to prove this, we need to show that the application of $\mathcal{H}_{n+1}^{-1}$ onto any divergence-free vector $\mathbf{z}$ (here, the gradient) is also divergence-free. To do so, we use induction, where we suppose it is true at some iteration $n$, i.e., $\mathcal{D}\mathcal{H}_n^{-1} \mathbf{z} = 0 $ (true for $n=0$ since $\mathcal{H}_0 = \mathcal{I}$), where $\mathcal{D}$ is the discrete-version of the divergence operator. We notice that the linear transformation involving the Cholesky decomposition presented before does not change this argument since it can be incorporated in the definition of $\mathcal{D}$. If we suppose that $\mathcal{D} \mathbf{s}_n = 0$, we have that:
\begin{equation}
    \begin{split}
        \mathcal{D} \mathcal{H}_{n+1}^{-1} \mathbf{z} & = \mathcal{D} \left( I - \frac{\mathbf{s}_n \mathbf{y}_n^T}{\mathbf{y}_n^T \mathbf{s}_n} \right) \mathcal{H}_n^{-1} \left( \mathcal{I} - \frac{\mathbf{y}_n \mathbf{s}_n^T}{\mathbf{y}_n^T \mathbf{s}_n} \right) \mathbf{z} + \mathcal{D} \frac{\mathbf{s}_n \mathbf{s}_n^T}{\mathbf{y}_n^T \mathbf{s}_n} \mathbf{z} \\
        & = \mathcal{D} \mathcal{H}^{-1}_n \left( \mathcal{I} - \frac{\mathbf{y}_n \mathbf{s}_n^T}{\mathbf{y}_n^T \mathbf{s}_n} \right) \mathbf{z} \\
        & = \mathcal{D} \mathcal{H}^{-1}_n \mathbf{z} - \mathcal{D} \mathcal{H}_n^{-1} \frac{\mathbf{y}_n \mathbf{s}_n^T}{\mathbf{y}_n^T \mathbf{s}_n} \mathbf{z} = 0
    \end{split}
\end{equation}
which is zero since both terms on the last equation are applications of $\mathcal{D} \mathcal{H}_n^{-1}$ onto divergence-free vectors.

\bibliography{mainbib}

\end{document}